\documentclass[10pt]{article}
\usepackage{amssymb,geometry,graphicx,setspace, scalefnt, verbatim, amsfonts,bm,float,booktabs,multirow,amsmath,mathtools,color,arydshln,subcaption}
\usepackage{amsthm}
\usepackage[left]{lineno}
\usepackage[round]{natbib}
\usepackage[colorlinks=true,
    linkcolor=blue,
    citecolor=blue,
    filecolor=magenta,
    urlcolor=blue,
    pdfborder={0 0 0}]{hyperref}
\usepackage{enumerate}
\usepackage{dsfont}
\usepackage{cancel}
\usepackage{mathrsfs}
\usepackage{authblk}
\usepackage{pifont}

\usepackage[ruled,vlined,linesnumbered]{algorithm2e}

\newcommand{\dd}{\mathop{}\!\mathrm{d}}

\allowdisplaybreaks
\newcommand{\SetAlgoItemize}{\setlength{\leftmargini}{1em}}

\title{\vspace{-1em}Bayesian nonparametric models for zero-inflated count-compositional data using ensembles of regression trees}
\author[1]{Andr\'{e} F. B. Menezes}
\author[2]{Andrew C. Parnell}
\author[1]{Keefe Murphy}

\affil[1]{\small Hamilton Institute and Department of Mathematics and Statistics, Maynooth University}
\affil[2]{\small School of Mathematics and Statistics, Insight Centre for Data Analytics, University College Dublin, Ireland}

\date{}

\begin{document}
\maketitle
\onehalfspacing

\begin{center}
\vspace{-1em}
\textbf{Abstract}\vspace{0.1cm}
\end{center}
Count-compositional data arise in many different fields, including high-throughput sequencing experiments, ecological surveys, and palaeoclimate studies, where a common, important goal is to understand how covariates relate to the observed compositions.
Existing methods often fail to simultaneously address key challenges inherent in such data, namely: overdispersion, an excess of zeros, cross-sample heterogeneity, and complex covariate effects.
To address these concerns, we propose two novel Bayesian models based on ensembles of regression trees. Specifically, we leverage the recently introduced zero-and-$N$-inflated multinomial distribution and assign independent nonparametric Bayesian additive regression tree (BART) priors to both the compositional and structural zero probability components of the model, to flexibly capture covariate effects.
We further extend this by adding latent random effects to capture overdispersion and more general dependence structures among the categories.
We develop an efficient inferential algorithm combining recent data augmentation schemes with established BART sampling
routines.
We evaluate our proposed models in simulation studies and illustrate their applicability through a case study of palaeoclimate modelling.\medskip


\section{Introduction}

Count-compositional data arise in many scientific fields where multivariate counts, which represent the frequencies of outcomes across several mutually exclusive categories, are constrained to sum to a total.
Such data are common in applications ranging from ecology \citep{Billheimer2001} and palaeoclimate research \citep{Vasko2000} to high-throughput sequencing experiments in biology, including microbiome \citep{Fernandes2014} and single-cell \citep{Buttner2021} studies.
In this paper, we use a palynology study as a running example, where counts represent the frequencies with which different pollen species appear in sediment cores \citep{Sweeney2018}.

A common goal in the analysis of count-compositional data is to understand how sample-specific covariates are associated with the observed compositions.
In our case study, there are three covariates which can be seen as proxies for the climate in the locations of the sediment cores at which the pollen counts were collected.
Similar questions arise in other domains where the compositional count responses depend on environmental or experimental variables.
In real-world applications, count-compositional data often present several key challenges.
They are typically high-dimensional, involving a large number of categories and/or covariates and, relative to the multinomial sampling distribution, exhibit an excess of zeros, more complex dependence structures, and overdispersion.
Moreover, it may be desirable to make minimal assumptions about the functional form of the covariates' effects, while still capturing nonlinearities and interactions.

This article introduces two count-compositional models based on flexible Bayesian nonparametric priors that simultaneously address these challenges.
We begin by reviewing existing methods and highlighting their limitations, which motivates our development of novel modelling strategies.
We then summarise our main contributions.

\subsection{Literature review}\label{sec:lit_review}

To account for the aforementioned features of count-compositional data, several methods have been proposed using compound multinomial models, i.e., treating the compositional probabilities $\bm{\theta}$ as random variables.
The most well-known approach assumes a conjugate Dirichlet prior for $\bm{\theta}$, which yields the Dirichlet-multinomial distribution \citep[DM;][]{Mosimann1962}.
To overcome the restrictive negative dependence structure imposed by the covariance matrix of the DM distribution, a common strategy is to assign a logistic-normal distribution for $\bm{\theta}$ \citep{Aitchison1980}, defined via the log-ratio transformation of $\bm{\theta}$.
This formulation leads to the multinomial logistic-normal (MLN) model.

Motivated by microbiome studies with a large number of covariates,
\citet{Chen2013} and \citet{Xia2013} proposed sparse group penalty methods for variable selection using log-linear DM and MLN regression models, respectively.
\citet{Wadsworth2017} proposed a Bayesian log-linear DM regression model with spike-and-slab priors.
To deal with having many taxa,
\citet{Grantham2020} proposed an MLN regression model with both fixed and random effects and a common factor analysis hyperprior.
\citet{Ren2017} further defined a dependent Dirichlet process as a prior for the multinomial probabilities and introduced dependence through the shrinkage-inducing factor analysis hyperprior of \citet{Bhattacharya2011}.
\citet{Ren2020} subsequently introduced an extension to accommodate linear covariate effects. \citet{Pedone2023} proposed a hierarchical subject-specific log-linear DM regression model using varying coefficients, which can capture nonlinear and interaction effects, but still requires pre-specification of the spline bases.
More recently, \citet{Ascari2025} introduced the extended flexible Dirichlet-multinomial regression model by compounding the
compositional probabilities with a finite mixture of Dirichlet components, sharing some constraints on their parameters.

Despite these developments, a common limitation of the above methods is that the assumed relationships between the covariates and the compositional probabilities must be pre-specified through parametric linear functional forms.
While convenient, these assumptions can be restrictive and fail to capture nonlinear and complex covariate effects, leading to model misspecification.
Methods developed within palaeoclimate research aiming to model the relationship between climate covariates on compositional pollen counts have by contrast adopted smooth functional forms.
\citet{Haslett2006} employed a log-linear DM model with Gaussian process (GP) priors defined on a two-dimensional climate space.
\citet{Tipton2019} builds on the \citet{Haslett2006} framework, but approximates the GPs as linear combinations of basis function expansions.
However, the current \textsf{R} implementation of the method of \citet{Tipton2019} supports only one covariate, thereby limiting broader applicability.
In the microbiome context, \citet{Aijo2018} considered the MLN model with GP priors for modelling the microbiome compositions over time.
However, their model does not otherwise account for covariate effects.
\citet{Silverman2022} subsequently proposed scalable Bayesian inference strategies for a broader class of MLN models, including the MLN-GP with covariates as a special case.

All of the above methods implicitly assume that all categories (e.g., microbes or pollen taxa) are present and that zeros occur only due to sampling variability.
To account for sparsity, some of these models have latent variables which shrink the individual-level probabilities towards zero when the observed counts are zero.
Consequently, the estimated probabilities are never exactly zero and remain strictly positive, even when the true probability of occurrence is zero.
However, not all zeros are the same, and ignoring them can have an impact on inference \citep{Moreno2019,Koslovsky2023}.
Rather than implicitly assuming that excess zeros always arise from sampling variability,
a common strategy is to employ zero-inflated mixture models, which introduce latent variables that classify the observed zeros into structural zeros, corresponding to cases where the event cannot occur, and at-risk (sampling) zeros, corresponding to cases where the event may occur but a zero count is observed \citep{Moreno2019}.

There are many models within this paradigm with explicit components to capture structural zeros in multivariate count data.
In the field of palaeoclimate research, \citet{Salter-Townshend2012} and \citet{Sweeney2012} considered a nested hierarchical structure for pollen taxa, using univariate zero-inflated distributions for the lowest hierarchical levels.
This approach, however, relies on the correct specification of an externally imposed hierarchical structure.
This sensitivity limits its applicability when such information is unavailable or unreliable.
Other researchers, instead, ignore the compositional nature of the data and fit a series of independent univariate zero-inflated models \citep[see, e.g.,][among others]{Jiang2019,Shuler2021}.
Another common solution to modelling multivariate count data is to combine univariate zero-inflated models through shared latent random effects that capture the dependence structure among categories \citep[see, e.g.,][]{Lee2018}.
However, these methods are inappropriate for count-compositional data, as they fail to account for the fact that multinomial counts are inherently constrained by a total.

Recently, there has been a growing interest in developing zero-inflated models which respect the compositional nature of multivariate counts.
\citet{Zeng2023} proposed a zero-inflated MLN with probabilistic principal component analysis (ZIMLN-PCA) model by incorporating a latent indicator to account for zero-inflation in the MLN model with a PCA covariance structure.
\citet{Koslovsky2023} proposed the zero-inflated Dirichlet-multinomial (ZIDM) model by modifying the Dirichlet prior formulation via normalised independent gamma random variables augmented with point masses at zero.
An appealing feature of ZIDM is that covariates are linked to both the compositional and structural zero probabilities. However, under the framework of \citet{Koslovsky2023}, linear functional forms are assumed for both parts of the model, albeit with an extension incorporating spike-and-slab priors for variable selection.
The zero-inflated generalised Dirichlet-multinomial regression model \citep{Tang2019}, which predates the ZIDM model but omits the spike-and-slab priors, relaxes the restrictive assumption of negative dependence between categories that characterises the multinomial and DM distributions.
Despite these advances, the existing zero-inflated count-compositional models generally require pre-specification of the parametric functional forms for the covariate effects on both the compositional and structural zero probabilities, potentially limiting their flexibility when such assumptions are misspecified.

\subsection{Our contributions}

While existing methods address problem-specific aspects of count-compositional data motivated by their respective application domains, they are incapable of addressing all aforementioned key challenges simultaneously, which may limit their broader applicability.
More specifically, existing approaches that flexibly accommodate covariate effects through GP priors \citep[see, e.g.,][]{Haslett2006,Tipton2019,Silverman2020} fail to explicitly model structural zeros, which commonly arise in microbiome and palaeoclimate applications. On the other hand, currently available zero-inflated mixture count-compositional models, such as the ZIDM model \citep{Koslovsky2023}, assume linear functional forms for the covariate effects on both the count and structural zero components.
Although attractive due to interpretability, this may be a restrictive assumption rarely satisfied in practice.
Our contribution fills this gap by developing Bayesian nonparametric models based on ensembles of regression trees that explicitly model structural zeros, accommodate overdispersion and more complex dependence structures, and flexibly capture the effects of covariates.

Our approach builds on the zero-and-$N$-inflated multinomial (ZANIM) distribution which was introduced and characterised as a finite mixture distribution by \citet{Menezes2025}.
This probabilistic framework is designed to account for zero-inflation in count-compositional data and the related phenomenon of $N$-inflation, which refers to zeros co-occurring in all but one category.
The ZANIM distribution has two sets of category-specific parameters, which describe the population-level count and structural zero probabilities.
In this paper, our proposed models extend ZANIM by assigning independent nonparametric  Bayesian additive regression trees (BART) priors to both components, to flexibly incorporate covariate effects.
Specifically, we employ the multinomial logistic BART prior of \citet{Murray2021} for the former and the probit BART prior of
\citet{Chipman2010} for the latter.
Our framework thus accommodates nonlinear covariate effects and interactions within both population-level components without requiring pre-specification of the functional forms, thereby capturing a wide range of data generating processes.

We refer to our proposed extension as the ZANIM logistic BART (ZANIM-BART) model.
Conceptually, this model extends the multinomial logistic BART (ML-BART) model of \citet{Murray2021} through the inclusion of category-specific BART priors for the structural zero probabilities.
Doing so helps to mitigate bias in the estimated covariate effects on the compositional probabilities.
To further accommodate overdispersion and capture more complex dependencies among the categories and across the samples, while still capturing zero-inflation, we also develop the ZANIM logistic-normal BART (ZANIM-LN-BART) model, which incorporates multivariate logistic-normal random effects on the compositional probabilities.
A special case of particular interest is when the structural zero components are omitted from ZANIM-LN-BART; in doing so, we obtain the MLN-BART model, which to the best of our knowledge has not previously appeared in the literature.

As far as we are aware, our use of BART is novel in count-compositional settings.
The success of BART in recovering unknown smooth functions and low-order interactions has been demonstrated empirically and theoretically \citep{Linero2018b,Rockova2020}.
In addition, \citet{Linero2017} established theoretical guarantees showing that BART approaches a GP prior as the number of trees tends to infinity.
This connection is appealing, as models with GP priors are more computationally demanding and scale poorly with increasing sample sizes.
In fact, as discussed by \citet{Linero2017}, BART may achieve equivalent or better empirical performance than a GP with less computational time.
Supported by these results, we thus adopt BART as a nonparametric prior in our novel ZANIM-BART and ZANIM-LN-BART models to provide flexible frameworks suitable for different count-compositional data analysis settings, such as our case study in palaeoclimate research.
The complexity of jointly using distinct ensembles of trees for both the compositional and structural zero probabilities, and doing so for each category, is mitigated through our adaptation of the data augmentation scheme developed for the ZANIM distribution \citep{Menezes2025}.
This methodological contribution enables the development of an efficient Markov chain Monte Carlo (MCMC) algorithm with minor modifications of the established BART sampling routines of \citet{Chipman2010} and \citet{Murray2021}.

The rest of this paper is structured as follows.
Section \ref{sec:methods} reviews relevant background and describes the novel methodological aspects of our models, along with their associated inference schemes.
Section \ref{sec:simulations} presents several simulation studies, which demonstrate that our proposed models accurately recover the compositional and structural zero probabilities in the presence of complex covariate effects, even when the data generating process is misspecified.
We illustrate our models by analysing the count-compositional data from a palynology experiment in Section \ref{sec:application}.
We show that ZANIM-LN-BART provides the best fit to the data, capturing its main features while also uncovering rich insights into the effects of climate on both population-level components of the model.
We conclude in Section \ref{sec:discussion} with a discussion and defer additional results to the Supplementary Material.
In particular, we also present an additional analysis of a benchmark data set from a study of the human gut microbiome.
Code with the implementations of our models is available from \url{https://github.com/AndrMenezes/zanicc}, along with code to reproduce our simulation studies and real data analyses.

\section{Methods}\label{sec:methods}

We assume that we observe a random sample of size
$n$ collected in the count matrix $\mathbf{Y}\in \mathbb{N}_0^{n\times d}$, where
$n$ denotes the number of samples and $d$ the number of categories. Each row
$\mathbf{Y}_i = (Y_{i1}, \ldots, Y_{id})$ records the observed frequencies of the $d$
mutually exclusive outcomes across $N_i$ trials in sample $i$.
Here, we consider the case where the total $N_i$ is fixed.
Formally, the random vector $\mathbf{Y}_i \in \mathbb{S}^d_{N_i}$ lies in the discrete simplex space
constrained to an individual-specific total count $N_i$, where
$\smash{\mathbb{S}^d_{N_i} = \{\mathbf{Y}_i \in (0, \ldots, N_i)^d; \sum_{j=1}^d Y_{ij} = N_i\}}$.
The multinomial distribution is the natural probabilistic model for each
observation, i.e., $\mathbf{Y}_i \sim \operatorname{Multinomial}_d\left\lbrack N_i, \theta_1, \ldots, \theta_d\right\rbrack$,
where the parameter vector
$\bm{\theta} = (\theta_1, \ldots, \theta_d)$ corresponds to the
population-level count (compositional) probabilities and lies on the continuous simplex space
$\smash{\mathbb{S}^d=\{\bm{\theta}\in\mathbb{R}^d; \theta_j > 0, \sum_{j=1}^d \theta_j=1\}}$.

Estimation of $\bm{\theta}$ is crucial as it provides the  information underlying the observed counts.
Researchers often seek to estimate the associations between these compositional probabilities and the available covariates, $\mathbf{x}_i = (x_{i1}, \ldots, x_{ip})$.
Typically, each $\theta_j$ is related to covariates through a suitable link function with a parametric linear functional form. Although convenient, this assumption imposes strong restrictions that limit the range of data generating processes the model can capture.
Moreover, count-compositional data with even a moderate number of categories $d$ (e.g., our palynology case study, where $d=28$) are usually sparse and overdispersed, and exhibit more complex dependencies than the multinomial distribution allows.

To address these challenges, we introduce two nonparametric ensemble of regression trees models based on the ZANIM distribution.
Section \ref{sec:zanim_distribution} reviews the ZANIM distribution and proposes an extension, which we refer to as the ZANIM logistic-normal (ZANIM-LN) distribution, to further capture overdispersion and complex dependences.
We briefly review nonparametric BART priors, on which both of our models rely, in Section \ref{sec:bart_review}.
Our novel models are then formulated in Section \ref{sec:zanim_ln_bart}, where we assign BART priors on the compositional and structural zero probability parameters of the ZANIM and ZANIM-LN distributions.
We develop data augmentation techniques for efficient sampling in Section \ref{sec:augmented_likelihood}, discuss prior specifications in Section \ref{sec:priors}, and provide a novel MCMC algorithm to perform posterior inference in Section \ref{sec:posterior_inference}.

\subsection{The ZANIM and ZANIM-LN distributions}\label{sec:zanim_distribution}

The main challenge in dealing with excess zeros in multivariate count-compositional settings, compared to univariate cases, is that the zero-inflation can occur in a single category or across multiple categories.
Extreme cases where zeros co-occur in all but one category are said to exhibit $N$-inflation.
In our recent research \citep{Menezes2025}, we provided new probabilistic characterisations of zero-inflated extensions of the multinomial and DM distributions, deriving their probability mass functions (PMF) and other key statistical properties.
The proposed framework represents both distributions as finite mixture distributions with $2^d$ components, which share a total of $2d$ parameters. In particular, the zero-and-$N$-inflated multinomial (ZANIM) distribution introduced by \citet{Menezes2025},
is characterised by the following stochastic representation:
\begin{align}
(z_{ij} \mid \zeta_j) &\overset{\operatorname{ind.}}{\sim} \operatorname{Bernoulli}\lbrack 1 - \zeta_j \rbrack, \quad j\in\{1,\ldots, d\},
\label{eq:zanim_stochastic_representation}
\\ \nonumber
(\mathbf{Y}_i \mid N_i, \bm{\theta}, \mathbf{z}_i) &\sim
\begin{cases} \delta_{\mathbf{0}_d}(\cdot), & \textrm{if} \: z_{ij} = 0 \: \forall\: j,\\
\operatorname{Multinomial}_d\left\lbrack N_i,\vartheta_{i1}, \ldots, \vartheta_{id}\right\rbrack,
& \textrm{otherwise},
\end{cases}
\end{align}
where $\delta_{\mathbf{0}_d}(\cdot)$ denotes the Dirac measure with unit mass at $\mathbf{0}_d$ and
\[\vartheta_{ij} = \frac{z_{ij}\theta_j}{\sum_{k=1}^d z_{ik}\theta_k}, \quad j\in \{1,\ldots,d\},\]
such that the individual-level count probabilities $\bm{\vartheta}_i=(\vartheta_{i1}, \ldots, \vartheta_{id})$ are
functions of the category-specific parameters $\theta_j$ and $\zeta_j$, which represent the population-level count
and structural zero probabilities, respectively. Notably, $\bm{\vartheta}_i$
exhibits spikes at zero induced by the latent at-risk indicators $\mathbf{z}_i$,
which in turn depend \textit{a priori} on the $\zeta_j$ parameters.
We stress that $\bm{\vartheta}_i$ describe within- and between-subject heterogeneity, while $\bm{\theta}$ characterises the counts at a global level.

The ZANIM distribution, due to its finite mixture nature, can handle overdispersion and (unlike the multinomial distribution) allow for both positive and negative covariances between the categories \citep[see][for further details on its PMF and its statistical properties]{Menezes2025}.
We note that \citet{Menezes2025} also provided a finite mixture representation and derived
the PMF and statistical properties for the zero-and-$N$-inflated Dirichlet-multinomial (ZANIDM) distribution.
This distribution was first introduced, in the form of a stochastic representation only, by
\citet{Koslovsky2023}, who referred to it as the zero-inflated Dirichlet-multinomial (ZIDM).

Moving beyond \citet{Menezes2025}, we propose the ZANIM logistic-normal (ZANIM-LN) distribution as an extension of ZANIM, which incorporates multivariate logistic-normal random effects \citep{Aitchison1980}.
Specifically, we introduce the latent variables $\mathbf{u}_i \sim \operatorname{Normal}_d\left\lbrack \bm{0}, \bm{\Sigma}_U \right\rbrack$ through the parameterisation $\vartheta_{ij} = z_{ij}\alpha_{j} e^{u_{ij}}/\sum_{k=1}^dz_{ik}\alpha_{k} e^{u_{ik}}$, where $\alpha_j > 0$ is a category-specific concentration parameter. The corresponding population-level count probabilities are obtained by normalising $\alpha_j$, i.e., $\theta_{j}=\alpha_j/\sum_{k=1}^d \alpha_k$.
Clearly, ZANIM-LN is richer than ZANIM since $\vartheta_{ij}$ inherits the additional subject-specific variability of the random effects $u_{ij}$.
We note that the formulation of the ZANIM-LN distribution has similarities with the ZIMLN-PCA model of \citet{Zeng2023}.
However, as we will discuss in Section \ref{sec:priors}, we address identifiability of the covariance matrix in a different fashion, using a factor-analytic hyperprior for the random effects that allows for parsimonious specification of the covariance matrix $\bm{\Sigma}_U$, rather than the PCA approach of \citet{Zeng2023}.
This enables the model to more flexibly capture category-specific overdispersion, which can help to account for sampling zeros, in addition to the explicit mixture components in ZANIM for handling structural zeros.

The ZANIM-LN distribution has some special cases of interest.
When $\bm{\zeta}=\mathbf{0}_d$, we recover the MLN model \citep{Grantham2020,Silverman2022}, albeit under a slightly different parameterisation which separates the linear predictor into fixed, category-specific concentration parameters $\alpha_j$, which enter multiplicatively on the probability scale, and Gaussian random effects $u_{ij}$ which are centered at zero, rather than having the location of the linear predictor be absorbed into the random effects.

When $\mathbf{u}_i=\mathbf{0}_d\:\forall\:i$, we obtain the aforementioned ZANIM distribution, and when both the random effects and zero-inflation parameters are all fixed at zero, we recover the usual multinomial distribution.
In Supplementary Material \ref{supp:zanim_ln_distribution}, we compare the marginal PMFs and theoretical moments of the ZANIM, ZANIM-LN, and ZANIDM distributions under parameter settings where all three distributions share the same expectations.
We note that ZANIM-LN accommodates more overdispersed counts while capturing zero- and $N$-inflation under the finite mixture framework of \citet{Menezes2025}.

\subsection{Overview of the nonparametric BART priors}\label{sec:bart_review}

We now present a brief review and establish notation for the nonparametric BART prior introduced by
\citet{Chipman2010} and the log-linear BART extension proposed by \citet{Murray2021}.
These priors underpin the novel Bayesian models for count-compositional developed in the subsequent sections.
In BART, an unknown function of interest, $f(\mathbf{x}_i)$, is represented as a sum of $m$ decision trees
$f(\mathbf{x}_i) = \sum_{h=1}^{m}g(\mathbf{x}_i, \mathcal{T}_h, \mathcal{M}_h)$,
where each decision tree function $g(\mathbf{x}_i, \mathcal{T}_h, \mathcal{M}_h)$ is parametrised
by a binary tree structure $\mathcal{T}_h$ consisting of the tree topology
with terminal and internal nodes denoted by $\mathcal{L}_h$ and $\mathcal{B}_h$, respectively,
and terminal node parameters $\mathcal{M}_h$.
For each internal node $b\in \mathcal{B}_h$, there is a spitting rule of the form
$\lbrack x_{j_{b}} \leq c_b\rbrack$. The terminal node parameters are defined as
$\mathcal{M}_h = \{\mu_{ht}\colon t \in \mathcal{L}_h\}$, where $\mu_{ht} \in \mathbb{R}$.
The tree structure $\mathcal{T}_h$, represented by a sequence of decision rules,
induces a partition $\{\mathcal{A}_{h1}, \ldots, \mathcal{A}_{hb_h}\}$ of the covariate space
$\mathcal{X}$, each element of which corresponds to a terminal node of the tree.
Given $(\mathcal{T}_h, \mathcal{M}_h)$, the regression tree function can thus be expressed as a step function given by
$g(\mathbf{x}_i, \mathcal{T}_h, \mathcal{M}_h)= \mathds{1}(\mathbf{x}_i \in \mathcal{A}_{ht})\mu_{ht},$
for $t\in \mathcal{L}_h$.

The number of trees $m$ in the ensemble is considered as a specified hyperparameter with typical large values
of $50$, $100$, or $200$ often giving reasonable performance
\citep[see, e.g.,][]{Chipman2010,Bleich2014,Sparapani2023}.
The BART prior on $f(\mathbf{x}_i)$
is formulated to impose regularisation and prevent overfitting.
It is assumed that the trees are independent and the terminal nodes are conditionally independent given the trees, i.e.,
$\pi(\bm{\mathcal{T}}, \bm{\mathcal{M}}) = \prod_{h=1}^m\pi_\mathcal{T}(\mathcal{T}_h)\pi_{\mathcal{M}}(\mathcal{M}_h \mid \mathcal{T}_h)$.
For $\pi_\mathcal{T}(\mathcal{T}_h)$, the most common choice to control the tree depth, which we adopt here, is the
branching process prior proposed by \citet{Chipman1998}. For the splitting rules $\lbrack x_{j_{b}} < c_b\rbrack$ associated to each internal node $b \in \mathcal{B}_h$,
the following procedure is suggested:
(i) sample a covariate index $j_{b}$ uniformly from $\{1,\ldots, p\}$ and then
(ii) sample a split value $c_b$ uniformly from those that yield a valid partition.
If no such valid splitting rule exists, the node is forced to be terminal.
For the terminal node parameters $\mathcal{M}_h$, it is assumed that
$\pi_{\mathcal{M}}(\mathcal{M}_h \mid \mathcal{T}_h)=\prod_{t \in \mathcal{L}_h} \pi_\mu(\mu_{ht})$,
where $\pi_\mu$ is chosen so that it is conditionally conjugate.

\citet{Chipman2010} proposed an inference procedure for the BART parameters $\{(\mathcal{T}_h, \mathcal{M}_h)\}_{h=1}^m$
using MCMC methods with a tailored version of Bayesian backfitting \citep{Hastie2000}.
Conditional on $\{(\mathcal{T}_\ell, \mathcal{M}_\ell)\}_{\ell \neq h}$,
a block update on $(\mathcal{T}_h, \mathcal{M}_h)$ is performed,
which consists of two steps: (i) update the tree topology
$\mathcal{T}_{h}$ from the conditional distribution of
$(\mathcal{T}_{h} \mid \mathcal{T}_{(h)}, \mathcal{M}_{(h)}, \ldots)$
using a Metropolis-Hastings (MH) step with transition kernels proposed by \citet{Chipman1998},
(ii) sample the terminal node parameters $\mathcal{M}_h$ from their
full conditional distribution $\pi(\mathcal{M}_h \mid \mathcal{T}_{h}, \mathcal{T}_{(h)}, \mathcal{M}_h, \ldots)$.
See \citet{Chipman2010} for further details of the sampling strategy and an extension which incorporates a probit link function for binary responses.

The log-linear BART prior of \citet{Murray2021} reparameterises the terminal node parameters via $\lambda_{ht} = e^{\mu_{ht}}$, where $\lambda_{ht}>0$, such that
$\Lambda_h = \{\lambda_{ht}\colon t \in \mathcal{L}_h\}$.
Thus, the prior can be expressed as
$\log\left\lbrack f(\mathbf{x}_i) \right\rbrack = \sum_{h=1}^{m}\log\left\lbrack g\left(\mathbf{x}_i; \mathcal{T}_h, \Lambda_h\right)\right\rbrack
= \sum_{h=1}^{m}\log(\lambda_{ht})\mathds{1}(\mathbf{x}_i \in \mathcal{A}_{ht})$, where $t \in \mathcal{L}_h$.
According to
\citet{Murray2021}, this prior can be used for any probability distribution which has a (possibly augmented) likelihood
of the form
\begin{equation}\label{eq:likelihood_form_log_bart}
\mathscr{L}\left(\bm{\mathcal{T}}, \bm{\Lambda}; \mathbf{y}, \mathbf{x}, \ldots\right)
= \prod_{i=1}^n \upsilon_i f(\mathbf{x}_i)^{\kappa_i}\exp\left\lbrack \nu_i f(\mathbf{x}_i)\right\rbrack,
\end{equation}
where $\upsilon_i$, $\kappa_i$, and $\nu_i$ are functions of data $\mathbf{y}$ or latent variables.
In particular, \citet{Murray2021} shows that, after appropriate data augmentation, the multinomial
distribution admits an augmented likelihood that factorises
across the categories and takes the form
in \eqref{eq:likelihood_form_log_bart} for each category.
\citet{Murray2021} refers to this model as multinomial logistic BART (ML-BART) and evaluates its performance in
multi-classification settings, where $N_i=1$.
Under the wider log-linear BART framework,
the prior for the parameters $\{(\mathcal{T}_h, \Lambda_h)\}_{h=1}^m$ has
the same independence assumption across the trees as \citet{Chipman2010} and thus factorises via
$\pi(\mathcal{T}_h, \Lambda_h) = \pi_\mathcal{T}(\mathcal{T}_h)\pi_{\Lambda}(\Lambda_h \mid \mathcal{T}_h)$,
where $\pi_{\Lambda}(\Lambda_h \mid \mathcal{T}_h)=\prod_{t \in \mathcal{L}_h} \pi_\lambda(\lambda_{ht})$
and $\pi_\Lambda$ is chosen to maintain conditional conjugacy, in the spirit of the standard BART.

\subsection{The ZANIM-BART and ZANIM-LN-BART models}\label{sec:zanim_ln_bart}

Building on the BART frameworks above, and the zero-inflated count-compositional distributions described in Section \ref{sec:zanim_distribution}, we adopt the log-linear BART prior of \citet{Murray2021} and the probit BART prior of \citet{Chipman2010} to propose two novel nonparametric models: ZANIM-BART and ZANIM-LN-BART.
Specifically, we assume the log-linear BART prior of \citet{Murray2021} for the logistic transformation of $\bm{\theta}_i$, for both models, such that the population-level count probability of category $j$ becomes covariate-dependent via
\begin{equation}\label{eq:zanim_ln_bart_theta}
\theta_{ij} = \frac{f_j^{(\mathrm{c})}(\mathbf{x}_i)}{\sum_{k=1}^d f_k^{(\mathrm{c})}(\mathbf{x}_i)},
\quad
\log f_j^{(\mathrm{c})}(\mathbf{x}_i) =
\sum_{h=1}^{m_\theta}\log\left\lbrack g\left(\mathbf{x}_i; \mathcal{T}^{(\mathrm{c})}_{hj},
\Lambda_{hj}\right)\right\rbrack, \end{equation}
where $\smash{\mathcal{T}^{(\mathrm{c})}_{hj}}$ denotes the $h$-th binary tree topology for the $j$-th category and $\Lambda_{hj} = \{\lambda_{htj}\colon \mathcal{L}^{(\mathrm{c})}_{hj} \}$ is the corresponding set of terminal node parameters.
Identifiability of the category-specific regression trees $f_j^{(\operatorname{c})}$ can be obtained by setting $f_k^{(\operatorname{c})} = 1$ for a given reference category $k$.
However, we instead use proper priors on $f_j^{(\operatorname{c})}$ and work in the unidentified parameter space, which \citet{Murray2021} showed to have some computational benefits for the ML-BART model.

We further consider the probit BART prior of \citet{Chipman2010} for $\bm{\zeta}$, such that the covariate-dependent structural zero probability for category $j$ is given by
\begin{equation}\label{eq:zanim_ln_bart_zeta}
\zeta_{ij} = \Phi\left( f^{(0)}_j(\mathbf{x}_i) \right) = \Phi\left\lbrack \sum_{h=1}^{m_\zeta} g\left(\mathbf{x}_i; \mathcal{T}^{(0)}_{hj}, \mathcal{M}_{hj}\right)
\right\rbrack, \end{equation}
where $\Phi(\cdot)$ is the standard normal cumulative distribution function, $\smash{\mathcal{T}^{(0)}_{hj}}$ denotes the $h$-th tree for the structural zero component of category $j$, and
$\smash{\mathcal{M}_{hj} = \{\mu_{htj}\colon \mathcal{L}^{(0)}_{hj} \}}$ is the corresponding set of terminal node parameters.
It is worth noting that our models assign independent BART priors $\smash{f^{(c)}_j}$ and $\smash{f^{(0)}_j}$, for all categories $j\in\{1,\ldots,d\}$, to both the population-level count probability and structural zero probability parameters of the ZANIM and ZANIM-LN distributions.
This has the effect of making the individual-level count probabilities covariate-dependent also.
Under ZANIM-LN-BART, these are given by $\vartheta_{ij} \propto z_{ij}e^{u_{ij}}f_j^{(\mathrm{c})}(\mathbf{x}_i)$, where $z_{ij} \overset{\operatorname{ind.}}{\sim} \operatorname{Bernoulli}\lbrack 1 - f_j^{(0)}(\mathbf{x}_i) \rbrack$ and the normalisation is with respect to a sum over the categories.
As before, such quantities can be obtained under ZANIM-BART by setting $\mathbf{u}_i=\mathbf{0}_d\:\forall\:i$.
Consequently, our proposed models generalise the ML-BART model of \citet{Murray2021} in two directions.
The ZANIM-BART model introduces independent BART priors $\smash{f_j^{(0)}}$ for the structural zero probabilities and the ZANIM-LN-BART model further extends this framework by incorporating subject-specific Gaussian random effects.
Another special case of interest arising from ZANIM-LN-BART is the multinomial logistic-normal BART (MLN-BART) model, which is obtained by setting $\bm{\zeta}_i=\mathbf{0}_d \:\forall\: i$.

These novel extensions of the ML-BART model are important to help capture different features of real count-compositional data, as we will demonstrate in the simulations and the application in Sections \ref{sec:simulations} and \ref{sec:application}, respectively.
Crucially, the novel models allow the individual-level count probabilities to be decomposed into different population-level components.
The proposed ZANIM-BART and ZANIM-LN-BART models both have two population-level parameters, $\theta_{ij}$ and $\zeta_{ij}$, which each depend on their respective category-specific regression trees, $f_j^{(\mathrm{c})}(\mathbf{x}_i)$ and $f_j^{(0)}(\mathbf{x}_i)$, and govern the category-specific count and structural zero probabilities, respectively.
Under the ZANIM-BART model, $\vartheta_{ij}$ are explained by explicit structural zero components, $z_{ij}$, and both sets of category-specific regression trees.
Under the ZANIM-LN-BART model, $\vartheta_{ij}$ is further explained by unobserved latent characteristics, $u_{ij}$.
In contrast, the ML-BART model is limited to providing inference only on $\theta_{ij}$, at the population-level, through the category-specific regression trees $f_j^{(\mathrm{c})}(\mathbf{x}_i)$.
Unlike the MLN-BART, ZANIM-BART, and ZANIM-LN-BART models, ML-BART evidently does not have subject-specific latent variables of any kind; neither the structural zero indicators, $z_{ij}$, nor the random effects, $u_{ij}$.

\subsection{Likelihood formulations}\label{sec:augmented_likelihood}

We now derive the augmented likelihood function for the ZANIM-LN-BART model.
This derivation plays a fundamental role in our methodological contribution, as it enables the development of an efficient MCMC algorithm for sampling the model parameters.
The corresponding augmented likelihood function for the ZANIM-BART model can be obtained by setting the random effect terms $u_{ij}$ to $0$ in the expressions which follow.

Let $\smash{\bm{f}^{(\mathrm{c})} = (f_1^{(\mathrm{c})}, \ldots, f_d^{(\mathrm{c})})}$ and $\smash{\bm{f}^{(0)} = (f_1^{(0)}, \ldots, f_d^{(0)})}$ denote the independent BART priors which characterise the ZANIM-LN-BART model defined in \eqref{eq:zanim_ln_bart_theta} and \eqref{eq:zanim_ln_bart_zeta}.
We build on \eqref{eq:zanim_stochastic_representation} using two steps of data augmentation.
First, we adapt the data augmentation scheme proposed by \citet[see Supplementary Material S.1 for details]{Menezes2025} for the ZANIM distribution, by introducing the sample-specific latent variables
\begin{equation}\label{eq:full_conditional_phi_i}
\left(\phi_i \mid \mathbf{y}_i, \mathbf{z}_i, \mathbf{u}_i, \bm{f}^{(\mathrm{c})} \right) \overset{\operatorname{ind.}}{\sim}
\operatorname{Gamma}\left\lbrack N_i, \sum_{j=1}^d z_{ij}f_j^{(\mathrm{c})}(\mathbf{x}_i)e^{u_{ij}}\right\rbrack.
\end{equation}
This construction yields the following augmented likelihood for $\smash{\bm{f}^{(\mathrm{c})}}$ and $\smash{\bm{f}^{(0)}}$:
\begin{align}
\mathscr{L}\left(\bm{f}^{(\mathrm{c})}, \bm{f}^{(0)}; \mathbf{y}, \mathbf{x}, \mathbf{u}, \mathbf{z}, \bm{\phi}\right)
&\propto
\prod_{i=1}^n
\prod_{j=1}^d
\left\{
\left\lbrack \Phi\left(f^{(0)}_j(\mathbf{x}_i)\right)\right\rbrack^{1-z_{ij}}
\left\lbrack 1 - \Phi\left( f^{(0)}_j(\mathbf{x}_i)\right)\right\rbrack^{z_{ij}}
\right. \label{eq:augmented_likelihood_zanim_ln_bart} \\[-1em] \nonumber
&\phantom{\propto\prod_{i=1}^n\prod_{j=1}^d}\times
\left.
\left\lbrack f_j^{(\mathrm{c})}(\mathbf{x}_i)\right\rbrack^{y_{ij}}
e^{-\phi_i z_{ij}e^{u_{ij}}f_j^{(\mathrm{c})}(\mathbf{x}_i)}
\right\},
\end{align}
which notably factorises over the categories $j\in\{1,\ldots,d\}$. Importantly, the BART priors $\smash{f_j^{(\mathrm{c})}}$ and $\smash{f_j^{(0)}}$ are conditionally independent given the data and the latent variables and the terms involving $f_j^{(\mathrm{c})}$ admit a form analogous to the generic likelihood function of the log-linear BART prior in \eqref{eq:likelihood_form_log_bart}.

In the second step, to simplify the terms of the zero-inflation component in \eqref{eq:augmented_likelihood_zanim_ln_bart}, we follow the
probit BART
model of \citet{Chipman2010} and apply the data augmentation scheme of \citet{Albert1993}.
However, we stress that, unlike \citet{Chipman2010}, the binary response itself, $z_{ij}$, is also latent in our case.
Specifically, we introduce further latent variables $w_{ij}$, where $\smash{w_{ij} \sim \operatorname{TN}_{\lbrack -\infty, 0 \rbrack}\lbrack f_j^{(0)}(\mathbf{x}_i), 1 \rbrack}$ when $z_{ij}=1$ and
$\smash{w_{ij} \sim \operatorname{TN}_{\lbrack 0, \infty\rbrack}\lbrack f_j^{(0)}(\mathbf{x}_i), 1 \rbrack}$ when $z_{ij}=0$ (which corresponds to a structural zero), where $\operatorname{TN}_{\lbrack a, b\rbrack}\lbrack\mu,\sigma^2\rbrack$ denotes the truncated normal distribution on the interval $\lbrack a, b \rbrack$ with location $\mu$ and scale parameter $\sigma$.
The resulting augmented likelihood takes the form
\begin{equation}\label{eq:augmented_likelihood_zanim_ln_bart_further}
\mathscr{L}\left(\bm{f}^{(\mathrm{c})}, \bm{f}^{(0)}; \mathbf{y}, \mathbf{x}, \mathbf{u}, \mathbf{z}, \mathbf{w}, \bm{\phi}\right)
\propto
\prod_{i=1}^n
\prod_{j=1}^d
\left\{
\varphi\left(w_{ij}; f^{(0)}_j(\mathbf{x}_i), 1\right)
\left\lbrack f_j^{(\mathrm{c})}(\mathbf{x}_i)\right\rbrack^{y_{ij}}
e^{-\phi_i z_{ij}e^{u_{ij}}f_j^{(\mathrm{c})}(\mathbf{x}_i)}
\right\},
\end{equation}
where $\varphi(x; \mu, \sigma^2)$ is the Gaussian probability density function
with mean $\mu$ and variance $\sigma^2$.
Finally, we note that the full conditional distribution of the latent variables $z_{ij}$, given
$y_{ij}$, $u_{ij}$, $\phi_{i}$, $\smash{f^{(\mathrm{c})}_j}$, and $\smash{f^{(0)}_j}$, follows
\begin{equation}\label{eq:full_conditional_z_ij}
\left(z_{ij} \mid \ldots\right) \overset{\operatorname{ind.}}{\sim}
\begin{cases}
\operatorname{Bernoulli}\left\lbrack
\dfrac{
\left\lbrack1- \Phi\left( f^{(0)}_j(\mathbf{x}_i)\right) \right\rbrack e^{-\phi_iu_{ij}f^{(\mathrm{c})}_j(\mathbf{x}_i)}
}
{
\Phi\left( f^{(0)}_j(\mathbf{x}_i)\right) + \left\lbrack 1- \Phi\left( f^{(0)}_j(\mathbf{x}_i)\right) \right\rbrack
e^{-\phi_i u_{ij}f^{(\mathrm{c})}_j(\mathbf{x}_i)}}\right\rbrack, & \textrm{if}~y_{ij} = 0,\\
1, & \textrm{otherwise}.
\end{cases}
\end{equation}
\indent These factorisations play a central role in our MCMC schemes, discussed later in Section  \ref{sec:posterior_inference}, as they allow us to sample the parameters $\smash{\{\mathcal{T}^{(\mathrm{c})}_{jh}, \Lambda_{jh}\}}$ and $\smash{\{\mathcal{T}^{(0)}_{jh}, \mathcal{M}_{jh}\}}$ using established BART sampling routines. Further details are provided in Supplementary Material \ref{supp:inference_zanim_ln}.

\subsection{Prior specifications}\label{sec:priors}

A key aspect of the BART framework is the adoption of regularisation priors which shrink the  contribution of each terminal node to the final prediction produced by the overall additive ensemble.
This prevents any one tree from dominating and helps to mitigate against overfitting.
This principle also applies to the extensions of BART which we use as nonparametric priors in our ZANIM-BART and ZANIM-LN-BART models.

Recall that both models incorporate category-specific regression trees for both the compositional and structural zero probabilities, with corresponding parameters $\{\mathcal{T}^{(\mathrm{c})}_{hj}, \Lambda_{hj}\}_{h=1}^{m_{\theta}}$ and $\{\mathcal{T}_{hj}^{(0)}, \mathcal{M}_{hj}\}_{h=1}^{m_{\zeta}}$.
For the priors on the tree topologies $\smash{\mathcal{T}^{(\mathrm{c})}_{hj}}$ and $\smash{\mathcal{T}^{(\mathrm{c})}_{hj}}$, we adopt the branching process prior introduced by \citet{Chipman1998}, with the hyperparameter values recommended by \citet{Chipman2010}.
To sample a covariate index for forming the splitting rules, we sample uniformly from the $p$ available predictors with probability $\mathbf{s} = (s_1, \ldots, s_p)$, where $s_k = 1/p\:\forall\:k\in\{1\ldots,p\}$.
Alternatively, in the additional microbiome application in Supplementary Material \ref{supp:human_gut_microbiome}, which incorporates a large number of covariates, we instead adopt the sparsity-inducing prior of \citet{Linero2018} which assumes $\mathbf{s} \sim \operatorname{Dirichlet}\lbrack \omega/p, \ldots, \omega/p \rbrack$, with an additional hyperprior on $\omega/(\omega + \rho) \sim \operatorname{Beta}\lbrack a_\omega, b_\omega\rbrack$, where we follow \citet{Linero2018} and set $a_\omega=0.5$, $b_\omega=1$, and $\rho=p$.

Given the tree topologies $\smash{\mathcal{T}^{(\mathrm{c})}_{hj}}$ and $\smash{\mathcal{T}^{(0)}_{hj}}$, we follow \citet{Murray2021} and \citet{Chipman2010} by assuming conditionally conjugate priors for $\{\Lambda_{hj}\}_{h=1}^{m_\theta}$ and $\{\mathcal{M}_{hj}\}_{h=1}^{m_\zeta}$.
For the terminal node parameters $\{\Lambda_{hj}\}_{h=1}^{m_\theta}$ of the BART priors associated with the compositional probabilities, we consider $\smash{\lambda_{htj} \overset{\operatorname{ind.}}{\sim} \operatorname{Gamma}\lbrack c_0, d_0\rbrack}$ and calibrate this in accordance with the recommendations of \citet{Murray2021} for the multinomial logistic BART model.
Specifically, we consider that each $\smash{f_j^{(\mathrm{c})}}$ has the same number of trees $m_\theta$ and, through the normal approximation of the log-odds between two distinct categories, obtain the hyperparameters $c_0$ and $d_0$ in such a way that $\mathbb{E}\lbrack \lambda_{htj} \rbrack = 0$ and $\operatorname{Var}\lbrack \lambda_{htj}\rbrack = a^2_\lambda / m_\theta$, where $a_\lambda$ is a tuning parameter.
Solving for $c_0$ and $d_0$, we obtain expressions in terms of the digamma and trigamma functions: $d_0 = \exp\{ (\ln \Gamma)^\prime(c_0)\}$ and $(\ln \Gamma)^{\prime\prime}(c_0) = a^2_\lambda / m_{\theta}$, where the last equation is computed using the Newton-Raphson algorithm.
Although the choice of $a_\lambda = 3.5/\sqrt{2}$ suggested by \citet{Murray2021} works reasonably well in practice, we follow \citet{Linero2020} by assuming the hyperprior $a_\lambda \sim \operatorname{half-Cauchy}\lbrack 0, 1\rbrack$ and updating $a_\lambda$ using slice sampling, in order to allow the model to determine the amount of variability according to the data.

Regarding the terminal node parameters $\{\mathcal{M}_{hj}\}_{h=1}^{m_\zeta}$ in the BART priors describing the structural zero probabilities, the conjugate priors are $\smash{\mu_{htj} \overset{\operatorname{ind.}}{\sim} \operatorname{Normal}\lbrack 0, \sigma^2_\mu\rbrack}$.
We proceed as per \citet{Chipman2010} by shrinking each category-specific $\smash{f_j^{(0)}}$ towards zero by setting $\sigma_\mu = 0.5 / (k\sqrt{m_\zeta})$, where $k=2$ is such that $\smash{f_j^{(0)}}$ has high prior probability of lying in $(-3.0, 3.0)$.
For the numbers of trees related to the population-level count probabilities, $m_\theta$, we follow the recommendation of \citet{Murray2021} and set $m_\theta=100$ trees per category.
Our default recommendation for the the number of trees for the structural zero probabilities is $m_\zeta=100$.
This is justified as a balance between computational efficiency and overall performance.
In practice, we observed a negligible effect on the parameter recovery in the simulations presented in Section \ref{sec:simstudy2} when using smaller number of trees for the structural zero components.
Consequently, when computational efficiency is a priority we recommend using a smaller value, such as $m_\zeta=20$.

The ZANIM-LN-BART model has multivariate Gaussian random effects,
$\mathbf{u}_i$, with $d$-dimensional covariance matrix $\mathbf{\Sigma}_U$.
However, $\mathbf{\Sigma}_U$ is singular because the parameters $\theta_{ij}$ in \eqref{eq:zanim_ln_bart_theta}
lie on the $d$-dimensional continuous simplex space $\mathbb{S}_d$.
To address this identifiability issue, we use a sum-to-zero constraint and transform
$\mathbf{u}_i$ into the $(d-1)$-dimensional vector $\mathbf{v}_i$ by setting
$\mathbf{u}_i=\mathbf{B}\mathbf{v}_i$, where $\mathbf{B}$ is $d\times (d-1)$ orthogonal matrix,
such that $\operatorname{Var}\left\lbrack \mathbf{u}_i \right\rbrack = \bm{\Sigma}_U = \mathbf{B} \bm{\Sigma}_V \mathbf{B}^\top$.
We conclude the specification of ZANIM-LN-BART model by placing a factor-analytic hyperprior on the latent variables
$\mathbf{v}_i$ via the decomposition
$\mathbf{v}_i = \bm{\Gamma} \bm{\eta}_i + \bm{\epsilon}_i$,
where $\bm{\Gamma} = \left\lbrack \gamma_{jk}\right\rbrack_{j=1,\ldots,d-1}^{k=1,\ldots,q}$ is a
$(d-1) \times q$ factor loadings matrix,
$\bm{\eta}_i \sim \operatorname{Normal}_q\lbrack \mathbf{0}_q, \mathbf{I}_q\rbrack$, and
$\bm{\epsilon}_i \sim \operatorname{Normal}_{d-1}\left\lbrack\mathbf{0}_{d-1}, \bm{\Psi}\right\rbrack$, with
$\bm{\Psi}=\operatorname{diag}(\psi_1, \ldots, \psi_{d-1})$.
Regarding the specification of $q$, we fix an upper limit for the number of loadings columns as the Ledermann bound
\citep{Anderson1956} and shrink the contributions of the redundant loadings columns by employing the nonparametric
multiplicative gamma process shrinkage prior of \citet{Bhattacharya2011}.
By doing so, we obviate the need to pre-specify a fixed number of latent factors. This prior has been adopted in a wide range of settings,
including model-based clustering \citep{Murphy2020}, Bayesian partial least squares regression \citep{Urbas2024}, and the modelling of count-compositional data \citep{Ren2017}.
See the Supplementary Material \ref{supp:inference_zanim_ln} for further details of the corresponding full conditional distributions.

Notably, dimension-reduction of latent multivariate Gaussian random effects has also been considered in other modelling frameworks designed for count-compositional data.
In particular, the ZIMLN-PCA model of \citet{Zeng2023} employs probabilistic principal component analysis, which can be recovered under our approach by setting $\Psi = \psi\mathbf{I}_{d-1}$. Our allowance for uncommon idiosyncratic variances is designed to better capture category-specific levels of excess variability.
Our approach also differs via the imposition of a sum-to-zero constraint, which treats all categories equally, whereas \citet{Zeng2023} address the non-identifiability of the covariance matrix via an inherently asymmetric additive log-ratio transformation, which requires an arbitrary reference category.

\subsection{Posterior inference}\label{sec:posterior_inference}

With the likelihood and priors formulated, we now derive our MCMC algorithm by leveraging
the Bayesian backfitting and generalised Bayesian backfitting algorithms of \citet{Chipman2010} and \citet{Murray2021}
to sample $\smash{\{\mathcal{T}^{(\mathrm{c})}_{hj}, \Lambda_{hj}\}_{h=1}^{m_\theta}}$
and $\smash{\{\mathcal{T}^{(0)}_{hj}, \mathcal{M}_{hj}\}_{h=1}^{m_\zeta}}$, respectively.
This is facilitated by the factorisation of the augmented likelihood in
\eqref{eq:augmented_likelihood_zanim_ln_bart_further}, which allows conditionally independent updates to be carried out across the categories $j\in\{1,\ldots,d\}$, for each $\smash{f_j^{(\mathrm{c})}}$ and $\smash{f_j^{(0)}}$.
Again, we derive the algorithm for ZANIM-LN-BART and recall that ZANIM-BART is
recovered as a special case by removing the random effects.
To implement these Bayesian backfitting schemes, we require the integrated likelihood functions for the trees and
the full conditional distributions of the corresponding terminal node parameters.

We begin by deriving these quantities for the tree ensembles $\smash{f_j^{(\mathrm{c})}}$ associated with the compositional probabilities.
Let $\smash{\mathcal{T}^{(\mathrm{c})}_{(h)j}}$ and $\Lambda_{(h)j}$ denote the tree topologies and the terminal node parameters for all trees of category $j$, excluding the $h$-th.
Following \citet{Murray2021}, we denote by
$\smash{f^{(\mathrm{c})}_{(h)j}}(\mathbf{x}_i) = \prod_{\ell \neq h} g(\mathbf{x}_i; \mathcal{T}^{(\mathrm{c})}_{\ell j}, \Lambda^{(\mathrm{c})}_{\ell j})$ the fit from all but the $h$-th tree of category $j$.
By integrating out $\Lambda_{hj}$ from the augmented likelihood in \eqref{eq:augmented_likelihood_zanim_ln_bart} using their
conditionally conjugate priors, it can be shown that
\begin{equation}\label{eq:integrated_likelihood_lambdas}
\pi\left(\mathcal{T}^{(\mathrm{c})}_{hj} \mid \mathcal{T}^{(\mathrm{c})}_{(h)j}, \Lambda_{(h)j}, \mathbf{y},
\mathbf{z}, \mathbf{u}, \bm{\phi}\right)
\propto
\pi_{\mathcal{T}}\left(\mathcal{T}^{(\mathrm{c})}_{hj}\right)
\prod_{t \in \mathcal{L}^{(\mathrm{c})}_{hj}}
\frac{d_0^{c_0}}{\Gamma(c_0)}
\frac{\Gamma\left(r^{(\mathrm{c})}_{htj} + c_0\right)}{(s^{(\mathrm{c})}_{htj} + d_0)^{r^{(\mathrm{c})}_{htj}+c_0}},
\end{equation}
where $\smash{r^{(\mathrm{c})}_{htj}}=\sum_{i\colon\mathbf{x}_i \in \mathcal{A}^{(\mathrm{c})}_{htj}} y_{ij}$
and $\smash{s^{(\mathrm{c})}_{htj}}=\sum_{i\colon\mathbf{x}_i \in \mathcal{A}^{(\mathrm{c})}_{htj}}\phi_i z_{ij} e^{u_{ij}} f^{(\mathrm{c})}_{(h)j}(\mathbf{x}_i)$.
Likewise, it follows that the full conditional distributions of the terminal node parameters
$\Lambda_{hj}$, are
\begin{equation}\label{eq:full_conditional_lambdas}
\left(
\lambda_{thj} \mid
\mathcal{T}^{(\mathrm{c})}_{hj},  \mathcal{T}^{(\mathrm{c})}_{(h)j}, \Lambda_{(h)j}, \mathbf{y}, \mathbf{z}, \bm{\phi}
\right)
\overset{\operatorname{ind.}}{\sim}\operatorname{Gamma}\left\lbrack r^{(\mathrm{c})}_{htj} + c_0, s^{(\mathrm{c})}_{htj} + d_0\right\rbrack,
\quad t \in  \mathcal{L}^{(\mathrm{c})}_{hj}.
\end{equation}

Similarly, for the tree ensembles $\smash{f_j^{(0)}}$ associated with the structural zero probabilities,
we let $\smash{\mathcal{T}^{(0)}_{(h)j}}$ and $\mathcal{M}_{(h)j}$ denote the tree topologies
and corresponding terminal node parameters for all trees, except for
the $h$-th, for category $j$. Following \citet{Chipman2010},
let $r_{(h)ij} \equiv w_{ij} - \sum_{\ell \neq h}\smash{g(\mathbf{x}_i; \mathcal{T}^{(0)}_{\ell j}, \mathcal{M}_{\ell j})}$ be the
partial residuals, such that $\smash{(\mathcal{T}^{(0)}_{(h)j}, \mathcal{M}_{(h)j})}$ only depends on the data through
$\mathbf{r}_{(h)j}=(r_{(h)1j}, \ldots, r_{(h)nj})$, where
$\smash{r_{(h)ij} \sim \operatorname{Normal}\lbrack g(\mathbf{x}_i; \mathcal{T}^{(0)}_{hj}, \mathcal{M}_{hj} ), 1\rbrack}$.
After integrating out $\mathcal{M}_{hj}$ from the augmented likelihood in \eqref{eq:augmented_likelihood_zanim_ln_bart}, using their conditionally conjugate priors, we obtain
\begin{equation}\label{eq:integrated_likelihood_mus}
\pi\left(\mathcal{T}^{(0)}_{hj} \mid \mathbf{r}_{(h)j}\right)
\propto
\pi_{\mathcal{T}}\left(\mathcal{T}^{(0)}_{h}\right)
\prod_{t \in \mathcal{L}^{(0)}_{hj}}
\left(\dfrac{1}{n_{htj}\sigma_\mu^2 + 1}\right)^{1/2}\!
\exp\left\lbrack \dfrac{1}{2} \left(  \dfrac{\sigma^2_\mu\left(s^{(0)}_{htj}\right)^2}{2(n_{htj}\sigma^2_\mu + 1)} \right)\right\rbrack,
\end{equation}
where $\smash{s^{(0)}_{htj}} = \sum_{i\colon \mathbf{x}_i \in \mathcal{A}^{(0)}_{htj}}r_{(h)ij}$
and $n_{htj}$ is the total number of observations in the partition $\smash{\mathcal{A}^{(0)}_{htj}}$.
It can similarly be shown that the full conditional distributions of the terminal node parameters are
\begin{equation}\label{eq:full_conditional_mus}
\left(\mu_{htj} \mid \mathcal{T}^{(0)}_{hj}, \mathbf{r}_{(h)j} \right) \overset{\operatorname{ind.}}{\sim} \operatorname{Normal}\left\lbrack
\dfrac{s^{(0)}_{htj}}{n_{htj} + \sigma^2_\mu}, 1 / (n_{htj} + 1 / \sigma_\mu^2)\right\rbrack, \quad t \in \mathcal{L}^{(0)}_{hj}.
\end{equation}
\indent Sampling from the tree topologies, $\smash{\mathcal{T}^{(\mathrm{c})}_{h}}$ and $\smash{\mathcal{T}^{(0)}_{h}}$,
is carried out via MH using the respective integrated likelihood functions
in \eqref{eq:integrated_likelihood_lambdas} and \eqref{eq:integrated_likelihood_mus}.
The tree proposals are generated through transition kernels
defined as mixtures of birth, death, and change moves \citep[for more details, see][]{Kapelner2016}. Finally, we emphasise that the updates from the full conditional distributions of the latent variables $\phi_i$ given in \eqref{eq:full_conditional_phi_i}, the latent zero-inflation indicators $z_{ij}$ given in \eqref{eq:full_conditional_z_ij}, and
the truncated normal latent variables $w_{ij}$, form part of our MCMC algorithm.
Additional sampling steps are required for ZANIM-LN-BART, where we update the
random effects using elliptical slice sampling \citep{Murray2010} and update
the parameters of the factor-analytic hyperprior via closed-form Gibbs steps.
Further details, along with pseudocode describing the MCMC algorithm for ZANIM-LN-BART, are provided in Supplementary Material \ref{supp:inference_zanim_ln}.
Convergence assessment is also discussed in Supplementary Material \ref{supp:convergence}, where it is shown that all models evaluated throughout this paper exhibit satisfactory convergence.

\section{Simulation studies}\label{sec:simulations}

We perform extensive simulations to compare the performance of the proposed ZANIM-BART and ZANIM-LN-BART models against existing methods under different data generating mechanisms.
The first scenario illustrates the ability of our models to recover nonlinear relationships between covariates and both the category-specific compositional and structural zero probabilities.
This scenario also demonstrates that models which ignore the structural zeros can lead to biased estimates of the compositional probabilities.
The second scenario, in which the number of compositional elements ($d$) and the sample size ($n$) are varied, stresses parameter estimation and the scalability of our models in realistic count-compositional settings.
We compare our ZANIM-BART and ZANIM-LN-BART models with existing Bayesian models, including BART-based special cases of our models:
ML-BART \citep{Murray2021}, MLN-BART, DM regression (DM-reg), ZANIDM regression (ZANIDM-reg) \citep{Koslovsky2023}, and MLN-GP \citep{Silverman2022}.

To fit the MLN-GP model, we use the function \texttt{basset} from the \textsf{R} package \texttt{fido} with its default arguments and the squared exponential kernel.
For the remaining models, we provide implementation via our own \textsf{R} package \texttt{zanicc}, which is available from the repository at \url{https://github.com/AndrMenezes/zanicc}. For efficient computation, the core parts of the MCMC algorithms to fit the models are written in \textsf{C++}.
We use our own implementation of the ZIDM model of \citet{Koslovsky2023} without the spike-and-slab priors, which we refer to as the ZANIDM-reg model in reference to its finite mixture representation and modified inference scheme developed in \citet{Menezes2025}.
Our implementation differs from that of \citet{Koslovsky2023} in several other aspects. In particular: for the zero-inflation part, we employ the probit link function along with the data augmentation approach of \citet{Albert1993}, rather than the logit link function; for the the count part, we sample from the regression coefficients using elliptical slice sampling \citep{Murray2010}, instead of Metropolis-Hastings steps with univariate random walk proposals; finally, we implement conjugate Gibbs updates for the latent variables as proposed by \citet{Menezes2025}.
For all models, MCMC chains were run for $15{,}000$ iterations, and the first $10{,}000$ iterations were discarded as burn-in.
All hyperparameters of the ZANIM-BART and ZANIM-LN-BART models are specified as described in Section \ref{sec:priors} and we adopt the same hyperparameter values, where applicable, for the ML-BART and MLN-BART models.

We assess the estimation performance of the methods by quantifying the discrepancy between the true values of the population-level count probabilities,
structural zero probabilities,
and individual-level count probabilities,
and their corresponding posterior mean estimates $\widehat{\theta}_{ij}, \widehat{\zeta}_{ij}$ and $\widehat{\vartheta}_{ij}$. Given that $\bm{\theta}_{i}$ lies on the simplex space, we compute the average Kullback–Leibler (KL) divergence:
$\operatorname{KL}(\bm{\theta}) = n^{-1} \sum_{i=1}^n \sum_{j=1}^d \theta_{ij} \log(\theta_{ij} / \widehat{\theta}_{ij})$.
As the individual-level probabilities $\bm{\vartheta}_{i}$ also lie on the simplex, but may take values equal to zero, we carefully modify the contribution of some terms when calculating the analogous $\operatorname{KL}(\bm{\vartheta})$.
When $\vartheta_{ij}=0$, we set its contribution equal to zero.
When $\vartheta_{ij} > 0$ and $\widehat{\vartheta}_{ij}=0$, we replace the value of $\widehat{\vartheta}_{ij}$ with $1.0$.
For $\zeta_{ij}$, we compute the overall KL divergence by averaging over the corresponding category-specific KL divergences,  i.e., we define $\operatorname{KL}(\bm{\zeta}) = d^{-1} \sum_{j=1}^d \operatorname{KL}(\bm{\zeta}_j)$, where $\operatorname{KL}(\bm{\zeta}_j)=n^{-1}\sum_{i=1}^n\left\lbrack \zeta_{ij} \log(\zeta_{ij}/\widehat{\zeta}_{ij}) + (1-\zeta_{ij}) \log((1-\zeta_{ij}) / (1-\widehat{\zeta}_{ij}))\right\rbrack$.
Additionally, we assess the uncertainty of the methods using the empirical coverage probability (CP) of the $95\%$ credible intervals for the same set of parameters, which we also average over the categories.
We stress that some of the competing models adopt slightly different definitions for the population-level count, individual-level count, and population-level structural zero probabilities, depending on the type of random effects (Dirichlet, logistic-normal, or none) and whether or not they incorporate structural zero components.
However, we continue to use the notation $\bm{\theta}$, $\bm{\vartheta}$, and $\bm{\zeta}$ to refer to these respective quantities, for consistency.
See Table \ref{tab:compound_multinomial_models} in Supplementary Material \ref{supp:inference_zanim_ln} for further details.
Furthermore, we note that some of these models lack one or more of these quantities. In particular, the ML-BART model only estimates population-level count probabilities, $\theta_{ij}$, and the models without explicit structural zero components clearly do not estimate $\bm{\zeta}$.

\subsection{Simulations: Scenario 1} \label{sec:simstudy1}

To obtain preliminary insights and illustrate key features of our methods, we present a first simulated scenario which shows the efficacy of ZANIM-BART and ZANIM-LN-BART in terms of parameter recovery when both the population-level count probabilities $(\theta_{ij})$ and structural zero probabilities $(\zeta_{ij})$ depend on a covariate through nonlinear functional forms.

Specifically, we consider $d=4$ categories, $n=400$ observations, and, for maximum simplicity, only one covariate $x_i$, with values placed uniformly over $\lbrack -1, 1\rbrack$.
The compositional probabilities, $\theta_{ij}=\alpha_{ij}/\sum_{k=1}^d\alpha_{ik}$, and structural zero probabilities, $\zeta_{ij}$, were related to the covariate $x_i$ through the following nonlinear functional forms:
$\log \alpha_{i1} = 5\cos(\pi x_i)$, $\log \alpha_{i2} = 1.5\cos(2 \pi x_i)$, and $\log \alpha_{i3} = -2 x_i^2$,
along with
$\zeta_{i1}=\Phi(e^{-5 x^2_{i}} - 1.5)$, $\zeta_{i2}=\Phi( x_i - 2 (x_i -0.5)^2- 1.5)$, $\zeta_{i3}=\Phi(-2x_i + 3x_i^3- 1.5)$, and $\zeta_{i4}=\Phi(3x_i - 2 x_i^3- 1.5)$.
The total counts, $N_i$, were sampled from a discrete uniform distribution defined on the interval $\lbrack 100, 500\rbrack$.
Conditional on the parameters $N_i$, $\theta_{ij}$, and $\zeta_{ij}$, we generate three data sets by simulating the compositional counts, $\mathbf{Y}_i$, under the ZANIM, ZANIDM, and ZANIM-LN distributions.
Under these DGPs, the proportion of zeros across the four categories are approximately $0.2600, 0.1325, 0.1150$, and $0.2075$, of which each split into structural and sampling zeros with the proportions $(0.1375, 0.1225)$, $(0.0700, 0.0625)$, $(0.0800, 0.0350)$, and $(0.1650, 0.0425)$, respectively.
For the ZANIM-LN DGP, we set the covariance matrix through the factor-analytic decomposition $\bm{\Sigma}_U = \bm{\Gamma}\bm{\Gamma}^\top + \Psi$, where the factor loadings matrix $\bm{\Gamma}$ has $q=2$ factors, with entries simulated from the standard uniform distribution, and the covariance of the idiosyncratic errors is $\bm{\Psi} = \operatorname{diag}(0.32, 0.33, 0.34, 0.35)$.
Further details on how to simulate counts under these distributions using their stochastic representations are given in Supplementary Material \ref{supp:zanim_ln_distribution}.
We further include a fourth DGP, where the counts are simulated from a distribution without structural zero components.
In particular, we generate data from the MLN distribution with all parameters, excluding $\bm{\zeta}=\mathbf{0}_d$, set as per the ZANIM-LN DGP described above.
The proportions of sampling zeros across the four categories are $0.1400$, $0.0800$, $0.0675$, and $0.0600$.
Our aim with this DGP is to verify the robustness of our models under misspecification, when there are only sampling zeros.

Results for this scenario are presented in Table \ref{tab:summary_sim_1}.
Overall, we find that the ZANIM-LN-BART model exhibits consistently lower KL values and CP values close to the nominal level under all four DGPs.
While the ZANIM-BART model outperforms the competing models under the ZANIM DGP, which is expected as the assumed underlying distribution aligns with the true data generating mechanism, its performance slightly declines under the DGPs with extra latent heterogeneity (ZANIDM and ZANIM-LN).
Notably, when structural zeros are present in the data, the models which do not explicitly account for them (ML-BART, MLN-BART, MLN-GP, and DM-reg) perform poorly in recovering $\theta_{ij}$.
This is reflected in substantially large $\operatorname{KL}(\bm{\theta})$ values relative to the ZANIM-BART and ZANIM-LN-BART models, as well as $\operatorname{CP}(\bm{\theta})$ values that deviate from the nominal level.
Although these models (excluding ML-BART) can, to some extent, provide reasonable point estimates for the individual-level probabilities --- as indicated by their relatively small values of $\operatorname{KL}(\bm{\vartheta})$ --- their uncertainty quantification remains inadequate.
Specifically, since these models cannot produce exactly zero estimates for $\vartheta_{ij}$, their $\operatorname{CP}(\bm{\vartheta})$ values are below the nominal value.
Although the ZANIDM-reg model also allows for covariate-dependent, category-specific structural zero components, its performance is consistently worse than our proposed models, as it assumes linear functional forms for the covariate effects.
In contrast, the flexibility induced by the BART priors allows our models to adapt to nonlinear structures without requiring explicit functional form specifications.
Finally, under the DGP without structural zeros (MLN), the ZANIM-LN-BART model performs comparably to its special case, the MLN-BART model.
This result suggest that the additional mixture zero-inflation components in ZANIM-LN-BART, introduced to accommodate structural zeros, does not compromise accuracy and uncertainty quantification when structural zeros are absent, while providing substantial gains when they are present.

\begin{table}[!ht]
\centering
\caption{Simulation results for Scenario 1 with $d=4$, $n=400$, and $p=1$ under four different DGPs, each corresponding to a different distributional assumption.
Accuracy in the estimation methods are assessed via the Kullback-Leibler divergence (KL), while uncertainty quantification is
evaluated via the coverage probabilities (CP) of the $95\%$ credible intervals.
The runtimes of the methods are reported in seconds (sec).
Smaller KL values indicate improved recovery, while coverage values close to the nominal level $(0.95)$ indicate well-calibrated
posterior uncertainty. Values in bold indicate the best performing method under the given metric within each DGP, while missing values correspond to non-existence of the given parameter under the given method or DGP.}
\label{tab:summary_sim_1}
\scriptsize
\begin{tabular}{llllllllr}
  \toprule
DGP & Model & $\operatorname{KL}\left(\bm{\theta}\right)$ & $\operatorname{CP}\left(\bm{\theta}\right)$ & $\operatorname{KL}\left(\bm{\vartheta}\right)$ & $\operatorname{CP}\left(\bm{\vartheta}\right)$ & $\operatorname{KL}\left(\bm{\zeta}\right)$ & $\operatorname{CP}\left(\bm{\zeta}\right)$ & Time (sec) \\
  \midrule
\multirow{7}{*}{ZANIM} & ZANIM-BART & \textbf{0.0009} & \textbf{0.9425} & \textbf{0.0011} & \textbf{0.9506} & 0.0073 & 0.9113 & 117.4 \\
   & ZANIM-LN-BART & 0.0010 & 0.9912 & 0.0018 & 0.9850 & \textbf{0.0067} & \textbf{0.9156} & 127.1 \\
   & ML-BART & 0.0827 & 0.4981 &  &  &  &  & 77.8 \\
   & MLN-BART & 0.0917 & 0.5606 & 0.0066 & 0.8419 &  &  & 90.3 \\
   & ZANIDM-reg & 0.5365 & 0.1044 & 0.0058 & 0.9394 & 0.1614 & 0.3881 & 15.6 \\
   & DM-reg & 0.5966 & 0.0806 & 0.0055 & 0.8413 &  &  & 9.7 \\
   & MLN-GP & 0.0637 & 0.2900 & 0.0075 & 0.8425 &  &  & 1.0 \\
   \hdashline
\multirow{7}{*}{ZANIDM} & ZANIM-BART & 0.0018 & 0.8638 & 0.0031 & 0.7788 & 0.0169 & 0.8881 & 105.4 \\
   & ZANIM-LN-BART & \textbf{0.0014} & \textbf{0.9725} & \textbf{0.0026} & \textbf{0.9556} & \textbf{0.0161} & \textbf{0.8900} & 114.3 \\
   & ML-BART & 0.1288 & 0.3919 &  &  &  &  & 70.5 \\
   & MLN-BART & 0.0996 & 0.5500 & 0.0062 & 0.8344 &  &  & 82.5 \\
   & ZANIDM-reg & 0.5207 & 0.1006 & 0.0052 & 0.9419 & 0.1532 & 0.3794 & 11.5 \\
   & DM-reg & 0.6216 & 0.0744 & 0.0052 & 0.8406 &  &  & 9.7 \\
   & MLN-GP & 0.0777 & 0.2675 & 0.0074 & 0.8419 &  &  & 1.0 \\
   \hdashline
\multirow{7}{*}{ZANIM-LN} & ZANIM-BART & 0.0312 & 0.3563 & 0.0514 & 0.4106 & 0.0241 & 0.6687 & 114.6 \\
   & ZANIM-LN-BART & \textbf{0.0052} & \textbf{0.9850} & \textbf{0.0043} & 0.9550 & \textbf{0.0129} & \textbf{0.8988} & 129.8 \\
   & ML-BART & 0.1382 & 0.2712 &  &  &  &  & 76.8 \\
   & MLN-BART & 0.0989 & 0.5700 & 0.0061 & 0.8381 &  &  & 89.2 \\
   & ZANIDM-reg & 0.5466 & 0.0975 & 0.0055 & \textbf{0.9475} & 0.1062 & 0.4644 & 15.5 \\
   & DM-reg & 0.6203 & 0.0719 & 0.0054 & 0.8363 &  &  & 12.0 \\
   & MLN-GP & 0.0825 & 0.3231 & 0.0072 & 0.8456 &  &  & 1.1 \\
   \hdashline
\multirow{7}{*}{MLN} & ZANIM-BART & 0.0230 & 0.3994 & 0.0596 & 0.2888 &  &  & 106.4 \\
   & ZANIM-LN-BART & \textbf{0.0042} & 0.9906 & 0.0047 & \textbf{0.9481} &  &  & 112.3 \\
   & ML-BART & 0.0225 & 0.4250 & & &  &  & 69.8 \\
   & MLN-BART & 0.0051 & \textbf{0.9831} & \textbf{0.0046} & 0.9456 &  &  & 78.4 \\
   & ZANIDM-reg & 0.5474 & 0.0725 & 0.0061 & 0.9275 &  &  & 11.3 \\
   & DM-reg & 0.5835 & 0.0581 & 0.0058 & 0.9444 &  &  & 9.5 \\
   & MLN-GP & 0.0136 & 0.3387 & 0.0050 & 0.9300 &  &  & 1.0 \\
   \bottomrule
\end{tabular}
\end{table}

Regarding the computational cost, for this particular scenario with small sample size $(n=400)$ and few categories $(d=4)$, MLN-GP achieves the fastest runtime (around one second).
This is expected, as its inference scheme employs a bespoke `collapsed-uncollapsed' sampler with a Laplace approximation step \citep{Silverman2020}, while the inference for all other models is fully based on MCMC.
The next fastest methods are the regression-based models, with runtimes ranging from 9.5 seconds (DM-reg) to 15.6 seconds (ZANIDM-reg).
The use of BART priors, particularly in combination with logistic-normal random effects and/or structural zero components, evidently increases the runtime.
However, the most computationally-intensive method here (ZANIM-LN-BART) still remains feasible, with a reasonable maximum runtime of approximately 130 seconds.

\begin{figure}[!ht]
    \centering
    \includegraphics[width=.99\linewidth]{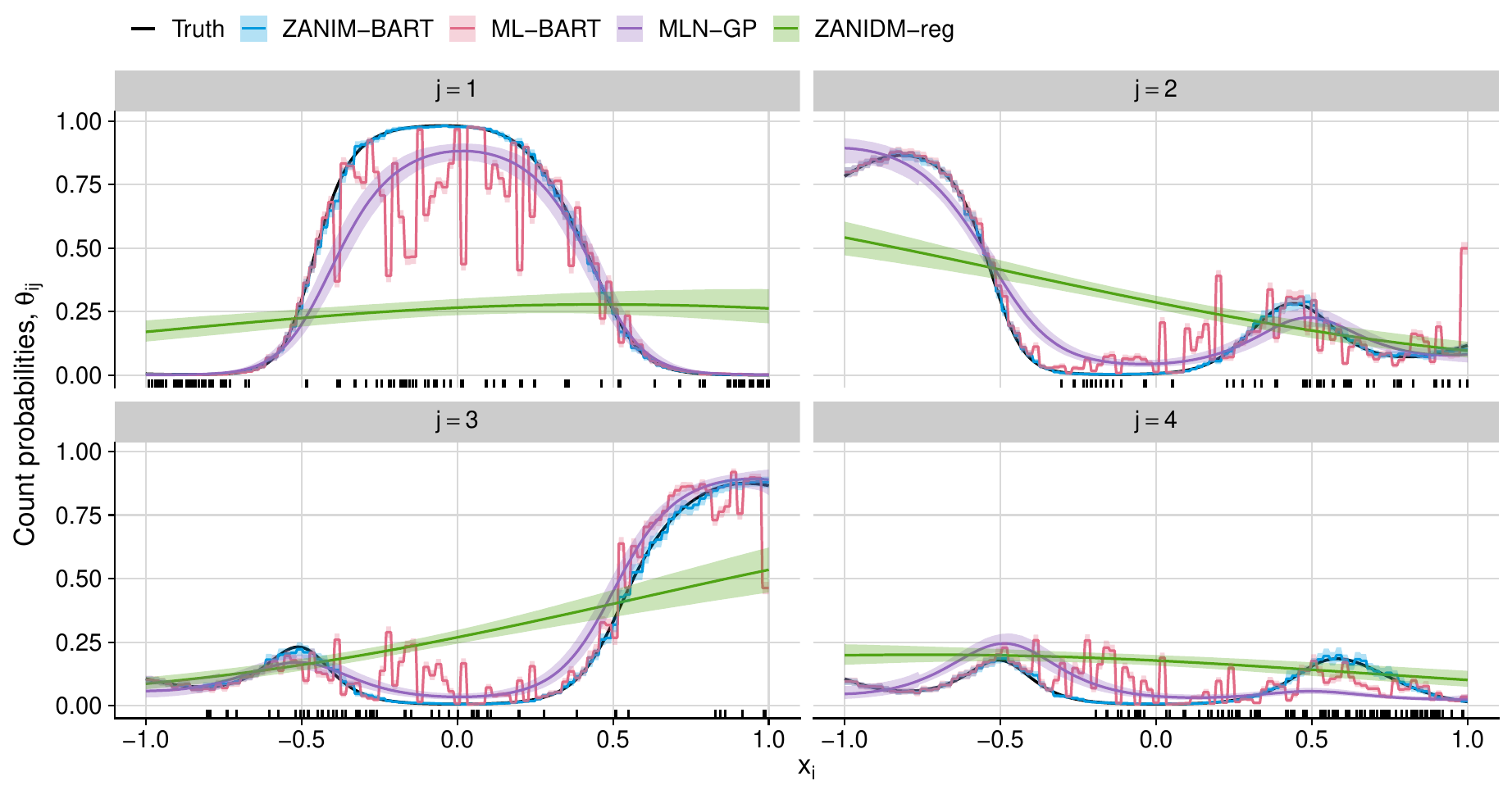}
    \caption{ZANIM-BART, ML-BART, MLN-GP, and ZANIDM-reg estimates of the true population-level count probabilities $\theta_{ij}$ (black lines) for $d=4$ categories under the ZANIM DGP. The posterior median and $95\%$ credible intervals are given in each case. The rugs along the $x$-axes represent samples where the observed counts are zero.}
    \label{fig:count_prob_scenario_1_zanim}
\end{figure}

\begin{figure}[!hb]
    \centering
    \includegraphics[width=.99\linewidth]{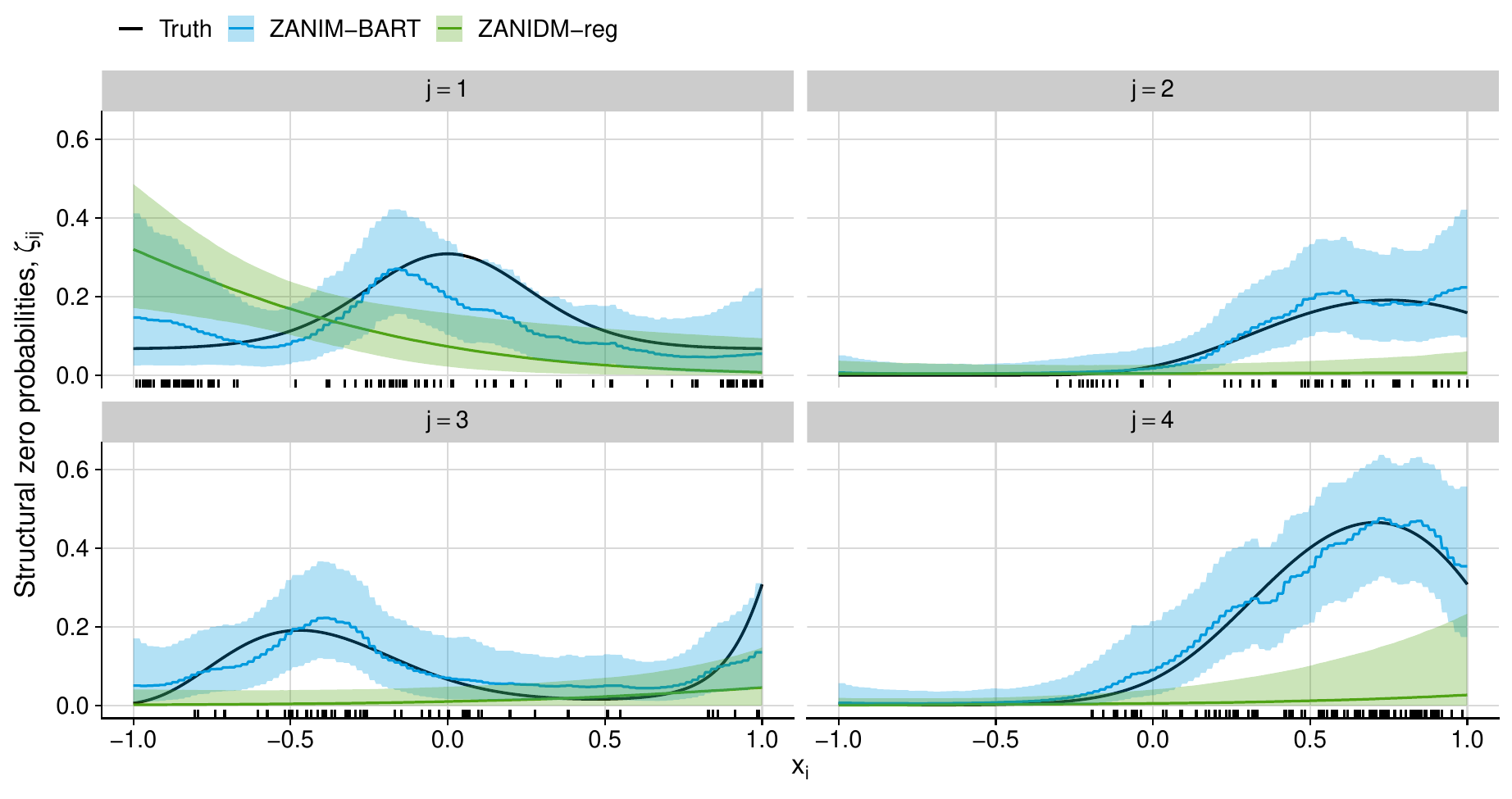}
    \caption{ZANIM-BART and ZANIDM-reg estimates of the true structural zero probabilities $\zeta_{ij}$ (black lines) for $d=4$ categories under the ZANIM DGP. The posterior median and $95\%$ credible intervals are given in each case. The rugs along the $x$-axes represent samples where the observed counts are zero.}
\label{fig:structural_prob_scenario_1_zanim}
\end{figure}

Figures \ref{fig:count_prob_scenario_1_zanim} and \ref{fig:structural_prob_scenario_1_zanim} illustrate the posterior estimates, under selected different models, of the population-level count and structural zero probabilities, respectively, at given predictor values $x_i$, together with the corresponding true values under the ZANIM DGP.
For visualisation purposes, we present results for the ZANIM-BART, ML-BART, MLN-GP, and ZANIDM-reg models only.
Additional results for other models under the ZANIM-LN DGP are deferred to Supplementary Material \ref{supp:add_simstudy1}.
From Figure \ref{fig:count_prob_scenario_1_zanim}, it is evident that ML-BART, MLN-GP, and ZANIDM-reg fail to capture the underlying true behaviour of the compositional probabilities.
The ML-BART model exhibits oscillating peaks as it attempts to account for the structural zeros present in the data.
Although the MLN-GP model provides smooth estimates of the compositional probabilities, they are effectively biased, especially in region of the covariate space characterised by an excess of zeros.
In fact, given the compositional constraints, the presence of excess zeros, in even one category, may have an adverse affect on the quality of the estimates for all categories.
Among the models which explicitly incorporate structural zero components, the posterior estimates of $\theta_{ij}$ and $\zeta_{ij}$ under the ZANIDM-reg model, in Figures \ref{fig:count_prob_scenario_1_zanim} and \ref{fig:structural_prob_scenario_1_zanim}, respectively, illustrate the restrictiveness of the parametric linear functional forms imposed on the corresponding regressions.
In contrast, the proposed ZANIM-BART model provides accurate estimates of the true behaviour of $\theta_{ij}$ and $\zeta_{ij}$.
Finally, although not shown in Figures \ref{fig:count_prob_scenario_1_zanim} and \ref{fig:structural_prob_scenario_1_zanim}, the ZANIM-LN-BART model provides similar results to the ZANIM-BART model, albeit with wider credible intervals due to its additional random-effects structure. The same can be said of the MLN-BART model, relative to ML-BART, though it exhibits fewer peaks.


\subsection{Simulations: Scenario 2}\label{sec:simstudy2}

We next evaluate the performance of our proposed ZANIM-BART and ZANIM-LN-BART models and competing methods in more challenging settings designed to reflect characteristics commonly observed in real count-compositional data.
We consider $d \in \{20, 40\}$ categories, varied sample sizes $n\in\{200, 500, 1000\}$, total counts $N_i$ drawn uniformly from $\lbrack 1000, 5000\rbrack$, and $p=6$ covariates drawn from the standard normal distribution.
In order to avoid simulating the compositional counts from the underlying distributional assumptions of our proposed models, we generate $\mathbf{Y}_i$ using the ZANIDM distribution.
We further allow the true covariate effects to contain nonlinearities and interactions that are not represented by additive regression trees.
Our DGP thus deviates from both of our BART-based models and the parametric linear assumptions of the ZANIDM-reg model.
In particular, to accommodate different types of complexity in the relationships between the covariates, $\mathbf{x}_i$, and the population-level count probabilities, $\theta_{ij}=\alpha_{ij}/\sum_{k=1}^d\alpha_{ik}$, and structural zero probabilities, $\zeta_{ij}$, we specify the functional forms, for each category $j\in\{1,\ldots,d\}$, as follows:
\begin{align}
\log \alpha_{ij} = \beta^{(\alpha)}_{0j} + \beta^{(\alpha)}_{1j} x^{(j)}_{i1} + \sin\left(2\pi x^{(j)}_{i2}\right) + x^{2(j)}_{i2} + x^{(j)}_{i1}x^{(j)}_{i2},
\label{eq:functional_form_alpha_sim2} \\
\operatorname{logit} \zeta_{ij} = \beta^{(\zeta)}_{0j} + \beta^{(\zeta)}_{1j} x^{(j)}_{i1} + \sin\left(2\pi x^{(j)}_{i2}\right) + x^{2(j)}_{i2} + x^{(j)}_{i1}x^{(j)}_{i2},\label{eq:functional_form_zeta_sim2}
\end{align}
where superscripts with the $j$ index are used to indicate that the covariates, which are both randomly chosen from the $p=6$ available covariates, are both category- and parameter-specific.
To vary the degree of abundance across the categories, the intercepts for the compositional probabilities were drawn from
$\smash{\beta^{(\alpha)}_{0j}} \sim \operatorname{Uniform}\lbrack -2.3, 2.3\rbrack$, while the baseline structural zero probabilities were sampled from $\smash{\beta^{(\zeta)}_{0j}} \sim \operatorname{Uniform}\lbrack -0.1, 1.5\rbrack$.
The coefficients $\smash{\beta^{(\zeta)}_{1j}}$ and $\smash{\beta^{(\alpha)}_{1j}}$ were sampled from the interval
$\lbrack-1.8,-1.2\rbrack \cup\lbrack 1.2,1.8\rbrack$, ensuring both positive and negative covariate effects.
Thus, the same covariates may affect both the compositional and structural zero probabilities, the categories may share the same covariates, and the functional forms for both parameters contain linear, nonlinear, and interaction effects.
Although two covariates are used in each regression equation above, \eqref{eq:functional_form_alpha_sim2} and \eqref{eq:functional_form_zeta_sim2}, we use all $p=6$ available covariates across the category-specific regression structures when fitting the competing models.
For each combination of $n$ and $d$, we generate six replicate data sets, and average the models' performance across the replicates.
Under these settings, the simulated counts exhibit a considerable degree of sparsity due to either sampling or structural zeros.
Specifically, when $d=20$, approximately $35\%$ of all observations are zero, with $9\%$ sampling and $26\%$ structural zeros.
For the $d=40$ settings, nearly $40\%$ of all observations are zero, with $14\%$ sampling and $26\%$ structural zeros.

The results of the simulation experiments in Table \ref{tab:summary_sim_2__zanidm} are based on six replicate data sets.
The findings clearly indicate that our proposed ZANIM-BART and ZANIM-LN-BART models perform well overall, exhibiting lower values of the KL divergences and improved CP values for all three quantities, compared to competing methods, regardless of the dimension, $d$, and sample size, $n$.
Across the different settings, ZANIM-BART achieves the lowest $\operatorname{KL}(\bm{\theta})$, followed closely by ZANIM-LN-BART.
The other competing models all perform poorly in estimating the true compositional probabilities, as indicated by their substantially larger $\operatorname{KL}(\bm{\theta})$ values relative to our proposed models.
As demonstrated in the previous scenario of Section \ref{sec:simstudy1}, this result can be explained by the bias induced by the fact that the DM-reg, ML-BART, MLN-BART, and MLN-GP models do not explicitly account for the structural zeros present in the data.
Although ZANIDM-reg can accommodate covariate-dependent structural zero probabilities, its assumed functional form is linear.
It therefore fails to capture the true covariate effects, leading to poor recovery of $\theta_{ij}$.

Our proposed models have similar performance in terms of recovering the true structural zero probabilities, with the ZANIM-LN-BART model having slightly better values of $\operatorname{KL}(\bm{\zeta})$ and $\operatorname{CP}(\bm{\zeta})$.
These improvements can be explained by the extra latent random effects present in the ZANIM-LN-BART model, which help to distinguish between structural and sampling zeros.
While all models provide $\operatorname{CP}(\bm{\theta})$ and $\operatorname{CP}(\bm{\zeta})$ values below the nominal level, where applicable, it is notable that the underestimation is far less pronounced under our ZANIM-BART and ZANIM-LN-BART models.
With respect to the individual-level probabilities, $\vartheta_{ij}$, the ZANIM-LN-BART model consistently achieves small $\operatorname{KL}(\bm{\vartheta})$ values and $\operatorname{CP}(\bm{\vartheta})$ value close to the nominal value.
Although the competing models can provide reasonably accurate estimates of the individual-level count probabilities, as indicated by their small values of $\operatorname{KL}(\bm{\vartheta})$, their corresponding $\operatorname{CP}(\bm{\vartheta})$ values are quite below the nominal level.
The only slight exception in this regard is the ZANIDM-reg model --- although it still underestimates $\operatorname{CP}(\bm{\vartheta})$, relative to the ZANIM-LN-BART model --- which matches the underlying distributional assumption of the ZANIDM DGP with which the counts were simulated.
This occurs because these models either cannot estimate exactly zero probabilities, which are present in the true values of $\vartheta_{ij}$, or rely on parametric linear functional forms, as in the case of the ZANIDM-reg and DM-reg models.

In terms of computational time, the BART-based models have the longest runtimes, as expected.
There is a negligible difference in runtime between ZANIM-BART and ZANIM-LN-BART and their corresponding special cases without structural zero components (ML-BART and MLN-BART, respectively).
The regression-based models, DM-reg and ZANIDM-reg, generally have shorter runtimes. However, as discussed above, they perform poorly across all evaluation metrics.
Similarly, MLN-GP has the fastest runtime for small sample sizes $(n=200)$, but fails to recover both $\theta_{ij}$ and $\vartheta_{ij}$, exhibiting particularly large $\operatorname{KL}(\bm{\theta})$ values relative to our proposed models.
Moreover, its runtime grows substantially as the sample size increases.
For the most challenging setting with $n=1000$ and $d=40$, its runtime becomes comparable to ZANIM-BART and ZANIM-LN-BART, due to the computational demands associated with the use of the GP prior.

We further examine the estimation performance of the methods under an experiment where the counts are simulated from the ZANIM-LN distribution. The results are detailed in Supplementary Material \ref{supp:add_simstudy2}.
Briefly, ZANIM-LN-BART performs better than ZANIM-BART in recovering $\theta_{ij}$, as expected.
The relative performance of the competing methods was quite similar to the results presented in \autoref{tab:summary_sim_2__zanidm}, albeit with slightly worse metrics due to the additional random effects induced by the ZANIM-LN DGP.

\begin{table}[!ht]
\centering
\caption{
Simulation results for Scenario 2 with varying $d \in \{20, 40\}$, $n \in {200, 500, 1000}$, and $p=6$.
The counts are simulated from the ZANIDM distribution with true functional forms given in \eqref{eq:functional_form_alpha_sim2} and \eqref{eq:functional_form_zeta_sim2}.
The metrics are averaged over six replicate data sets.
Accuracy in the estimation methods are assessed via the Kullback-Leibler divergence (KL), while uncertainty quantification is evaluated via the coverage probabilities (CP) of the $95\%$ credible intervals.
The runtime of the methods are reported in seconds (sec).
Smaller KL values indicate improved recovery, while coverage values close to the nominal level $(0.95)$ indicate well-calibrated posterior uncertainty.
Values in bold indicate the best performing method under the given metric within each $(n,d)$ setting, while missing values correspond to non-existence of the given parameter under the given method.}
\label{tab:summary_sim_2__zanidm}
\begingroup
\scriptsize
\begin{tabular}{lllllllllr}
  \toprule
$d$ & $n$ & Model & $\operatorname{KL}\left(\bm{\theta}\right)$ & $\operatorname{CP}\left(\bm{\theta}\right)$ & $\operatorname{KL}\left(\bm{\vartheta}\right)$ & $\operatorname{CP}\left(\bm{\vartheta}\right)$ & $\operatorname{KL}\left(\bm{\zeta}\right)$ & $\operatorname{CP}\left(\bm{\zeta}\right)$ & Time (sec) \\
  \midrule
\multirow{7}{*}{$20$} & \multirow{7}{*}{$200$} & ZANIM-BART & \textbf{0.1183} & \textbf{0.6844} & 0.0027 & 0.9236 & 0.0969 & 0.6791 & 193.3 \\
   &  & ZANIM-LN-BART & 0.2041 & 0.6483 & \textbf{0.0025} & \textbf{0.9560} & \textbf{0.0965} & \textbf{0.6821} & 257.8 \\
   &  & ZANIDM-reg & 0.9018 & 0.1477 & 0.0030 & 0.9339 & 0.2743 & 0.5866 & 55.8 \\
   &  & DM-reg & 1.1760 & 0.1089 & 0.0031 & 0.6882 &  &  & 39.8 \\
   &  & ML-BART & 1.9382 & 0.3940 &   &  &  &  & 165.9 \\
   &  & MLN-BART & 1.3236 & 0.4250 & 0.0037 & 0.6684 &  &  & 240.5 \\
   &  & MLN-GP & 0.7910 & 0.5257 & 0.0051 & 0.6725 &  &  & 9.5 \\
   \hdashline
\multirow{7}{*}{$20$} & \multirow{7}{*}{$500$} & ZANIM-BART & \textbf{0.0548} & 0.6327 & 0.0029 & 0.8735 & 0.0664 & 0.6904 & 476.6 \\
   &  & ZANIM-LN-BART & 0.0828 & \textbf{0.6545} & \textbf{0.0022} & \textbf{0.9556} & \textbf{0.0663} & \textbf{0.6910} & 498.2 \\
   &  & ZANIDM-reg & 0.9308 & 0.0899 & 0.0030 & 0.9120 & 0.3712 & 0.3959 & 149.8 \\
   &  & DM-reg & 1.1791 & 0.0650 & 0.0031 & 0.6893 &  &  & 110.3 \\
   &  & ML-BART & 2.0379 & 0.3632 &  &  &  &  & 414.3 \\
   &  & MLN-BART & 1.2443 & 0.3355 & 0.0037 & 0.6629 &  &  & 472.0 \\
   &  & MLN-GP & 0.7183 & 0.4389 & 0.0049 & 0.6814 &  &  & 184.2 \\
   \hdashline
\multirow{7}{*}{$20$} & \multirow{7}{*}{$1000$} & ZANIM-BART & \textbf{0.0298} & 0.5955 & 0.0029 & 0.8283 & 0.0449 & 0.7154 & 796.6 \\
   &  & ZANIM-LN-BART & 0.0489 & \textbf{0.6165} & \textbf{0.0021} & \textbf{0.9528} & \textbf{0.0435} & \textbf{0.7240} & 845.2 \\
   &  & ZANIDM-reg & 0.9320 & 0.0635 & 0.0030 & 0.9057 & 0.4272 & 0.2980 & 309.5 \\
   &  & DM-reg & 1.1834 & 0.0450 & 0.0030 & 0.6898 &  &  & 217.9 \\
   &  & ML-BART & 2.0838 & 0.3471 &   &  &  &  & 645.4 \\
   &  & MLN-BART & 1.1289 & 0.2828 & 0.0037 & 0.6594 &  &  & 830.0 \\
   &  & MLN-GP & 0.6573 & 0.3696 & 0.0047 & 0.6871 &  &  & 538.0 \\
   \hdashline
\multirow{7}{*}{$40$} & \multirow{7}{*}{$200$} & ZANIM-BART & \textbf{0.1251} & \textbf{0.7570} & 0.0051 & 0.9324 & 0.1042 & 0.6707 & 395.8 \\
   &  & ZANIM-LN-BART & 0.2600 & 0.6329 & \textbf{0.0050} & \textbf{0.9548} & \textbf{0.1042} & \textbf{0.6781} & 558.0 \\
   &  & ZANIDM-reg & 1.0703 & 0.1419 & 0.0061 & 0.9252 & 0.3027 & 0.6221 & 111.9 \\
   &  & DM-reg & 1.2877 & 0.1013 & 0.0061 & 0.6916 &  &  & 78.8 \\
   &  & ML-BART & 1.5297 & 0.4831 &   &   &  &  & 335.6 \\
   &  & MLN-BART & 1.1495 & 0.5067 & 0.0074 & 0.6579 &  &  & 485.1 \\
   &  & MLN-GP & 0.7216 & 0.4846 & 0.0122 & 0.6173 &  &  & 119.6 \\
   \hdashline
\multirow{7}{*}{$40$} & \multirow{7}{*}{$500$} & ZANIM-BART & \textbf{0.0548} & \textbf{0.7161} & 0.0048 & 0.8912 & 0.0744 & 0.6776 & 990.4 \\
   &  & ZANIM-LN-BART & 0.0801 & 0.7087 & \textbf{0.0041} & \textbf{0.9569} & \textbf{0.0736} & \textbf{0.6825} & 1054.9 \\
   &  & ZANIDM-reg & 1.0722 & 0.0863 & 0.0060 & 0.9080 & 0.3831 & 0.4438 & 307.0 \\
   &  & DM-reg & 1.3040 & 0.0619 & 0.0061 & 0.6903 &  &  & 211.5 \\
   &  & ML-BART & 1.5232 & 0.4361 &  & &  &  & 859.4 \\
   &  & MLN-BART & 1.0498 & 0.4168 & 0.0074 & 0.6425 &  &  & 1064.0 \\
   &  & MLN-GP & 0.5862 & 0.3961 & 0.0108 & 0.6405 &  &  & 606.8 \\
   \hdashline
\multirow{7}{*}{$40$} & \multirow{7}{*}{$1000$} & ZANIM-BART & \textbf{0.0289} & 0.6734 & 0.0046 & 0.8473 & 0.0504 & 0.7037 & 1846.8 \\
   &  & ZANIM-LN-BART & 0.0380 & \textbf{0.6923} & \textbf{0.0036} & \textbf{0.9553} & \textbf{0.0496} & \textbf{0.7096} & 3853.8 \\
   &  & ZANIDM-reg & 1.0733 & 0.0593 & 0.0060 & 0.8898 & 0.5126 & 0.3147 & 632.3 \\
   &  & DM-reg & 1.2962 & 0.0437 & 0.0061 & 0.6927 &  &  & 434.2 \\
   &  & ML-BART & 1.5816 & 0.3969 &   &   &  &  & 1431.5 \\
   &  & MLN-BART & 0.9373 & 0.3467 & 0.0075 & 0.6361 &  &  & 3677.0 \\
   &  & MLN-GP & 0.5402 & 0.3444 & 0.0105 & 0.6492 &  &  & 3180.0 \\
   \bottomrule
\end{tabular}
\endgroup
\end{table}

Overall, our simulation studies suggest that the established practical benefits of the standard BART model extend to our use of BART as a nonparametric prior in the ZANIM-BART and ZANIM-LN-BART models.
In particular, the simulation results demonstrate that the both models effectively recover unknown smooth functions and low-order interaction effects on the population-level count probabilities as well as the structural zero probabilities across a range of scenarios.
Additionally, when the data are generated without structural zeros, our proposed models yields results comparable to those of their corresponding special cases, ML-BART and MLN-BART, indicating that they do not introduce bias when structural zeros are absent.
Despite the ability of ZANIDM-reg to accommodate structural zeros and overdispersion, it is fundamentally limited by the need to correctly specify the functional form, as it relies on parametric linear regression assumptions.
While the MLN-GP model can capture complex relationships without pre-specification, due to its GP prior, we show that it performs poorly when structural zeros are present in the data. Furthermore, its runtime scalability becomes similar to our models for large sample sizes.
Indeed, the overall performance of existing approaches highlights the importance of simultaneously accounting for nonlinearities, interaction effects, overdispersion, and explicit modelling of structural zeros, as we do in our models.

\section{Modern pollen-climate data analysis}\label{sec:application}

We now illustrate the usefulness of our proposed ZANIM-BART and ZANIM-LN-BART models in modelling real-world sparse count-compositional data.
Our application belongs to the broad field of palaeoclimate reconstruction.
A key step in quantitative palaeoclimate reconstruction is to build a model to describe the relationships between the present vegetation and climatic drivers \citep{Sweeney2018}.
In this context, such a model is referred to as a forward model.
Counts of pollen grains are commonly used to reflect the vegetation for a specific climate, such that changes in pollen composition provide information about the climate at that location.
As discussed in \citet{Sweeney2018}, the specification of effective forward models for pollen-climate data presents two important statistical challenges: the need for a proper count-compositional likelihood that accounts for excess zeros, and flexible functional forms allowing for complex, potentially multimodal pollen-climate relationships.
Our goals in this section are to demonstrate the flexibility and practical advantages of our proposed ZANIM-BART and ZANIM-LN-BART models in capturing pollen-climate relationships, while addressing these key challenges in forward model specification.

We consider an updated version of the data set used by \citet{Haslett2006} and \citet{Parnell2015}, containing $n=7832$ samples of $d=28$ pollen taxa collected at different locations in the Northern Hemisphere, along with three climate covariates: \textit{GDD5}, the growing degree days above $5^\circ \operatorname{C}$, calculated as the sum of daily temperatures above $5^\circ \operatorname{C}$ over a year; \textit{MTCO}, the mean temperature of the coldest month in degrees Celsius; and \textit{AET/PET}, the ratio of actual to potential evapotranspiration.
The total counts, $N_i$, range from $74$ to $1003$ and the pollen counts exhibit substantial variability, overdispersion, sparsity, and complex, cross-category dependencies (both positive and negative), as detailed by the descriptive analyses given in Supplementary Material \ref{supp:add_palaeoclimate}.
Notably, $63.21\%$ of the observations are zero and overdispersion and variability, as quantified by the multivariate coefficient of variation (MCV) of \citet{Albert1993} and the multiple marginal dispersion index (MDI) of \citet{Kokonendji2018}, are high.
So too is the excess of zeros relative to the multinomial distribution, as measured using a multivariate zero-inflation (ZI) index.
We note that the MDI and ZI indices are constructed by appropriately averaging the marginal dispersion and zero-inflation indices --- $\operatorname{DI}\lbrack Y_j\rbrack$ and $\operatorname{ZI}_b\lbrack Y_j\rbrack$, respectively --- where the latter is an adapted version of the index of \citet{Kim2018} which quantifies the excess of zeros with respect to the binomial distribution. Further definitions of these various indices are provided in Supplementary Material \ref{supp:add_palaeoclimate}.
For this data set, the empirical values of MCV, MDI, and ZI are $121.9, 215.5$, and $0.576$, respectively, indicating pronounced variability, overdispersion, and zero-inflation.

Previous studies suggested that the relationships between pollen composition and climate covariates are nonlinear \citep[e.g.,][]{Vasko2000,Haslett2006}.
As such, the BART-based models ---  ML-BART, MLN-BART, and our proposed ZANIM-BART and ZANIM-LN-BART models -- are suitable for this application, since these models are capable of addressing nonlinearities and interactions without the need to pre-specify the functional forms of the pollen-climate relationships.
The ZANIM-BART and ZANIM-LN-BART models further allow for explicit modelling of the zero-inflation mechanism through their taxa-specific sets of regression trees which account for covariate-dependence in the structural zero probabilities.
For comparison purposes, we also include the DM-reg and ZANIDM-reg models, despite knowing that their unrealistic parametric linear assumptions are likely implausible for these data. After demonstrating the superior fit of our ZANIM-LN-BART model to these data in the comparative evaluations in Section \ref{sec:application_comparison}, we present some inferential insights obtained using this model in Section \ref{sec:application_inference}.

\subsection{Model comparison}\label{sec:application_comparison}

We evaluate the same set of models used in the simulation studies in Section \ref{sec:simulations}, excluding the MLN-GP model which was not feasible to run, given the lack of computational scalability and the increased memory burden with respect to the sample size for methods based on GPs.
Furthermore, all hyperparameter settings were left unchanged from the simulation studies.
In particular, we recall that we use $m_\theta=m_\zeta=100$ trees per category for the count and structural zero components, respectively, where applicable.
For all models, MCMC chains were run with $10{,}000$ iterations, with the first $5{,}000$ iterations discarded as burn-in.
We randomly hold out $2{,}000$ observations and use the remaining $5{,}832$ observations for fitting the models.
Under these settings, the runtime of the ZANIM-BART, ZANIM-LN-BART, MLN-BART, and ML-BART models were $2.83$, $2.47$, $2.27$, and $1.76$ hours, respectively.
These runtimes are largely driven by the large training sample size, $n$, more so than the number of taxa, $d$, or number of covariates, $p$.
For the regression-based models, the runtime of DM-reg was $26.57$ minutes, while ZANIDM-reg took $33.23$ minutes.

Our comparison of models is conducted visually, using the holdout predictive check (HPC) approach proposed by \citet{Moran2023}, which splits the data into training and holdout samples, $\mathbf{Y}^{\operatorname{in}}$ and $\mathbf{Y}^{\operatorname{out}}$, respectively, in order to avoid using the same data twice.
The premise of the HPC is that if a model provides a good description of the data, the observations drawn from the posterior predictive distribution should resemble draws from the population distribution (i.e., the observed holdout data).
The HPC proceeds by choosing diagnostic statistics $d(\mathbf{Y})$ which capture relevant features of the data.
The observed holdout statistics $d(\mathbf{Y}^{\operatorname{out}})$ are then compared with the posterior predictive distribution of the diagnostic obtained under the observed training data, i.e., $p(d(\mathbf{Y}^{\operatorname{rep}}) \mid \mathbf{Y}^{\operatorname{in}})$, where $\mathbf{Y}^{\operatorname{rep}}$ is drawn from the posterior predictive density $p(\mathbf{Y}^{\operatorname{rep}} \mid \mathbf{Y}^{\operatorname{in}})$.
We have carefully chosen the diagnostic statistics to reflect key characteristics that the models should be able to capture; namely, overdispersion, zero-inflation, and the compositional structure of the counts.
Specifically, we consider the MDI, which summarises the overdispersion across the taxa, the average compositional entropy, given by $-n^{-1}\sum_{i=1}^n 1/\log(d)\sum_{j=1}^dy_{ij}/N_i \log(y_{ij}/N_i)$, the overall proportion of zeros, given by $(nd)^{-1}\sum_{i=1}^n\sum_{j=1}^d \mathds{1}(y_{ij}=0)$, and the multivariate ZI index for multinomial data described above.

\begin{figure}[!ht]
    \centering
\includegraphics[width=1.0\linewidth]{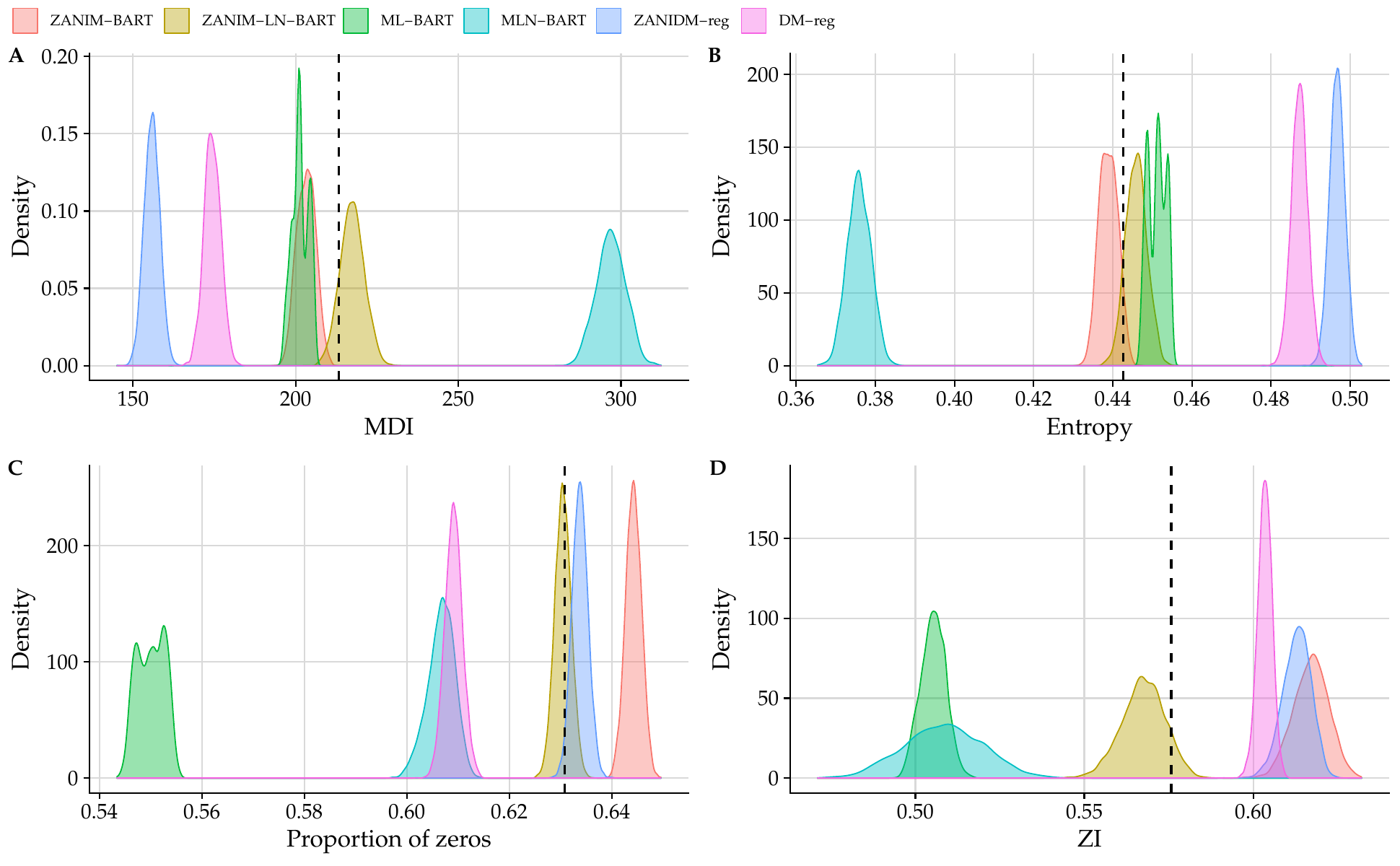}
    \caption{Holdout predictive checks using the MDI (\textbf{A}), entropy (\textbf{B}), proportion of zeros (\textbf{C}), and multivariate ZI index (\textbf{D}) from the ZANIM-BART, ZANIM-LN-BART, ML-BART, MLN-BART, ZANIDM-reg, and DM-reg models. The vertical dashed lines indicate the empirical values of each diagnostic statistic computed from the holdout data.}
    \label{fig:ppcs_holdout}
\end{figure}

\autoref{fig:ppcs_holdout} displays the HPCs of the four diagnostic statistics for each model with the vertical dashed lines indicating the empirical value obtained from the holdout data.
We observe that the ZANIM-LN-BART model provides the best overall fit, as the empirical values of the diagnostic statistics lie in regions of moderate-to-high posterior predictive probability.
In contrast, the other models produce HPCs that clearly deviate from the empirical diagnostics.
Regarding the overall dispersion in the data, the HPCs of the MDI indicate that the ML-BART and ZANIM-BART models underestimate dispersion, while MLN-BART overestimates it.
This can be explained by the fact that the BART prior alone in the ML-BART model is insufficient to explain the high level of overdispersion in the data.
In particular, the lack of explicit zero-inflation components and latent random effects limits its ability to adequately represent the variability in the data.
However, incorporating only zero-inflation components, as in ZANIM-BART, or only latent random effects, as in MLN-BART, is also insufficient for accurately describing the overdispersion.
On the other hand, the ZANIM-LN-BART model, which simultaneously captures nonlinear covariate effects, with explicit zero-inflation components and latent effects, is able to provide a better description of the observed overdispersion.
Similar conclusions are also observed in terms of capturing the high level of sparsity in the data.
The models without explicit mixture zero-inflation components, ML-BART and MLN-BART, substantially underestimate both the average proportion of zeros and the multinomial ZI index.
The regression-based models, DM-reg and ZANIDM-reg, perform poorly for all four  diagnostics statistics considered here.
Their deviations from the empirical values of each index are attributable to the fact that they cannot capture nonlinearities and interaction effects in the pollen-climate relationships.
While the ZANIM-BART model tends to overestimate the number of zeros, likely by misclassification of sampling and structural zeros, the ZANIM-LN-BART model accurately describes the sparsity structure in the data, as indicated by its HPCs for both the average proportion of zeros and the multinomial ZI index.
Consequently, ZANIM-LN-BART leads to markedly better HPC performance for the average compositional entropy.
Finally, of our two proposed models, it is notable that ZANIM-LN-BART evidently exhibits greater fidelity to the empirical data than ZANIM-BART, across all considered metrics.

\begin{figure}[!hb]
    \centering
    \includegraphics[width=1.0\linewidth]{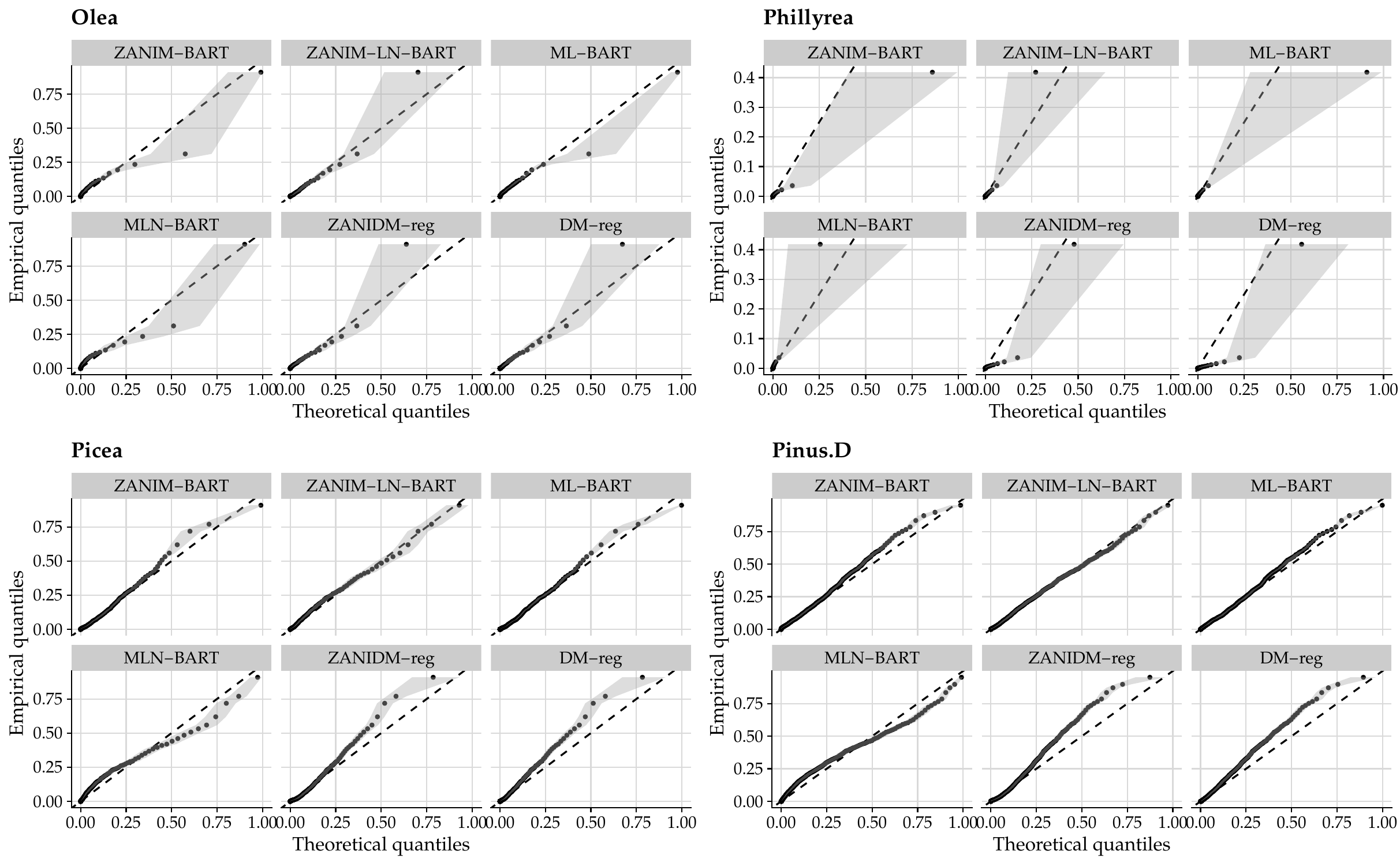}
    \caption{Quantile-quantile plots for the relative abundances, $y_{ij}/N_i$, computed for the holdout data from the ZANIM-BART, ZANIM-LN-BART, ML-BART, MLN-BART, ZANIDM-reg, and DM-reg  models for four representative taxa. \textbf{A}: \textit{Olea}, which has a high zero-inflation. \textbf{B}: \textit{Phillyrea}, the least abundant taxon. \textbf{C}: \textit{Pinus.D}, the most abundant taxon.
For each model, $q$-quantiles from the posterior predictive distribution of the relative abundances were computed, using $q=200$ subsets of equal sizes. The medians of the posterior predictive quantiles are used to draw the points and the $2.5\%$ and $97.5\%$ quantiles are used to draw the interpolated grey ribbon envelopes.}
    \label{fig:qqplots_holdout}
\end{figure}

Finally, we assess the goodness-of-fit of the models using marginal quantile-quantile plots computed for the holdout data using the posterior predictive distribution of the relative abundances, $y_{ij}/N_i$.
The results are given in \autoref{fig:qqplots_holdout} for four representative taxa, as determined by the descriptive statistics reported in Supplementary Material \ref{supp:add_palaeoclimate}.
The specific taxa are: \textit{Olea}, which has a high zero-inflation index of $\operatorname{ZI}_b(Y_j)=0.84$; \textit{Phillyrea}, the least abundant taxon, with $\mathbb{E}\lbrack Y_j\rbrack=1.12$; \textit{Pinus.D}, the most abundant taxon, with $\mathbb{E}\lbrack Y_j\rbrack=225.28$; \textit{Picea}, which has a relatively high dispersion index of $\operatorname{DI}\lbrack Y_j\rbrack=256.58$.
Consistent with the HPCs, these plots show that the theoretical quantiles under ZANIM-LN-BART closely follow the observed holdout quantiles, indicating the best overall fit to the data.
In contrast, all other models are clearly inadequate in reproducing the holdout quantiles.
These findings further highlight the flexibility of ZANIM-LN-BART in capturing different marginal count patterns, which we attribute to its ability to simultaneously account for zero-inflation, overdispersion, complex dependencies, and flexible covariate effects.

\subsection{Inferential results}\label{sec:application_inference}

\begin{figure}[!ht]
    \centering
\includegraphics[width=1.0\linewidth]{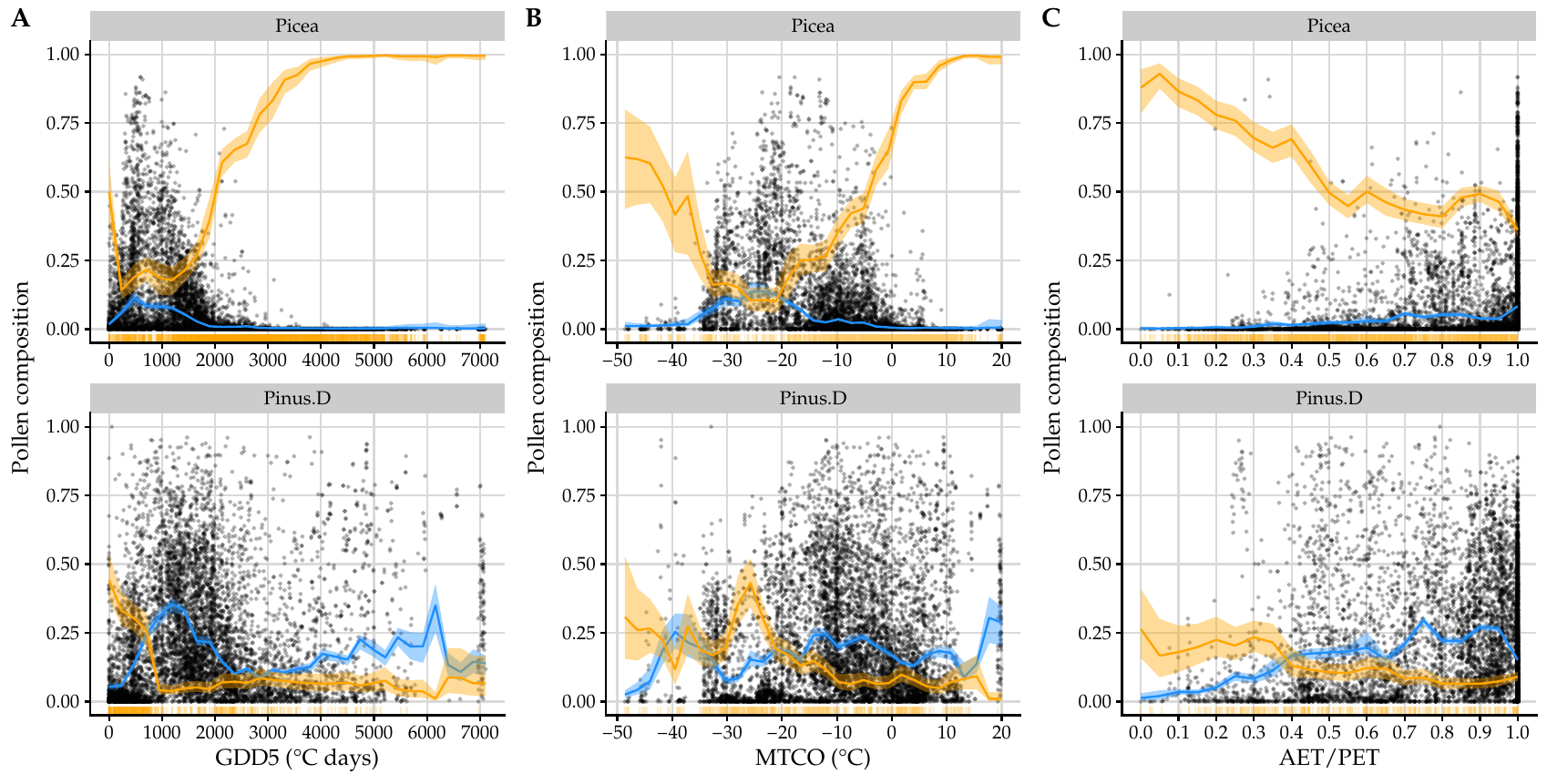}
    \caption{
     Partial dependence plots (PDPs) of the climate covariate effects for \textit{Picea} and \textit{Pinus.D}.
     Black dots represent the observed relative abundances, $y_{ij} / N_i$, and the orange rugs along the $x$-axes represent samples where the observed counts are zero. These PDPs show the effects of the climate covariates GDD5 (growing season warmth, Panel \textbf{A}), MTCO (harshness of winter, Panel \textbf{B}), and AET/PET (available moisture, Panel \textbf{C}), on the population-level count (blue) and structural zero probabilities (orange) via superimposed curves, along with associated $95\%$ credible interval bands.}
    \label{fig:pdps_picea_pinusd}
\end{figure}

Next, considering the ZANIM-LN-BART model, we conclude this analysis by illustrating the marginal effects that the climate covariates GDD5, MTCO, and AET/PET have on the compositional and structural zero probabilities using partial dependence plots \citep[PDPs;][]{Friedman2001}.
To do so, we create a three-dimensional grid which represents the observed climate sites, following the approach of \citet{Haslett2006}.
This yields $3{,}120$ observations, where each of the three covariates have $31$ unique values.
From these PDPs, researchers can obtain insights on how the climate conditions affect both the abundance and absence of pollen taxa.
\autoref{fig:pdps_picea_pinusd} shows the PDPs for \textit{Picea} and \textit{Pinus.D}, with one panel per covariate.
In general, the PDPs for the compositional probabilities (blue curves) smooth the observed compositions (black points) and the structural zero probabilities (orange curves) are consistent with the observed zeros (orange rugs).
Overall, it is apparent that ZANIM-LN-BART helps to prevent biased estimation of the covariate effects, such as downward (upward) bias under zero-inflation ($N$-inflation), and enables the identification of climatic regions with high or low structural zero probabilities.
Supplementary material \ref{supp:add_palaeoclimate} provides complementary PDPs for all taxa with the effects of MTCO, GDD5, and AET/PET.

We observe a unimodal climate response of \textit{Picea} to GDD5 in Panel \textbf{A}, with a mode around $1000$.
As GDD5 increases beyond this value, the compositional and structural zero probabilities decrease and increase, respectively. Consistent with the observed zeros indicated by the rug, this suggests that \textit{Picea} is less likely to be observed under warmer growing conditions.
In Panel \textbf{B}, MTCO exhibits a quadratic effect on the structural zero probabilities for \textit{Picea}, with higher probabilities under extremely cold and hot conditions (MTCO $< 30^\circ \operatorname{C}$ and MTCO $> 20^\circ \operatorname{C}$), and the effect on the compositional probabilities is again unimodal, with the mode located between $-30$ and $-20$.
Finally, regarding the PDPs for AET/PET and \textit{Picea} shown in Panel \textbf{C}, we see a nearly constant linear effect on the compositional probabilities with small peaks around AET/PET values between $0.7$ and $0.9$, with a further linear negative effect on the structural zero probabilities.

On the other hand, for \textit{Pinus.D}, we observe a multimodal effect of GDD5 on the compositional probabilities, with two notable modes around $1000$ and $6000$, where the latter displays more uncertainty due to the scarcity of observations with GDD5 around $6000$ (see Panel \textbf{A}).
The probability of structural zeros is elevated when GDD5 is near $0$, but the effect is insubstantial and largely flat at higher GDD5 values.
Similarly, Panels \textbf{B} and \textbf{C} indicate multimodal effects of MTCO and AET/PET on the compositional probabilities and modest (though still nonlinear) effects on the structural zero probabilities  for \textit{Pinus.D}.
As the estimated structural zero probabilities are much smaller than those for \textit{Picea}, this suggests that the zeros of \textit{Pinus.D} are primarily explained by unobserved heterogeneity captured through the random effects in the ZANIM-LN-BART model, rather than direct climatic influences.
This interpretation is consistent with the superior posterior predictive performance of ZANIM-LN-BART relative to ZANIM-BART for this taxon (see \autoref{fig:qqplots_holdout}).

\section{Discussion}\label{sec:discussion}

Our novel ZANIM-BART and ZANIM-LN-BART models represent advances over existing approaches for modelling count-compositional data by simultaneously addressing key challenges commonly encountered in such data, including an excess of zeros, overdispersion, and cross-sample heterogeneity, while flexibly modelling covariate effects without imposing any rigid parametric assumptions.
This is achieved by extending the ZANIM distribution \citep{Menezes2025} with independent nonparametric multinomial logistic BART and probit BART priors on the category-specific parameters that relate the covariates to both the population-level count and structural zero probabilities, respectively.
Our ZANIM-LN-BART model further incorporates multivariate logistic-normal random effects, which allows for excess variation to be explained by unobserved latent characteristics and captures more complex dependence among the categories.

We developed an efficient MCMC algorithm for conducting inference for both proposed models and demonstrated its efficacy through simulation studies.
We combine the data augmentation schemes introduced by \citet{Menezes2025} for the ZANIM distribution with the established BART sampling routines of \citet{Chipman2010} and \citet{Murray2021}.
Conceptually, our models extend the ML-BART framework of \citet{Murray2021} in two directions: ZANIM-BART introduces category-specific BART priors to accommodate excess zeros and the ZANIM-LN-BART model further generalises this by incorporating latent random effects to account for unobserved heterogeneity.
As a by-product, removing the zero-inflation mixture components from ZANIM-LN-BART yields the MLN-BART model, which appears to represent a new extension of the ML-BART model.

The simulation studies highlighted the importance of explicitly modelling structural zeros and showed that the ZANIM-BART and ZANIM-LN-BART models can accurately recover the true population-level count probabilities, structural zero probabilities, and individual-level count probabilities.
Notably, our models are unique in their ability to provide estimates for all three quantities without requiring pre-specification of the functional form related to the category-specific count and structural zero probabilities, which is an improvement over existing methods such as the ZIDM model.
Moreover, when structural zeros are absent from the data-generating process, the proposed models perform similarly to their corresponding special cases, indicating that the additional zero-inflated mixture components do not compromise goodness-of-fit and uncertainty quantification.

We further illustrated our models through the analysis of modern pollen-climate data from a palaeoclimate study.
Using posterior predictive holdout checks, we demonstrated that our ZANIM-LN-BART model provides the best fit to the data and effectively addresses key statistical challenges discussed by \citet{Sweeney2018}, namely: (i) a proper count-compositional model capable of capturing zero-inflation, and (ii) sufficient flexibility to capture different pollen-climate relationships, including multimodality, without imposing restrictive parametric assumptions.
Moreover, we presented inferential outputs that can help researchers to extract meaningful insights from their pollen count-compositional data.
In particular, we examined the effects of climate covariates on the pollen composition and identified climate conditions that can affect both the count probabilities and structural zero probabilities across pollen taxa.
We also demonstrated in Supplementary Material \ref{supp:human_gut_microbiome} that our models are useful in the analysis of a benchmark human gut microbiome data set.
We showed that ZANIM-LN-BART and ZANIM-BART provide superior fits compared to the alternative ZIDM and DM regression models, while also being able to identify important covariates and uncover nonlinear relationships between the intake nutrients and the microbial compositions.

A general assumption of our proposed models is that the structural zero indicators, $\mathbf{z}_i$, are independent across the categories.
Although ZANIM-BART and ZANIM-LN-BART can capture the co-occurrence of zeros across categories nonetheless, it may be of interest to relax the \textit{a priori} independence assumption, while retaining category-specific BART priors for the structural zero probabilities.
This could be addressed by assuming that $\mathbf{z}_i$ has a multivariate logistic-normal BART prior, rather than the present assumption of probit BART priors for each $z_{ij}$.
This would be similar in spirit to the multinomial logistic-normal BART proposed in this paper, which incorporates random effects to capture overdispersion and complex dependencies among the counts.
Another promising approach would be to explore the introduction of shared Gaussian latent factors for $\mathbf{z}_i$ within the nonparametric seemingly unrelated probit BART framework of \citet{Esser2025}.
Either extension would require only minor modifications in the updates of the regression tree parameters associated with the structural zero probabilities, along with an additional sampling step for the correlation matrix of the shared Gaussian random effects.

It is also of interest to reduce the computational cost of our models associated with the large total number of regression trees, $d \times (m_\theta + m_\zeta)$, while maintaining the flexibility of the BART priors.
A promising research direction in this line is to consider the shared trees framework of \citet{Linero2020}.
In this approach, the tree structure could be shared within each category by using the same tree topology for the compositional and structural zero probabilities, or shared across the categories for both sets of parameters.
Preliminary experiments in a high-$d$ setting indicate that the computational speed of our models is greatly improved when adopting a shared trees approach, at the expense of only a minor loss of performance relative to the ZANIM-BART and ZANIM-LN-BART models.
We intend to further investigate this approach in future work.

Lastly, we would also like to incorporate the ZANIM-BART and ZANIM-LN-BART models within the complete palaeoclimate reconstruction framework of \citet{Parnell2015}.
However, this might require an extension to overcome BART's inherent lack of smoothness, as smoothness is key to obtaining coherent palaeoclimate reconstructions \citep{Sweeney2018}.
The soft-BART of \citet{Linero2018b} and the GP-BART of \citet{Maia2024} are approaches that can be leveraged to enforce such smoothness, though both would add considerable computational cost.

Overall, we anticipate that our proposed models, especially ZANIM-LN-BART, will be particularly valuable in zero-inflated count-compositional settings where the goal is to uncover complex covariate effects on both the count and structural zero components, while simultaneously accounting for high degrees of overdispersion and complex dependencies among the categories.

\section*{Acknowledgments}
{\small
Andr{\'e} F. B. Menezes's work was supported by Taighde {\'{E}}ireann -- Research Ireland under Grant number 18/CRT/6049.
Andrew Parnell's work was additionally supported by: the UCD-Met \'{E}ireann Research Professorship Programme (28-UCDNWPAI);
Northern Ireland's Department of Agriculture, Environment and Rural Affairs (DAERA), UK Research and Innovation (UKRI) via the International Science Partnerships Fund (ISPF) under Grant number [22/CC/11103] at the Co-Centre for Climate + Biodiversity + Water; Decarb-AI, supported by AIB and Research Ireland: an Innovate for Ireland Centre (25/I4I-TC/13542); and a Research Ireland Research Centre award (12/RC/2289\_P2).}

\bibliographystyle{agsm}
\bibliography{references}

\section*{Supplementary Material}

The Supplementary Material includes further details on the ZANIM-LN distribution in Section \ref{supp:zanim_ln_distribution}, derivations for the data augmentations and posterior inference scheme for the ZANIM-LN-BART model in Section \ref{supp:inference_zanim_ln}, and additional results for both simulation studies (Section \ref{supp:add_simstudies}) and the case study presented in the main paper on  modern pollen-climate data analysis (Section \ref{supp:add_palaeoclimate}), along with an additional application to a benchmark human gut microbiome data set (Section \ref{supp:human_gut_microbiome}), and convergence assessments (Section \ref{supp:convergence}).

\setcounter{section}{0}
\setcounter{subsection}{0}
\setcounter{figure}{0}
\setcounter{table}{0}
\setcounter{equation}{0}
\renewcommand{\thetable}{S.\arabic{table}}
\renewcommand{\thefigure}{S.\arabic{figure}}
\renewcommand{\theequation}{S.\arabic{equation}}
\renewcommand{\thealgocf}{S.\arabic{algocf}}
\renewcommand{\thesection}{S.\arabic{section}}

\section{Additional details on the ZANIM-LN distribution}\label{supp:zanim_ln_distribution}

Section \ref{sec:zanim_distribution} introduced the ZANIM-LN distribution by incorporating logistic-normal random effects on the individual-level probabilities of the ZANIM distribution.
Here, we first give a brief overview of the ZANIM, ZANIM-LN, and ZANIDM distributions, then compare the theoretical properties of the ZANIM-LN distribution in relation to the ZANIM and ZANIDM distributions.

The ZANIM, ZANIM-LN, and ZANIDM distributions extend the multinomial distribution to account for structural zeros in count-compositional data.
In all three distributions, the compositional counts follow $\mathbf{Y}_i\sim \operatorname{Multinomial}_d\left\lbrack N_i, \bm{\vartheta}_{i}\right\rbrack$, where the individual-level count probabilities, $\bm{\vartheta}_i=(\vartheta_{i1}, \ldots, \vartheta_{id})$, are obtained through some function of the parameters which govern the count probabilities and latent structural zero indicators, $z_{ij}$.
Specifically, $\smash{(z_{ij} \mid \zeta_j) \overset{\operatorname{ind.}}{\sim} \operatorname{Bernoulli}\left\lbrack 1 - \zeta_{j}\right\rbrack}$ indicates whether observation $i$ from category $j$ is a structural zero or not, and the probabilities $\vartheta_{ij}$ are obtained by appropriately normalising the parameters which govern the count probabilities over the non-zero components across each category.
In particular, under the ZANIM distribution introduced by \citet{Menezes2025}, the individual-level probabilities are given by $\vartheta_{ij}=z_{ij}\theta_{j}/\sum_{k=1}^dz_{ik}\theta_{k}$, where $\theta_{j}$ is a category-specific population-level count probability parameter.
The ZANIM-LN distribution further extends the ZANIM distribution by incorporating logistic-normal random effects, $u_{ij}$, such that the individual-level count probabilities become $\vartheta_{ij}=z_{ij}e^{u_{ij}}\alpha_{j}/\sum_{k=1}^dz_{ik}e^{u_{ik}}\alpha_{k}$, where $u_{ij}\sim \operatorname{N}_d \left \lbrack \mathbf{0}_d, \bm{\Sigma}_U \right\rbrack$ and $\alpha_{j}>0$ is a category-specific concentration parameter.
The ZANIDM distribution, which was first introduced by \citet{Koslovsky2023} and later given a richer characterisation by \citet{Menezes2025}, instead incorporates Dirichlet random effects through the normalisation of independent gamma random variables, such that the individual-level probabilities are given by $\vartheta_{ij}=z_{ij}\lambda_{ij}/\sum_{k=1}^d z_{ik}\lambda_{ik}$, where $\lambda_{ij}\overset{\mathrm{ind.}}{\sim}\operatorname{Gamma}\left\lbrack\alpha_{j}, 1\right\rbrack$ and $\alpha_{j}>0$ is a category-specific concentration parameter.
In both the ZANIM-LN and ZANIDM distributions, their corresponding population-level count probabilities are obtained by normalising the concentration parameters $\alpha_{j}$, i.e., $\theta_{j}=\alpha_{j}/\sum_{k=1}^d \alpha_{k}$.

In light of the above definitions, all three distributions can be characterised as compound multinomial models.
Table \ref{tab:compound_multinomial_models} gives details on the various compound multinomial models employed throughout this paper.
This table highlights the different specifications of the random effects and structural zero components.
For clarity, the distributions are presented without explicitly displaying covariate effects.
Note, however, that the population-level count and structural zero probabilities ($\theta_{j}$ and $\zeta_{j}$, respectively) may depend on a set of covariates $\mathbf{x}_i$ through either parametric or nonparametric specifications. This further induces covariate-dependence in the individual-level count probabilities, $\vartheta_{ij}$.
For instance, in the DM-reg model, $\log \alpha_{ij} = \mathbf{x}_i^\top \bm{\beta}_j$, while in the proposed ZANIM-BART and ZANIM-LN-BART models, we specify $\theta_{ij} = f^{(\mathrm{c})}_j(\mathbf{x}_i)/\sum_{k=1}^d f^{(\mathrm{c})}_k(\mathbf{x}_i)$ and $\zeta_{ij} = \Phi(f^{(0)}_j(x_i))$ through multinomial logistic BART and probit BART priors, respectively.

\begin{table}[!ht]
\centering
\scriptsize
\caption{Compound multinomial distributions for count-compositional data.
The multinomial distribution is included as a baseline for comparison.
The ``Random effect'' column indicates the presence and the type of subject-specific random effects --- Dirichlet, logistic-normal, or none (indicated by \ding{55}).
Models with structural zero components, indicated by \ding{51}, include latent indicators $\smash{z_{ij} \overset{\mathrm{ind.}}{\sim} \operatorname{Bernoulli}\lbrack 1 - \zeta_{ij}\rbrack}$, where $\zeta_{j}$ is the probability of a structural zero.
The population-level and individual-level count probabilities under each model are given in the columns $\theta_{j}$ and $\vartheta_{ij}$, respectively.
Data can be simulated under each distribution using its hierarchical representation; i.e., conditional on the individual-level probabilities $\vartheta_{ij}$, counts can be  generated from a multinomial distribution.}
\label{tab:compound_multinomial_models}
\begin{tabular}{lcccc}\hline
  Distribution & Random effects & Structural zeros & $\theta_{j}$ & $\vartheta_{ij}$ \\\hline
  $\operatorname{Multinomial}\lbrack N_i, \bm{\theta}\rbrack$ & \ding{55}  & \ding{55} & $\theta_{j}$ & \ding{55} \\
 $\operatorname{ZANIM}\lbrack N_i, \bm{\theta}, \bm{\zeta}\rbrack$ & \ding{55} & \ding{51} & $\theta_{j}$ & $z_{ij}\theta_{j} / \sum_{k=1}^d z_{ik}\theta_{k}$ \\
  $\operatorname{DM}\lbrack N_i, \bm{\alpha}\rbrack$  & $\lambda_{ij} \overset{\operatorname{ind.}}{\sim} \operatorname{Gamma}\lbrack \alpha_{j}, 1\rbrack$ & \ding{55} & $\alpha_{j}/\sum_{k=1}^d\alpha_{k}$ & $\lambda_{ij}/\sum_{k=1}^d\lambda_{ik}$ \\
 $\operatorname{ZANIDM}\lbrack N_i, \bm{\alpha}, \bm{\zeta}\rbrack$ & $\lambda_{ij} \overset{\operatorname{ind.}}{\sim} \operatorname{Gamma}\lbrack \alpha_{j}, 1\rbrack$ & \ding{51} & $\alpha_{j}/\sum_{k=1}^d\alpha_{k}$ &  $z_{ij}\lambda_{ij}/\sum_{k=1}^d z_{ik}\lambda_{ik}$\\
 $\operatorname{MLN}\lbrack N_i, \bm{\alpha}, \bm{\Sigma}_U\rbrack$ & $\mathbf{u}_i \sim \operatorname{N}_{d}\lbrack \mathbf{0}, \bm{\Sigma}_U\rbrack$ & \ding{55} & $\alpha_{j}/\sum_{k=1}^d\alpha_{k}$ & $\alpha_{j}e^{u_{ij}}/\sum_{k=1}^d \alpha_{k}e^{u_{ik}}$\\
 $\operatorname{ZANIM-LN}\lbrack N_i, \bm{\alpha}, \bm{\zeta}, \bm{\Sigma}_U\rbrack$ & $\mathbf{u}_i \sim \operatorname{N}_{d}\lbrack \mathbf{0}, \bm{\Sigma}_U\rbrack$ & \ding{51} & $\alpha_{j}/\sum_{k=1}^d \alpha_{k}$ & $z_{ij} \alpha_{j} e^{u_{ij}}/\sum_{k=1}^d z_{ik} \alpha_{k} e^{u_{ik}}$\\
\hline
\end{tabular}
\end{table}

We now we compare the theoretical properties of the ZANIM-LN distribution in relation to the ZANIM and ZANIDM distributions.
We follow \citet{Menezes2025}, which compared the ZANIM and ZANIDM distributions only, by presenting marginal PMF plots and theoretical moments for an example with
$d=3$ categories and $N = 30$ trials.
We use similar parameter values as per \citet{Menezes2025} for the ZANIM and ZANIDM distributions with
common zero-inflation parameters and expectations given by $\bm{\zeta} = (0.05, 0.15, 0.10)$ and
$\mathbb{E}\lbrack \mathbf{Y}\rbrack=(2.320,18.496,9.161)$, respectively.
For the ZANIM-LN distribution, we specify the covariance matrix of the random effects as
\[\bm{\Sigma}_U = \begin{pmatrix} 0.5 & -0.4 & 0.1\\ -0.4 &  0.6 & 0.3\\ 0.1 & 0.3 & 0.7\\\end{pmatrix},\]
and set its concentration parameters as $\bm{\alpha} \in \{0.031, 0.770, 0.241\}$, in order to
match the same expectations.
We remind the reader that the ZANIM-LN distribution does not have a closed-form expression for its PMF.
However, we can approximate it using Monte Carlo simulation; in particular, we generate $\mathbf{u}_i \sim \operatorname{Normal}_3\left\lbrack \mathbf{0}_3, \bm{\Sigma}\right\rbrack$ for $i=1,\ldots, 5000$ random samples and evaluate the analytical expression of the marginal PMF of the ZANIM distribution, for all $d=3$ categories, conditional on $\theta_{ij}=\alpha_{j} e^{u_{ij}}/\sum_{k=1}^d\alpha_{k} e^{u_{ik}}$ and $\bm{\zeta} \in \{0.05, 0.15, 0.10\}$, then average it to obtain the marginal PMFs under the ZANIM-LN distribution.

\begin{figure}[!ht]
    \centering
    \includegraphics[width=1\linewidth]{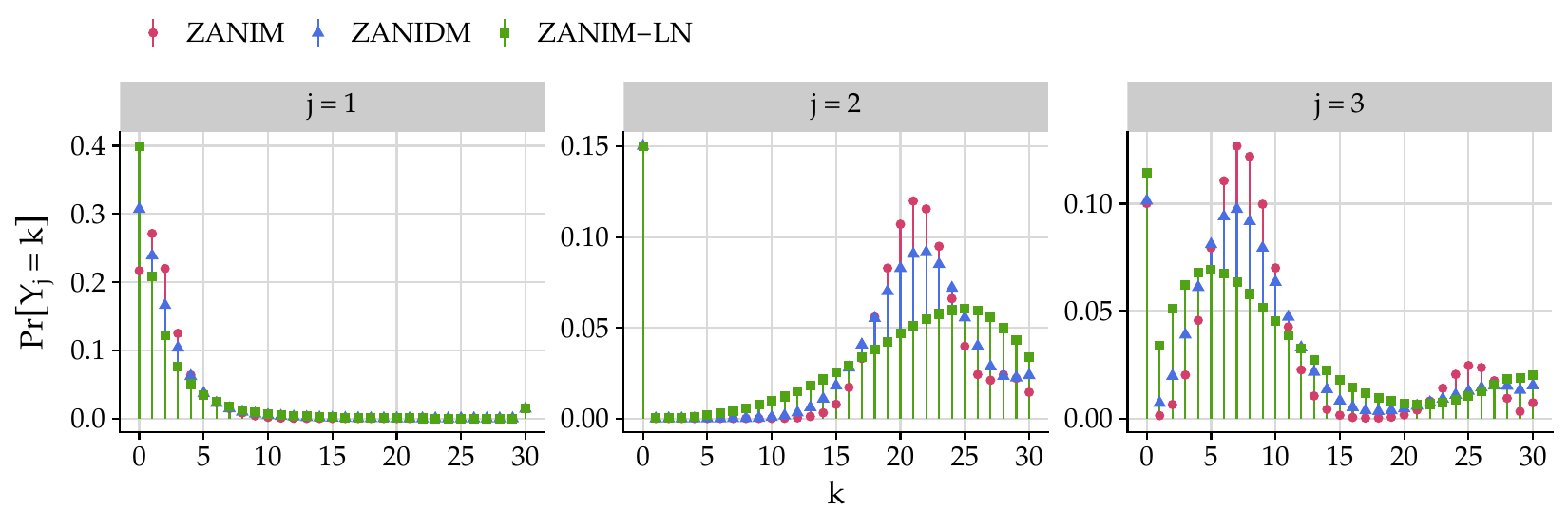}
    \caption{Comparison of the marginal PMFs of ZANIM (red circles), ZANIDM (blue triangles), and ZANIM-LN (green squares), with
    $\bm{\theta} \in \{0.05, 0.70, 0.25\}$ for ZANIM, $\bm{\alpha} \in \{2.0, 28.0, 10.0\}$ for ZANIDM, and
    $\bm{\alpha} \in \{0.031, 0.770, 0.241\}$ for ZANIM-LN,
    along with the category-specific probabilities of structural zeros
    $\bm{\zeta} \in \{0.05, 0.15, 0.10\}$ and $N = 30$ trials in each case. These parameter settings ensure that
    all three distributions have the same expectations, with $\mathbb{E}\lbrack \mathbf{Y}\rbrack=(2.320,18.496,9.161)$.}
    \label{fig:comparison_marginal_pmfs}
\end{figure}

The marginal PMF plots are given in \autoref{fig:comparison_marginal_pmfs}. We clearly see that the ZANIM-LN distribution
maintains the zero- and $N$-inflation characteristics while accommodating more overdispersed counts than both the ZANIM and ZANIDM
distributions. In the first marginal, $Y_1$, we observe a large spike at $k=0$, under all three distributions, which consists
not only of structural zeros, but also many sampling zeros. Interestingly, this spike is large in the ZANIM-LN distribution,
suggesting that it is capable of accounting for zeros due to overdispersion. The second marginal, $Y_2$, illustrates
the large presence of structural zeros for all three distributions, which is expected given that $\zeta_2 = 0.15$. It is also evident how the
ZANIM-LN distribution is more overdispersed than its counterparts. Finally, the multimodality evident in the plots of the third marginal, $Y_3$, shows the finite
mixture nature of all three distributions. We can also clearly see that the ZANIM-LN distribution has a heavy right tail consistent
with the high degree of overdispersion.

A similar Monte Carlo approximation using the laws of iterated expectations and
variances was employed to obtain the moments of the ZANIM-LN distribution. In \autoref{tab:moments}, we compare
the theoretical means and variances under ZANIM, ZANIDM, and ZANIM-LN for the above $d=3$ setting.
We also include the theoretical dispersion index,
$\operatorname{DI}\lbrack Y_j\rbrack = \operatorname{Var}\lbrack Y_j \rbrack / \mathbb{E}\lbrack Y_j \rbrack$,
and the following zero-inflation index:
\[\operatorname{ZI}_B\left\lbrack Y_j\right\rbrack= 1 + \left(\operatorname{Var}\left\lbrack Y_j \right\rbrack - \mathbb{E}\left\lbrack Y_j\right\rbrack\right) \frac{\log \Pr\left\lbrack Y_j = 0 \right\rbrack}{\mathbb{E}^2\left\lbrack Y_j \right\rbrack\log \operatorname{DI}\left\lbrack Y_j\right\rbrack},\]
which was proposed by \citet{Moreno2019}.
Note that the $\operatorname{ZI}_B\lbrack Y_j\rbrack$ index differs from our adaptation of the $\operatorname{ZI}_b\lbrack Y_j\rbrack$ index of \citet{Kim2018} employed in Section \ref{sec:application} of the main paper and Sections \ref{supp:descriptive_analysis_pollen} and \ref{supp:human_gut_microbiome} of the Supplementary Material.
In particular, $\operatorname{ZI}_B\lbrack Y_j\rbrack$ takes into account the empirical variance and the dispersion index, such that it measures the extent to which the zeros can be explained by the overdispersion.
In addition, when the marginal distribution of $Y_j$ follows a binomial or negative-binomial distribution,
then $\operatorname{ZI}_B\lbrack Y_j\rbrack=0$, indicating no zero-inflation,
while when $\operatorname{ZI}_B>0$, we have evidence of zero-inflation that cannot be described only by the overdispersion.
Consistent with the PMF plots from \autoref{fig:comparison_marginal_pmfs}, we can see that ZANIM-LN has the largest
variance in all three marginals, highlighting its greater flexibility in modelling overdispersion.
It is interesting to note that
$\operatorname{ZI}_B\lbrack Y_1\rbrack < 0$, suggesting that zero-inflation in this marginal could be explained by the
finite mixture nature of the distribution and its ability to capture overdispersion, which is consistent with the low values
for the structural zero probability $\zeta_1=0.05$ and also for the count probability parameters:
$\theta_1=0.05$ for ZANIM, $\alpha_1=2.0$ for ZANIDM, and $\alpha_1=0.031$ for ZANIM-LN.
In contrast, for the second and third marginals, the fact that $\operatorname{ZI}_B\lbrack Y_j\rbrack>0$ shows the ability
of the models to handle structural zeros.

\begin{table}[!ht]
\centering
\scriptsize
\caption{Comparison of the theoretical moments, dispersion indices, and zero-inflation indices of the ZANIM, ZANIDM, and ZANIM-LN distributions, with
$\bm{\theta} \in \{0.05, 0.70, 0.25\}$ for ZANIM, $\bm{\alpha} \in \{2.0, 28.0, 10.0\}$ for ZANIDM, and
    $\bm{\alpha} \in \{0.031, 0.770, 0.241\}$ for ZANIM-LN, along with $\bm{\zeta} \in \{0.05, 0.15, 0.10\}$ and $N=30$ trials in each case. These parameter settings ensure that
    all three distributions have the same expectations, i.e., $\mathbb{E}\lbrack \mathbf{Y}\rbrack=(2.320,18.496,9.161)$.}
    \label{tab:moments}
\begin{tabular}{llrrrr}
  \hline
 & Distribution & $\mathbb{E}\lbrack Y_j\rbrack$ & $\operatorname{Var}\lbrack Y_j\rbrack$ & $\operatorname{DI}\lbrack Y_j\rbrack$ & $\operatorname{ZI}_B\lbrack Y_j\rbrack$ \\
  \hline
\multirow{3}{*}{$j = 1$} & ZANIM & 2.320 & 14.326 & 6.174 & $-0.873$ \\
   & ZANIDM & 2.320 & 16.392 & 7.064 & $-0.577$ \\
   & ZANIM-LN  & 2.320 & 19.002 & 8.189 & $-0.351$ \\
   \hdashline
\multirow{3}{*}{$j = 2$} & ZANIM & 18.496 & 69.178 & 3.740 & 0.787 \\
   & ZANIDM & 18.496 & 72.723 & 3.932 & 0.780 \\
   & ZANIM-LN & 18.496 & 86.981 & 4.703 & 0.755 \\
   \hdashline
\multirow{3}{*}{$j = 3$} & ZANIM & 9.161 & 50.409 & 5.502 & 0.337 \\
   & ZANIDM & 9.161 & 54.658 & 5.966 & 0.305 \\
   & ZANIM-LN  & 9.161 & 65.776 & 7.180 & 0.258 \\
 \hline
\end{tabular}
\end{table}

 \section{Additional details on inference for the ZANIM-LN-BART model}\label{supp:inference_zanim_ln}

Following the notation of the main paper where $\smash{f_j^{(\mathrm{c})}(\mathbf{x}_i)}$ and $\smash{f_j^{(0)}(\mathbf{x}_i)}$
are the category-specific BART priors for the population-level count and structural zero probabilities
given in \eqref{eq:zanim_ln_bart_theta} and \eqref{eq:zanim_ln_bart_zeta}, respectively, we can thus
define the ZANIM-LN-BART model through the following stochastic representation:
\begin{align}
\left(z_{ij} \mid f_j^{(0)}(\mathbf{x}_i) \right) &\overset{\operatorname{ind.}}{\sim}
\operatorname{Bernoulli}\left\lbrack 1 - \Phi\left( f_j^{(0)}(\mathbf{x}_i) \right) \right\rbrack, \quad j\in\{1,\ldots, d\}, \nonumber\\ \nonumber
\mathbf{v}_i & \sim \operatorname{Normal}_{d-1}\left\lbrack \mathbf{0}_{d-1}, \bm{\Sigma}_V\right\rbrack, \\
\vartheta_{ij} &= \frac{z_{ij}f_j^{(\mathrm{c})}(\mathbf{x}_i)e^{u_{ij}}}{\sum_{k=1}^d z_{ij}f_k^{(\mathrm{c})}(\mathbf{x}_i)e^{u_{ik}}}
, \quad j\in\{1,\ldots, d\}, \label{eq:zanim_ln_bart_sr} \\ \nonumber
(\mathbf{Y}_i \mid N_i, \bm{\vartheta}_{i}) &\sim
\begin{cases} \delta_{\mathbf{0}_d}(\cdot), & \textrm{if} \: z_{ij} = 0 \: \forall\: j\in\{1,\ldots, d\}, \nonumber\\
\operatorname{Multinomial}_d\left\lbrack N_i, \vartheta_{i1}, \ldots, \vartheta_{id}\right\rbrack,
& \textrm{otherwise},
\end{cases}
\end{align}
where $\mathbf{u}_i=\mathbf{B}\mathbf{v}_i$ and $\mathbf{B}\in\mathbb{R}^{d\times(d-1)}$ is an orthogonal matrix,
such that $\sum_{j=1}^d u_{ij}=0$. Setting $\mathbf{u}_i=\mathbf{0}_d$ leads to the ZANIM-BART model.

\subsection{Augmented likelihood}

Combining the stochastic representation of the ZANIM-LN-BART model given in \eqref{eq:zanim_ln_bart_sr} with the data augmentation
in \eqref{eq:full_conditional_phi_i}, we can derive the following augmented likelihood contribution for the $i$-th observation:
\begin{align*}
\mathscr{L}_i & \propto \prod_{j=1}^d
\left\{
\left\lbrack \Phi\left(f^{(0)}_j(\mathbf{x}_i)\right)\right\rbrack^{1-z_{ij}}
\left\lbrack 1 - \Phi\left( f^{(0)}_j(\mathbf{x}_i)\right)\right\rbrack^{z_{ij}}
\left\lbrack \dfrac{z_{ij}f_j^{(\mathrm{c})}(\mathbf{x}_i)e^{u_{ij}}}{
\sum_{k=1}^dz_{ik}f_k^{(\mathrm{c})}(\mathbf{x}_i)e^{u_{ik}}}\right\rbrack^{y_{ij}}
\right\}\\
&\phantom{\propto}~\times
\left\lbrack
\sum_{j=1}^dz_{ij}f_j^{(\mathrm{c})}(\mathbf{x}_i)e^{u_{ij}}
\right\rbrack^{N_i} \exp\left\lbrack-\phi_i\sum_{j=1}^dz_{ij}f_j^{(\mathrm{c})}(\mathbf{x}_i)e^{u_{ij}}\right\rbrack\\
& \propto \prod_{j=1}^d
\left\{
\left\lbrack \Phi\left(f^{(0)}_j(\mathbf{x}_i)\right)\right\rbrack^{1-z_{ij}}
\left\lbrack 1 - \Phi\left( f^{(0)}_j(\mathbf{x}_i)\right)\right\rbrack^{z_{ij}}
\left\lbrack z_{ij}f_j^{(\mathrm{c})}(\mathbf{x}_i)e^{u_{ij}} \right\rbrack^{y_{ij}}
\right\}\\
&\phantom{\propto}~\times
\exp\left\lbrack-\phi_i\sum_{j=1}^dz_{ij}f_j^{(\mathrm{c})}(\mathbf{x}_i)e^{u_{ij}}\right\rbrack\\
& \propto \prod_{j=1}^d
\left\{
\left\lbrack \Phi\left(f^{(0)}_j(\mathbf{x}_i)\right)\right\rbrack^{1-z_{ij}}
\left\lbrack 1 - \Phi\left( f^{(0)}_j(\mathbf{x}_i)\right)\right\rbrack^{z_{ij}}
\left\lbrack z_{ij}f_j^{(\mathrm{c})}(\mathbf{x}_i)\right\rbrack^{y_{ij}}
e^{-\phi_i z_{ij} f_j^{(\mathrm{c})}(\mathbf{x}_i)e^{u_{ij}}}
\right\}\\
& \propto \prod_{j=1}^d
\left\{
\left\lbrack \Phi\left(f^{(0)}_j(\mathbf{x}_i)\right)\right\rbrack^{1-z_{ij}}
\left\lbrack 1 - \Phi\left( f^{(0)}_j(\mathbf{x}_i)\right)\right\rbrack^{z_{ij}}
\left\lbrack f_j^{(\mathrm{c})}(\mathbf{x}_i)\right\rbrack^{y_{ij}}
e^{-\phi_i z_{ij} f_j^{(\mathrm{c})}(\mathbf{x}_i)e^{u_{ij}}}
\right\},
\end{align*}
where we can drop the term $z_{ij}$ multiplying $\smash{f_j^{(\mathrm{c})}(\mathbf{x}_i)}$ in the third expression
because $z_{ij}=1$ when $y_{ij}>0$, by construction.
Evaluating the terms $\mathscr{L}_i$ across all $n$ observations yields the augmented likelihood in
\eqref{eq:augmented_likelihood_zanim_ln_bart}.

As stated, we clearly have that $z_{ij}=1$ when $y_{ij}>0$.
On the other hand, when $y_{ij}=0$, we have
\[p(z_{ij} = 0 \mid y_{ij}=0, \ldots) \propto \Phi\left(f^{(0)}_j(\mathbf{x}_i)\right)\]
and
\[
p(z_{ij} = 1 \mid y_{ij}=0, \ldots) \propto \left\lbrack 1-\Phi\left(f^{(0)}_j(\mathbf{x}_i)\right)\right\rbrack
e^{-\phi_i z_{ij} f_j^{(\mathrm{c})}(\mathbf{x}_i)e^{u_{ij}}},
\]
where the conditioning is with respect to the observed data and the parameters.
By normalising the above probabilities, we obtain the full conditional distribution of $z_{ij}\in\{0,1\}$ given in \eqref{eq:full_conditional_z_ij}.

Finally, to obtain the further augmented likelihood in \eqref{eq:augmented_likelihood_zanim_ln_bart_further},
we apply the data augmentation approach of \citet{Albert1993} to \eqref{eq:augmented_likelihood_zanim_ln_bart}.
We have that the conditional distribution of $w_{ij}$ given $z_{ij}$ is a truncated normal distribution
with probability density function given by
\[
p(w_{ij} \mid z_{ij}) = \left\lbrack\dfrac{\varphi(w_{ij}; f^{(0)}_j(\mathbf{x}_i), 1)}{1 - \Phi\left(f_0^{(j)}(\mathbf{x}_i) \right)} \right\rbrack^{z_{ij}}
\left\lbrack\dfrac{\varphi(w_{ij}; f^{(0)}_j(\mathbf{x}_i), 1)}{\Phi\left(f_0^{(j)}(\mathbf{x}_i) \right)} \right\rbrack^{1-z_{ij}},
\]
where $\varphi(x; \mu, \sigma^2)$ is Gaussian probability density function with mean $\mu$ and variance $\sigma^2$. Then, the augmented likelihood for the structural zero BART component of category $j$ takes the form
\begin{align*}
p(w_{ij} \mid z_{ij}) p(z_{ij}) &=
\left\lbrack 1 - \Phi\left(f_j^{(0)}(\mathbf{x}_i) \right)\right\rbrack^{z_{ij}}\Phi\left(f_j^{(0)}(\mathbf{x}_i) \right)^{z_{ij}}\\
&\phantom{=}\times
\left\lbrack\dfrac{\varphi\left(w_{ij}; f^{(0)}_j(\mathbf{x}_i), 1\right)}{1 - \Phi\left(f_j^{(0)}(\mathbf{x}_i) \right)} \right\rbrack^{1 - z_{ij}}
\left\lbrack\dfrac{\varphi\left(w_{ij}; f^{(0)}_j(\mathbf{x}_i), 1\right)}{\Phi\left(f_j^{(0)}(\mathbf{x}_i) \right)} \right\rbrack^{1-z_{ij}}\\
&=\varphi\left(w_{ij}; f^{(0)}_j(\mathbf{x}_i), 1\right),
\end{align*}
Using the above result for all categories $j\in\{1,\ldots,d\}$ and observations $i\in \{1,\ldots,n\}$
in conjunction with \eqref{eq:augmented_likelihood_zanim_ln_bart}, we obtain the desired
expression in \eqref{eq:augmented_likelihood_zanim_ln_bart_further}.

\subsection{Posterior inference}

Following the notation defined in Section \ref{sec:posterior_inference}, we
present the computations of the integrated likelihood functions of the trees and the full conditional
distributions of their corresponding terminal node parameters, which are given in \eqref{eq:integrated_likelihood_lambdas}
and \eqref{eq:full_conditional_lambdas} for the compositional probabilities $\smash{f_j^{(\mathrm{c})}}$ and
in \eqref{eq:integrated_likelihood_mus} and \eqref{eq:full_conditional_mus} for the
structural zero probabilities $\smash{f_j^{(0)}}$, respectively.

Following \citet{Murray2021}, let
$f^{(\mathrm{c})}_{(h)j}(\mathbf{x}_i) = \prod_{\ell \neq h}
g(\mathbf{x}_i; \mathcal{T}^{(\mathrm{c})}_{\ell j}, \Lambda^{(\mathrm{c})}_{\ell j})$.
From the augmented likelihood function of the ZANIM-LN-BART model given in \eqref{eq:augmented_likelihood_zanim_ln_bart_further},
the conditional likelihood function for $(\mathcal{T}^{(\mathrm{c})}_{hj}, \Lambda_{hj})$ can be derived as follows
\begin{align*}
\mathscr{L}\left(\mathcal{T}^{(\mathrm{c})}_{hj},\Lambda_{hj} \mid \mathcal{T}^{(\mathrm{c})}_{(h)j}, \Lambda_{(h)j}, \cdots\right)
&\propto
\prod_{i=1}^n \left\lbrack f_j^{(\mathrm{c})}(\mathbf{x}_i)\right\rbrack^{y_{ij}} \exp\left\{
-\phi_i z_{ij} e^{u_{ij}} f_j^{(\mathrm{c})}(\mathbf{x}_i)
\right\}\\
&\propto
\prod_{i=1}^n \left\lbrack
f^{(\mathrm{c})}_{(h)j}(\mathbf{x}_i) g(\mathbf{x}_i; \mathcal{T}^{(\mathrm{c})}_{hj}, \Lambda^{(\mathrm{c})}_{h})
\right\rbrack^{y_{ij}}\\
&\phantom{=}~\times \exp\left\{
-\phi_i z_{ij} e^{u_{ij}}
f^{(\mathrm{c})}_{(h)j}(\mathbf{x}_i) g(\mathbf{x}_i; \mathcal{T}^{(\mathrm{c})}_{hj}, \Lambda^{(\mathrm{c})}_{h})
\right\}\\
&\propto
\prod_{t \in \mathcal{L}^{(\mathrm{c})}_{hj}}
\prod_{i\colon \mathbf{x}_i \in \mathcal{A}^{(\mathrm{c})}_{htj}}\left\lbrack
f^{(\mathrm{c})}_{(h)j}(\mathbf{x}_i) \lambda_{hjt}\right\rbrack^{y_{ij}}
e^{
-\phi_i z_{ij} e^{u_{ij}} f^{(\mathrm{c})}_{(h)j}(\mathbf{x}_i) \lambda_{hjt}
}
\\
&\propto
\prod_{t \in\mathcal{L}^{(\mathrm{c})}_{hj}}\lambda_{hjt}^{r^{(\mathrm{c})}_{htj}}
e^{-s^{(\mathrm{c})}_{htj} \lambda_{hjt}},
\end{align*}
where $\smash{r^{(\mathrm{c})}_{htj}}=\sum_{i\colon\mathbf{x}_i \in \mathcal{A}^{(\mathrm{c})}_{htj}} y_{ij}$
and $\smash{s^{(\mathrm{c})}_{htj}}=\sum_{i\colon\mathbf{x}_i \in \mathcal{A}^{(\mathrm{c})}_{htj}}\phi_i z_{ij} e^{u_{ij}} \smash{f^{(\mathrm{c})}_{(h)j}(\mathbf{x}_i)}$
play the role of sufficient statistics.

Using the above expression for the conditional distribution of $\smash{(\mathcal{T}^{(\mathrm{c})}_{hj}, \Lambda_{hj})}$,
and recalling the conditionally conjugate priors,
$\smash{\lambda_{htj} \overset{\operatorname{ind.}}{\sim} \operatorname{Gamma}\lbrack c_0, d_0 \rbrack}$,
we obtain the full conditional posterior density for the terminal node parameters as follows:
\begin{equation*}
\pi\left(\Lambda_{(h)j} \mid \mathcal{T}^{(\mathrm{c})}_{hj}, \mathcal{T}^{(\mathrm{c})}_{(h)j}, \Lambda_{(h)j}, \cdots\right)
\propto
\prod_{t \in\mathcal{L}^{(\mathrm{c})}_{hj}}
\frac{d_0^{c_0}}{\Gamma(c_0)}
\lambda_{hjt}^{r^{(\mathrm{c})}_{htj} + c_0 - 1} \exp\left\{-(s^{(\mathrm{c})}_{htj}+d_0) \lambda_{hjt}\right\},
\end{equation*}
such that
$(\lambda_{hjt} \mid \cdot)\overset{\operatorname{ind.}}{\sim}\operatorname{Gamma}\left\lbrack r^{(\mathrm{c})}_{hjt} + c_0, s^{(\mathrm{c})}_{hjt} + d_0\right\rbrack$, as given in \eqref{eq:full_conditional_lambdas}.

The branching process prior proposed by \citet{Chipman1998} for the tree topologies is given by
\[
\pi_{\mathcal{T}}(\mathcal{T}_{h}) = \prod_{t \in \mathcal{B}_h} \left\lbrack a(1 + d_{ht})^{-b}\right\rbrack
\prod_{t \in \mathcal{L}_h} \left\lbrack 1 - a(1 + d_{ht})^{-b}\right\rbrack,
\]
where $d_{ht}$ is the depth of node $t$ in tree $h$ and $a \in (0,1)$ and $b\geq0$ are hyperparameters, with recommended values of $a = 0.95$ and $b = 2$. Note that $a (1 + d_{ht})^{-b}$ gives the probability of node $t$ of tree $h$ being an internal node at depth $d_{ht}$.
By considering this tree prior,
we obtain the integrated likelihood of tree $\smash{\mathcal{T}^{(\mathrm{c})}_{hj}}$ as follows:
\begin{align*}
\pi\left(\mathcal{T}^{(\mathrm{c})}_{hj} \mid \mathcal{T}^{(\mathrm{c})}_{(h)j}, \Lambda_{(h)j}, \cdots\right)
&\propto
\pi_{\mathcal{T}}\left(\mathcal{T}^{(\mathrm{c})}_{hj}\right)
\int
\prod_{t \in\mathcal{L}^{(\mathrm{c})}_{hj}}
\frac{d_0^{c_0}}{\Gamma(c_0)}
\lambda_{hjt}^{r^{(\mathrm{c})}_{htj} + c_0 - 1} \exp\left\{-(s^{(\mathrm{c})}_{htj}+d_0) \lambda_{hjt}\right\}\dd \lambda_{htj}\\
&\propto
\pi_{\mathcal{T}}\left(\mathcal{T}^{(\mathrm{c})}_{hj}\right)
\prod_{t \in \mathcal{L}^{(\mathrm{c})}_{hj}}
\frac{d_0^{c_0}}{\Gamma(c_0)}
\frac{\Gamma\left(r^{(\mathrm{c})}_{htj} + c_0\right)}{(s^{(\mathrm{c})}_{htj} + d_0)^{r^{(\mathrm{c})}_{htj}+c_0}}.
\end{align*}

Regarding the structural zero probabilities $\smash{f_j^{(0)}}$,
the steps of the derivation are straightforward and similar to those presented by \citet{Chipman2010}. Expressed in terms of the
vector of partial residuals $\smash{\mathbf{r}_{(h)j}}=(r_{(h)1j}, \ldots, r_{(h)nj})$,
the conditional likelihood function for $\smash{(\mathcal{T}^{(0)}_{hj},\mathcal{M}_{hj})}$ is given by
\begin{align*}
\mathscr{L}\left(\mathcal{T}^{(0)}_{hj},\mathcal{M}_{hj}; \mathbf{r}_{(h)j} \right) &\propto
\prod_{i=1}^n \exp \left\{-0.5 \left\lbrack r_{(h)ij} - g\left(\mathbf{x}_i, \mathcal{T}^{(0)}_{hj}, \mathcal{M}_{hj}\right)
\right\rbrack^2 \right\} \\
&\propto
\prod_{t\in \mathcal{L}^{(0)}_{hj}}
\prod_{i\colon \mathbf{x}_i \in \mathcal{A}^{(0)}_{htj}}
\exp\left\lbrack -0.5 \left(r_{(h)ij} - \mu_{hjt} \right)^2\right\rbrack \\
&\propto
\prod_{t\in \mathcal{L}^{(0)}_{hj}}\exp\left\lbrack -0.5 \left(n_{htj}\mu_{htj}^2 - 2\mu_{htj}s^{(0)}_{htj}\right)\right\rbrack,
\end{align*}
where $\smash{s^{(0)}_{htj}} = \sum_{i\colon \mathbf{x}_i \in \mathcal{A}^{(0)}_{htj}} r_{(h)ij}$ and $n_{htj}$ is the total number of observations in the partition $\smash{\mathcal{A}^{(0)}_{htj}}$.
The expressions in \eqref{eq:integrated_likelihood_mus} and \eqref{eq:full_conditional_mus} follow from
well-known calculations based on the conditional likelihood function for $\smash{(\mathcal{T}^{(0)}_{hj},\mathcal{M}_{hj})}$
given above, together with the conjugate normal prior assumed for the terminal node parameters, i.e.,
$\smash{\mu_{htj} \overset{\operatorname{ind.}}{\sim} \operatorname{Normal}\lbrack 0, \sigma^2_\mu\rbrack}$.

\subsubsection{Full conditional distributions for the random effects and factor analysis hyperparameters}

An additional step in the ZANIM-LN-BART model is to update the random effects
$\mathbf{u}_i$. As discussed in the main paper and explicitly written in \eqref{eq:zanim_ln_bart_sr}, we
consider a sum-to-zero constraint on the random effects $\mathbf{u}_i$ in order to ensure identifiability.
This implies that we effectively sample from the $(d-1)$ dimensional vector
$\mathbf{v}_i$ and map back through the transformation $\mathbf{u}_i=\mathbf{B}\mathbf{v}_i$. From \eqref{eq:zanim_ln_bart_sr}, the full conditional distribution of $\mathbf{v}_i$ is given by
\begin{equation}\label{eq:full_conditional_v}
\pi(\mathbf{v}_i \mid \mathbf{y}_i, \mathbf{z}_i, \bm{f}) \propto
\prod_{j=1}^d
\left(\frac{\exp\left\{ {(\mathbf{B}\mathbf{v}_i)}_{j} \right\} }{
\sum_{k=1}^dz_{ik}f_k^{(\mathrm{c})}(\mathbf{x}_i)
\exp\left\{ {(\mathbf{B}\mathbf{v}_i)}_{k} \right\}}\right)^{y_{ij}}\varphi\left(\mathbf{v}_i; \mathbf{0}_{d-1}, \bm{\Sigma}_V\right),
\end{equation}
where $\varphi\left(\mathbf{x}; \mathbf{\mu}_{d}, \bm{\Sigma}\right)$ denotes the density of the $d$-dimensional multivariate normal distribution
with mean $\bm{\mu}$ and covariance matrix $\bm{\Sigma}$.
We sample from \eqref{eq:full_conditional_v} using the elliptical slice sampling algorithm \citep{Murray2010}. As discussed in Section \ref{sec:priors}, we assume a factor-analytic hyperprior on $\mathbf{v}_i$
via the decomposition $\mathbf{v}_i = \bm{\Gamma} \bm{\eta}_i + \bm{\epsilon}_i$,
where
$\bm{\Gamma} = \left\lbrack \gamma_{jk}\right\rbrack_{j=1,\ldots,d-1}^{k=1,\ldots,q}$ is a
$(d-1) \times q$ factor loading matrix,
$\bm{\eta}_i \sim \operatorname{Normal}_q\lbrack \mathbf{0}_q, \mathbf{I}_q\rbrack$, and
$\bm{\epsilon}_i \sim \operatorname{Normal}_{d-1}\left\lbrack\mathbf{0}_{d-1}, \bm{\Psi}\right\rbrack$, with
$\bm{\Psi}=\operatorname{diag}\left\{ \psi_1, \ldots, \psi_{d-1}\right\}$.
We further adopt the shrinkage-inducing multiplicative gamma process prior (MGP) of \citet{Bhattacharya2011}, which is
defined as follows
\begin{align*}
(\gamma_{jk} \mid \rho_{jk}, \tau_k) \sim \operatorname{Normal}\left\lbrack 0, \rho^{-1}_{jk}\tau^{-1}_{k}\right\rbrack,
\quad
\rho_{jk} \sim \operatorname{Gamma}\left\lbrack \nu/2, \nu/2\right\rbrack,  \tau_k =\prod_{\ell=1}^k\varrho_\ell\\
\varrho_1 \sim \operatorname{Gamma}\left\lbrack a_1, 1\right\rbrack, \quad
\varrho_h \sim \operatorname{Gamma}\left\lbrack a_2, 1\right\rbrack,\:h \geq 2, \quad
\psi_j \sim \operatorname{Gamma}\left\lbrack a_\psi, b_\psi\right\rbrack,
\end{align*}
where $\tau_{k}$ is a global shrinkage parameter for the $k$-th column, and $\lbrack\rho_{jk}\rbrack_{j=1,\ldots,d-1}^{k=1,\ldots,q}$ are local shrinkage
parameters for each corresponding element of $\bm{\Gamma}$.

Let $\mathbf{V} \in \mathbb{R}^{n \times d-1}$ be the matrix whose rows are denoted by
$\{\mathbf{v}_i\}_{i=1}^n \in \mathbb{R}^{d-1}$ and let
$\mathbf{v}^{(j)} \in \mathbb{R}^n$ denote the $j$-th column of $\mathbf{V}$.
Similarly, let $\mathbf{H} \in \mathbb{R}^{n \times q}$ denote the factor score matrix
with rows $\bm{\eta}_1, \ldots, \bm{\eta}_n \in \mathbb{R}^q$.
From Bayesian linear regression results, the full conditional distribution of $\bm{\gamma}_j$ is
\begin{equation}\label{eq:full_conditional_fa_gamma}
\left(\bm{\gamma}_j \mid \mathbf{v}^{(j)}, \mathbf{H}\right) \sim \operatorname{Normal}_{q}\left\lbrack
\left( D_j^{-1} + \psi^{-1}_j \mathbf{H}^\top\mathbf{H}\right)^{-1}
\psi^{-1}_j\mathbf{H}^\top \mathbf{v}^{(j)},
\left(D_j^{-1} + \psi^{-1}_j \mathbf{H}^\top\mathbf{H}\right)^{-1}
\right\rbrack.
\end{equation}
where $D_j^{-1} = \operatorname{diag}\left(\rho_{j1}\tau_1, \ldots, \rho_{jq}\tau_q\right)$.
From conditional results of the multivariate normal distribution, the full conditional distribution
of $\bm{\eta}_i$ is
\begin{equation}\label{eq:full_conditional_fa_eta}
\left( \bm{\eta}_i \mid \mathbf{v}_i, \bm{\Gamma}, \bm{\Psi} \right)
\sim \operatorname{Normal}_q\left\lbrack
\left(\mathbf{I}_q + \bm{\Gamma}^\top \Psi^{-1}\bm{\Gamma}\right)^{-1}\bm{\Gamma}\Psi^{-1} \mathbf{v}_i,
\left(\mathbf{I}_q + \bm{\Gamma}^\top \Psi^{-1}\bm{\Gamma}\right)^{-1}\right\rbrack.
\end{equation}
As both \eqref{eq:full_conditional_fa_gamma} and \eqref{eq:full_conditional_fa_eta} are multivariate normal distributions, we note that both updates can be performed efficiently by employing block updates \citep{Rue2005}. It is also easy to see that the full conditional distribution of $\psi_j$ is
\begin{equation}\label{eq:full_conditional_fa_psi}
\!(\psi^{-1}_j \mid \mathbf{v}^{(j)}, \bm{\gamma}_j, \mathbf{H}) \sim \operatorname{Gamma}\left\lbrack n/2 + a_\psi,
\left(\mathbf{v}^{(j)} - \mathbf{H}\bm{\gamma}_j\right)^\top\left(
\mathbf{v}^{(j)} - \mathbf{H}\bm{\gamma}_j\right)/2+ b_\psi
\right\rbrack.
\end{equation}
Finally, the full conditional of the local shrinkage parameters $\rho_{jk}$ is given by
\begin{equation}\label{eq:full_conditional_local_shrinkage}
(\rho_{jk} \mid \cdots) \sim \operatorname{Gamma}\left\lbrack \frac{\nu + 1}{2}, \frac{\nu + \tau_k \gamma^2_{jk}}{2} \right\rbrack,
\end{equation}
for $j=1\ldots, (d-1)$ rows and  $k=1,\ldots, q$ columns.
The full conditional distributions of the global shrinkage parameters $\tau_k=\prod_{\ell=1}^k \varrho_\ell$
are updated sequentially, after sampling from the full conditional distribution of $\varrho_1$,
given by
\begin{equation}\label{eq:full_conditional_global_shrinkage_1}
(\varrho_1 \mid \cdots) \sim \operatorname{Gamma}\left\lbrack
a_1 + (d-1)q/2, 1 + 0.5 \sum_{\ell=1}^q
\left(\tau^{(1)}_\ell \sum_{j=1}^{d-1} \rho_{j\ell}\gamma^2_{j\ell}\right)
\right\rbrack,
\end{equation}
and, for $h \geq 2$, the full conditional distribution of $\varrho_h$, given by
\begin{equation}\label{eq:full_conditional_global_shrinkage_h_geq_2}
(\varrho_h \mid \cdots) \sim \operatorname{Gamma} \left\lbrack
a_2 + (d-1)/2 (q - h + 1), 1 + 0.5
\sum_{\ell=h}^q \left(\tau^{(h)}_{\ell} \sum_{j=1}^{d-1}\rho_{j\ell}\gamma^2_{j\ell}\right)
\right\rbrack,
\end{equation}
where $\tau^{(h)}_\ell = \varrho^{-1}_h\prod_{t=1}^\ell \varrho_t$.

\subsection{Inferential algorithm}
To present the algorithm, we denote the transition kernels of the tree proposals
by $\smash{K(\mathcal{T}^{(\mathrm{c})}_{h},\cdot)}$ and $\smash{K(\mathcal{T}^{(0)}_{h},\cdot)}$.
Our implementation uses the mixture of grow, prune, and change proposals  introduced in \citet{Chipman1998}.
We refer to \citet{Kapelner2016} for further descriptions of these transition kernels in the classical BART model of
\citet{Chipman2010}, which we adapt for the trees of our ZANIM-LN-BART model.
Algorithm \ref{alg:zanim_ln_bart} gives our implementation of the inference scheme for the ZANIM-LN-BART model in full.

\begin{algorithm}[!hb]
\caption{MCMC algorithm to fit the ZANIM-LN-BART model.\label{alg:zanim_ln_bart}}
\setstretch{1.125}
\SetAlgoLined
\SetAlgoItemize
\SetKwInput{KwInput}{Input}
\SetKwInput{KwOutput}{Output}
\SetKwInput{KwInit}{Initialise}
\KwInput{Data $\{\mathbf{y}_i, \mathbf{x}_i\}_{i=1}^n$, numbers of trees $m_\theta$ and $m_\zeta$, number of iterations $R$,
and all hyperparameters of the priors.}
\KwInit{
The BART priors parameters associated with $\smash{f^{(\mathrm{c})}_j(\mathbf{x}_i)}$ and $\smash{f^{(0)}_j(\mathbf{x}_i)}$ for all $j\in\{1,\ldots,d\}$ categories:
$\smash{\{\mathcal{T}^{(\mathrm{c})}_{hj}, \Lambda_{hj}\}_{h=1}^{m_\theta}}$ and
$\smash{\{\mathcal{T}^{(0)}_{hj}, \mathcal{M}_{hj}\}_{h=1}^{m_\zeta}}$, respectively.
The tree topologies are initialised as stumps and the terminal node parameters are set to $\lambda_{h1j}=1.0$ and $\mu_{h1j}=0.0$.
Draw from the priors of $z_{ij}$ and $\mathbf{v}_i$ and set $\mathbf{u}_i=\mathbf{B}\mathbf{v}_i$.}

\For{iterations $t$ from $1$ to $R$}{

    Update $(\phi_i \mid \mathbf{y}_i, \mathbf{z}_i, \mathbf{u}_i, \bm{f}^{(c)}(\mathbf{x}_i))$ from its full conditional in \eqref{eq:full_conditional_phi_i}.

    \For{categories $j$ from $1$ to $d$}{
    Update $(z_{ij} \mid y_{ij}, u_{ij}, \phi_{i}, f_j^{(\mathrm{c})}, f_j^{(0)})$ from its
    full conditional in \eqref{eq:full_conditional_z_ij}.

    Update $w_{ij} \sim \operatorname{TN}_{\lbrack -\infty, 0 \rbrack}\lbrack f_0^{(j)}(\mathbf{x}_i), 1\rbrack$
    if $z_{ij}=1$ and
    $\operatorname{TN}_{\lbrack 0, \infty\rbrack}\lbrack f_0^{(j)}(\mathbf{x}_i), 1 \rbrack$
    if $z_{ij}=0,\:\forall\:i \in \{1,\ldots, n\}$.

    \For{trees $h$ from $1$ to $m_\theta$}{

        Sample $\mathcal{T}^{\prime} \sim K(\cdot \mid \mathcal{T}^{(\mathrm{c})}_{hj})$.

        Compute the acceptance probability using the integrated likelihood in \eqref{eq:integrated_likelihood_lambdas} as follows:
        $$
        \alpha\left(\mathcal{T}^{(\mathrm{c})}_{hj}, \mathcal{T}^{\prime}\right) =
        1 \land \frac{\pi\left(\mathcal{T}^{\prime} \mid \mathcal{T}^{(\mathrm{c})}_{(h)j}, \Lambda_{(h)j}, \ldots\right)
        K\left(\mathcal{T}^{(\mathrm{c})}_{hj} \mid \mathcal{T}^{\prime}\right) }{
        \pi\left(\mathcal{T}^{(\mathrm{c})}_{hj} \mid \mathcal{T}^{(\mathrm{c})}_{(h)j}, \Lambda_{(h)j}, \ldots\right)
        K\left(\mathcal{T}^{\prime} \mid \mathcal{T}^{(\mathrm{c})}_{hj}\right)}.
        $$

        Set $\smash{\mathcal{T}^{(\mathrm{c})}_{hj}} = \mathcal{T}^{\prime}$ with probability
        $\alpha(\mathcal{T}^{(\mathrm{c})}_{hj}, \mathcal{T}^{\prime})$.

        Sample from the full conditionals of the terminal node parameters $\Lambda_{hj}$ in \eqref{eq:full_conditional_lambdas}.
    }

    \For{trees $h$ from $1$ to $m_\zeta$}{

        Sample $\mathcal{T}^{\prime} \sim K(\cdot \mid \mathcal{T}^{(0)}_{hj})$.

        Compute the acceptance probability using the integrated likelihood in \eqref{eq:integrated_likelihood_mus} as follows:
        $$
        \alpha\left(\mathcal{T}^{(0)}_{hj}, \mathcal{T}^{\prime}\right) =
        1 \land \frac{\pi\left(\mathcal{T}^{\prime} \mid \mathcal{T}^{(0)}_{(h)j}, \mathcal{M}_{(h)j}, \ldots\right)
        K\left(\mathcal{T}^{(0)}_{hj} \mid \mathcal{T}^{\prime}\right) }{
        \pi\left(\mathcal{T}^{(0)}_{hj} \mid \mathcal{T}^{(0)}_{(h)j}, \mathcal{M}_{(h)j}, \ldots\right)
        K\left(\mathcal{T}^{\prime} \mid \mathcal{T}^{(0)}_{hj}\right)}.
        $$

        Set $\smash{\mathcal{T}^{(0)}_{hj}} = \mathcal{T}^{\prime}$ with probability
        $\alpha(\mathcal{T}^{(0)}_{hj}, \mathcal{T}^{\prime})$.

        Sample from the full conditionals of the terminal node parameters $\mathcal{M}_{hj}$ in \eqref{eq:full_conditional_lambdas}.
    }

    Update $\mathbf{v}_i$ from its full conditional in \eqref{eq:full_conditional_v} using elliptical slice sampling,
    and then set $\mathbf{u}_i = \mathbf{B}\mathbf{v}_i$.

    Update the factor analysis hyperparameters $\bm{\gamma}_i$, $\bm{\eta}_i$, and $\psi_j$ from their full conditional
    distributions in \eqref{eq:full_conditional_fa_gamma}, \eqref{eq:full_conditional_fa_eta}, and \eqref{eq:full_conditional_fa_psi}, respectively.

    Update the shrinkage-inducing MGP hyperparameters $\rho_{jk}$, $\varrho_1$, and $\varrho_h,\:h\geq2$ from their respective full conditional
    distributions in \eqref{eq:full_conditional_local_shrinkage}, \eqref{eq:full_conditional_global_shrinkage_1}, and
    \eqref{eq:full_conditional_global_shrinkage_h_geq_2}, along with $\tau_k$.

}

}
\end{algorithm}

 \section{Additional results for the simulation studies}\label{supp:add_simstudies}

Here we present additional simulation results in Sections \ref{supp:add_simstudy1} and \ref{supp:add_simstudy2} for both scenarios discussed in Sections \ref{sec:simstudy1} and \ref{sec:simstudy2}, respectively.
We note that convergence assessments were also performed for both scenarios, but these have been deferred to Section \ref{supp:convergence}.

\subsection{Scenario 1}\label{supp:add_simstudy1}

\begin{figure}[!ht]
    \centering
    \includegraphics[width=1\linewidth]{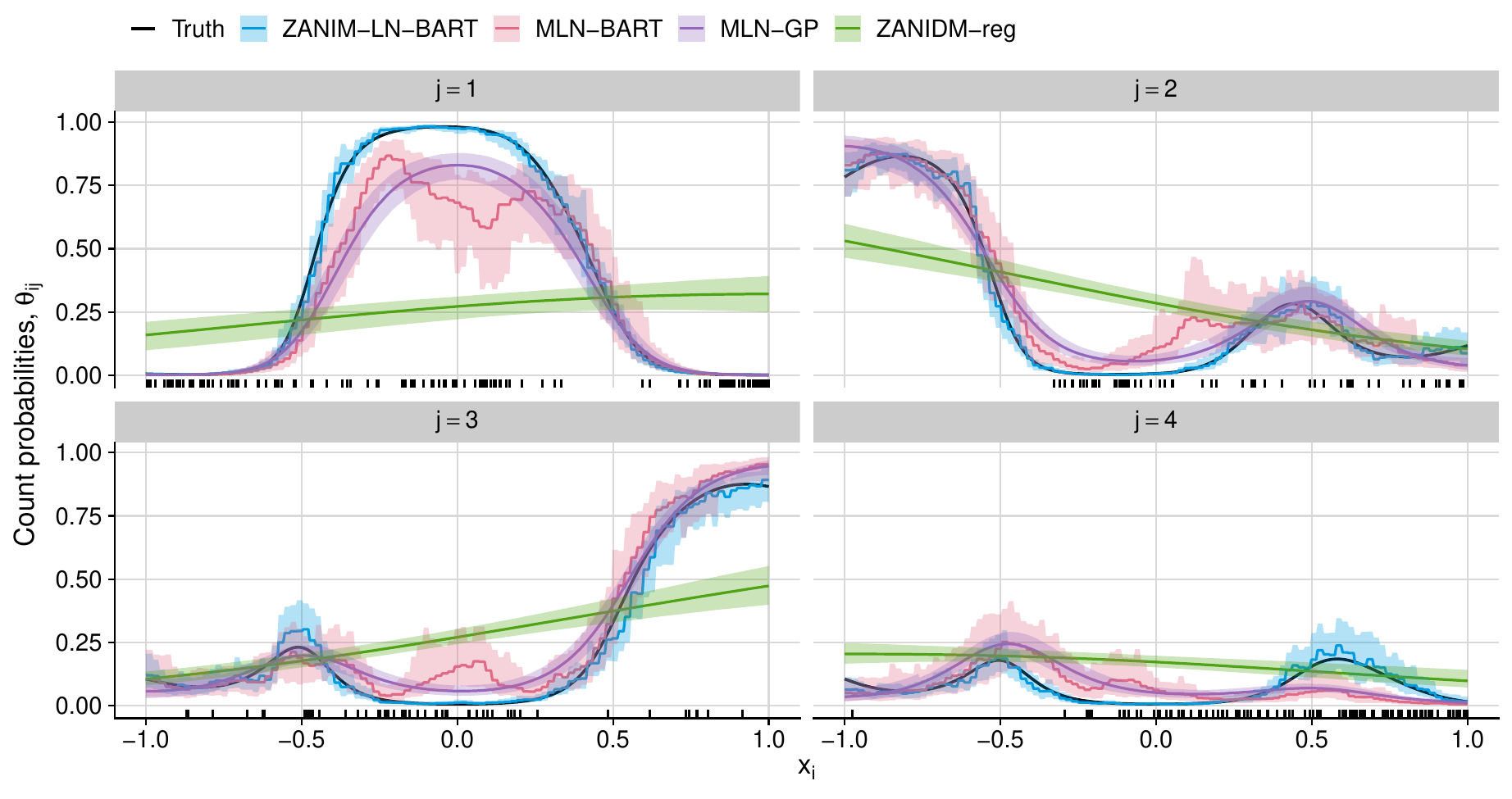}
    \caption{ZANIM-LN-BART, MLN-BART, MLN-GP, and ZANIDM-reg estimates of the true population-level count probabilities $\theta_{ij}$ (black lines) for $d=4$ categories under the ZANIM-LN DGP. The posterior median and $95\%$ credible intervals are given in each case. The rugs along the $x$-axes represent samples where the observed counts are zero.}
    \label{fig:theta_scenario_1_zanim_ln_fix}
\end{figure}

\begin{figure}[!hb]
    \centering
    \includegraphics[width=1\linewidth]{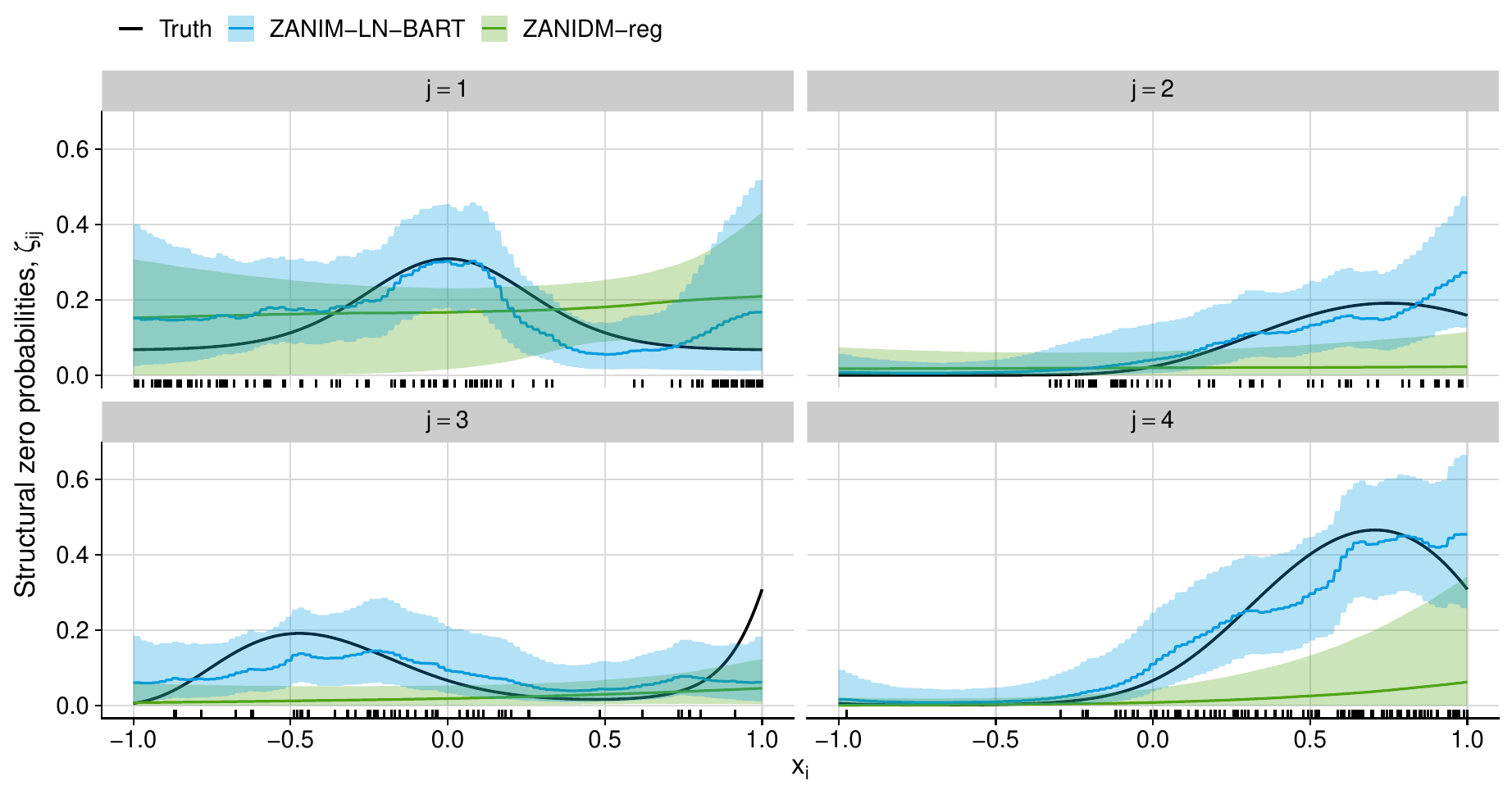}
    \caption{ZANIM-LN-BART and ZANIDM-reg estimates of the true structural zero probabilities $\zeta_{ij}$ (black lines) for $d=4$ categories under the ZANIM-LN DGP. The posterior median and $95\%$ credible intervals are given in each case. The rugs along the $x$-axes represent samples where the observed counts are zero.}
    \label{fig:zeta_scenario_1_zanim_ln_fix}
\end{figure}

In Section \ref{sec:simstudy1}, we presented Figures \ref{fig:count_prob_scenario_1_zanim} and \ref{fig:structural_prob_scenario_1_zanim}, which illustrate the posterior estimates for the population-level count and structural zero probabilities, respectively, under the ZANIM-BART, ML-BART, MLN-GP, and ZANIDM-reg models for the setting where the counts are simulated from the ZANIM distribution.
Here, we include complementary results when the data are generated from the ZANIM-LN distribution  in Figures \ref{fig:theta_scenario_1_zanim_ln_fix} and \ref{fig:zeta_scenario_1_zanim_ln_fix} for the ZANIM-LN-BART, MLN-BART, MLN-GP, and ZANIDM-reg models.
The ZANIM-BART and ML-BART models are not shown, as they have slightly worse performance than their equivalent models without logistic-normal random effects (see Table \ref{tab:summary_sim_1}).
Similar conclusions to those for Figures \ref{fig:count_prob_scenario_1_zanim} and \ref{fig:structural_prob_scenario_1_zanim} can be drawn from these plots.

\clearpage

\subsection{Scenario 2}\label{supp:add_simstudy2}

In Section \ref{sec:simstudy2}, we evaluated the performance of the proposed ZANIM-BART and ZANIM-LN-BART models along with several existing methods under a more challenging scenario, where the counts were simulated from the ZANIDM distribution.
Here, we present complementary simulation results, where the counts are simulated under the ZANIM-LN distribution with the same functional forms discussed in Section \ref{sec:simstudy2} for the population-level count probabilities $\theta_{ij}=\alpha_{ij}/\sum_{k=1}^d\alpha_{ik}$, and structural zero probabilities, $\zeta_{ij}$.
In this new distributional specification, the simulated counts have extra latent heterogeneity due to the Gaussian random effects, where the underlying covariance matrix has a factor-analytic representation, $\bm{\Sigma}_U = \bm{\Gamma}\bm{\Gamma}^\top + \bm{\Psi}$, with $q=10$ factors.
The entries of the factor loading matrix $\bm{\Gamma}$, with dimension $d\times q$, are simulated from a standard uniform distribution, while $\bm{\Psi}$ is a $d$-dimensional diagonal matrix with equally-spaced diagonal values from $0.25$ to $0.30$.
Under this scenario, the data have a similar degree of sparsity to the scenario reported in the main paper, but with more sampling zeros.
In particular, for the $d=20$ settings, nearly $37\%$ of all counts are zero, with $10\%$ sampling zeros and $27\%$ structural zeros.
When $d=40$, $41\%$ of all observations are zero, with $15\%$ sampling zeros and $26\%$ structural zeros.

Table \ref{tab:summary_sim_2__zanim_ln} summarises the simulation experiments based on six replicate data sets of the above scenario, where the counts are simulated from the ZANIM-LN distribution.
Here we find that the ZANIM-LN-BART model performs consistently better than the ZANIM-BART model in terms of recovering the true population-level count probabilities, as indicated by the lower $\operatorname{KL}(\bm{\theta})$ values, except in the setting with $d=40$ and
$n=200$, where ZANIM-BART has a marginally lower $\operatorname{KL}(\bm{\theta})$ value than ZANIM-LN-BART.

\begin{table}[!hb]
\centering
\caption{
Simulation results for Scenario 2 with varying $d \in \{20, 40\}$, $n \in {200, 500, 1000}$, and $p=6$. The counts are simulated from the ZANIM-LN distribution with true functional forms described in the main paper. The metrics are averaged over six replicate data sets.
Accuracy in the estimation methods are assessed via the Kullback-Leibler divergence (KL), while uncertainty quantification is
evaluated via the empirical coverage probabilities (CP) of the $95\%$ credible intervals. The runtime of the methods are reported in seconds (sec).
Smaller KL values indicate improved recovery, while coverage values close to the nominal level $(0.95)$ indicate well-calibrated
posterior uncertainty. Values in bold indicate the best performing method within each $(n,d)$ setting, while missing values correspond to non-existence of the given parameter under the given method.
}
\label{tab:summary_sim_2__zanim_ln}
\scriptsize
\begin{tabular}{lllllllllr}
  \toprule
$d$ & $n$ & Model & $\operatorname{KL}\left(\bm{\theta}\right)$ & $\operatorname{CP}\left(\bm{\theta}\right)$ & $\operatorname{KL}\left(\bm{\vartheta}\right)$ & $\operatorname{CP}\left(\bm{\vartheta}\right)$ & $\operatorname{KL}\left(\bm{\zeta}\right)$ & $\operatorname{CP}\left(\bm{\zeta}\right)$ & Time (sec) \\
  \midrule
\multirow{7}{*}{$20$} & \multirow{7}{*}{$200$} & ZANIM-BART & 0.3667 & 0.4318 & 0.0028 & 0.9182 & 0.0995 & 0.6797 & 240.6 \\
   &  & ZANIM-LN-BART & \textbf{0.3483} & \textbf{0.5928} & \textbf{0.0026} & \textbf{0.9565} & \textbf{0.0992} & \textbf{0.6840} & 325.8 \\
   &  & ZANIDM-reg & 0.9689 & 0.1499 & 0.0029 & 0.9354 & 0.3001 & 0.5854 & 77.6 \\
   &  & DM-reg & 1.2049 & 0.1062 & 0.0029 & 0.6810 &  &  & 57.1 \\
   &  & ML-BART & 2.1505 & 0.2704 &  &  &  &  & 208.5 \\
   &  & MLN-BART & 1.4367 & 0.4219 & 0.0035 & 0.6625 &  &  & 294.2 \\
   &  & MLN-GP & 0.7348 & 0.5350 & 0.0047 & 0.6827 &  &  & 11.0 \\
   \hdashline
\multirow{7}{*}{$20$} & \multirow{7}{*}{$500$} & ZANIM-BART & 0.2960 & 0.3529 & 0.0035 & 0.8627 & 0.0692 & 0.6868 & 615.4 \\
   &  & ZANIM-LN-BART & \textbf{0.1912} & \textbf{0.5689} & \textbf{0.0025} & \textbf{0.9540} & \textbf{0.0663} & \textbf{0.7008} & 705.1 \\
   &  & ZANIDM-reg & 0.9573 & 0.0950 & 0.0029 & 0.9248 & 0.3270 & 0.4151 & 229.8 \\
   &  & DM-reg & 1.1988 & 0.0669 & 0.0030 & 0.6796 &  &  & 172.1 \\
   &  & ML-BART & 2.2474 & 0.2417 &  &  &  &  & 520.9 \\
   &  & MLN-BART & 1.3250 & 0.3382 & 0.0036 & 0.6552 &  &  & 628.5 \\
   &  & MLN-GP & 0.7341 & 0.4303 & 0.0047 & 0.6748 &  &  & 146.5 \\
   \hdashline
\multirow{7}{*}{$20$} & \multirow{7}{*}{$1000$} & ZANIM-BART & 0.3068 & 0.3176 & 0.0041 & 0.8214 & 0.0493 & 0.6922 & 1081.5 \\
   &  & ZANIM-LN-BART & \textbf{0.1298} & \textbf{0.5219} & \textbf{0.0024} & \textbf{0.9545} & \textbf{0.0456} & \textbf{0.7206} & 1243.3 \\
   &  & ZANIDM-reg & 0.9716 & 0.0648 & 0.0029 & 0.9083 & 0.4360 & 0.3007 & 465.9 \\
   &  & DM-reg & 1.2104 & 0.0468 & 0.0029 & 0.6806 &  &  & 324.1 \\
   &  & ML-BART & 2.3351 & 0.2326 & &  &  &  & 889.0 \\
   &  & MLN-BART & 1.2373 & 0.2738 & 0.0035 & 0.6508 &  &  & 1106.6 \\
   &  & MLN-GP & 0.6938 & 0.3641 & 0.0046 & 0.6777 &  &  & 554.8 \\
   \hdashline
\multirow{7}{*}{$40$} & \multirow{7}{*}{$200$} & ZANIM-BART & \textbf{0.3433} & 0.5297 & 0.0059 & 0.9213 & 0.1082 & 0.6651 & 538.3 \\
   &  & ZANIM-LN-BART & 0.3611 & \textbf{0.6023} & \textbf{0.0055} & \textbf{0.9511} & \textbf{0.1070} & \textbf{0.6724} & 715.1 \\
   &  & ZANIDM-reg & 1.0645 & 0.1507 & 0.0063 & 0.9242 & 0.3203 & 0.6302 & 170.4 \\
   &  & DM-reg & 1.2539 & 0.1056 & 0.0063 & 0.6889 &  &  & 130.0 \\
   &  & ML-BART & 1.6209 & 0.3787 &  &  &  &  & 438.5 \\
   &  & MLN-BART & 1.2185 & 0.5089 & 0.0074 & 0.6577 &  &  & 658.6 \\
   &  & MLN-GP & 0.7136 & 0.5089 & 0.0113 & 0.6080 &  &  & 58.8 \\
   \hdashline
\multirow{7}{*}{$40$} & \multirow{7}{*}{$500$} & ZANIM-BART & 0.3167 & 0.4417 & 0.0066 & 0.8636 & 0.0802 & 0.6614 & 1293.3 \\
   &  & ZANIM-LN-BART & \textbf{0.1857} & \textbf{0.6252} & \textbf{0.0051} & \textbf{0.9502} & \textbf{0.0771} & \textbf{0.6785} & 1542.9 \\
   &  & ZANIDM-reg & 1.0837 & 0.0900 & 0.0061 & 0.9007 & 0.4267 & 0.4380 & 464.0 \\
   &  & DM-reg & 1.2715 & 0.0644 & 0.0061 & 0.6882 &  &  & 340.4 \\
   &  & ML-BART & 1.7620 & 0.3262 &  & &  &  & 1119.6 \\
   &  & MLN-BART & 1.1207 & 0.4161 & 0.0073 & 0.6417 &  &  & 1410.6 \\
   &  & MLN-GP & 0.5963 & 0.4198 & 0.0106 & 0.6403 &  &  & 556.8 \\
   \hdashline
\multirow{7}{*}{$40$} & \multirow{7}{*}{$1000$} & ZANIM-BART & 0.3131 & 0.3892 & 0.0080 & 0.8141 & 0.0564 & 0.6774 & 2205.8 \\
   &  & ZANIM-LN-BART & \textbf{0.1263} & \textbf{0.5855} & \textbf{0.0049} & \textbf{0.9469} & \textbf{0.0521} & \textbf{0.7090} & 2823.4 \\
   &  & ZANIDM-reg & 1.0638 & 0.0631 & 0.0061 & 0.8871 & 0.5220 & 0.3246 & 949.5 \\
   &  & DM-reg & 1.2696 & 0.0456 & 0.0061 & 0.6886 &  &  & 661.5 \\
   &  & ML-BART & 1.7902 & 0.3018 &  &  &  &  & 1896.9 \\
   &  & MLN-BART & 0.9970 & 0.3469 & 0.0073 & 0.6328 &  &  & 2426.9 \\
   &  & MLN-GP & 0.5422 & 0.3588 & 0.0100 & 0.6527 &  &  & 3854.3 \\
   \bottomrule
\end{tabular}
\end{table} \section{Additional results for the modern pollen-climate data analyses}\label{supp:add_palaeoclimate}

In Section \ref{sec:application} of the main paper, we analysed the modern pollen-climate data using several BART-based models designed for count-compositional data, including the proposed ZANIM-BART and ZANIM-LN-BART models, their special cases the ML-BART and MLN-BART models, and the regression-based models DM-reg and ZANIDM-reg.
We now provide additional results, organised as follows:
Section \ref{supp:descriptive_analysis_pollen} provides a descriptive analysis of the data,
Section \ref{supp:comparison_correlation} compares the empirical correlation of the holdout counts between the posterior mean correlations under the models, and
Section \ref{supp:add_pdps} presents complementary partial dependence plots based on the best-fitting ZANIM-LN-BART model for all taxa. Convergence assessments for ZANIM-LN-BART are deferred to Section \ref{supp:convergence}.

\subsection{Descriptive analysis}\label{supp:descriptive_analysis_pollen}

Since the pollen counts are multivariate and compositional, we carefully selected summary statistics to account for and describe its main features.
To quantity overdispersion, we compute the multivariate coefficient of variation (MCV) of \citet{Albert2010} and the multiple marginal dispersion index (MDI) of \citet{Kokonendji2018}.
For a random vector $\mathbf{Y}$, these indices are defined, respectively, as $\operatorname{MCV}(\mathbf{Y})=\left( \mathbb{E}\lbrack\mathbf{Y}\rbrack^\top \operatorname{Var} \lbrack \mathbf{Y}\rbrack \mathbb{E}\lbrack\mathbf{Y}\rbrack / (\mathbb{E}\lbrack\mathbf{Y}\rbrack^\top \mathbb{E}\lbrack\mathbf{Y}\rbrack) \right)^2$ and $\operatorname{MDI}(\mathbf{Y})=(\mathbb{E}\lbrack\mathbf{Y}\rbrack^\top \mathbb{E}\lbrack\mathbf{Y}\rbrack)^{-1}\sum_{j=1}^d  \mathbb{E}\lbrack Y_j\rbrack  \operatorname{Var}\lbrack Y_j \rbrack $.
Note that the $\operatorname{MDI}(\mathbf{Y})$ can be expressed as a weighted average of the marginal dispersion indices $\operatorname{DI}\lbrack Y_j\rbrack=\mathbb{E}\lbrack Y_j\rbrack/\operatorname{Var}\lbrack Y_j\rbrack$.
The empirical values of these indices are $\widehat{\operatorname{MCV}}\lbrack\mathbf{Y}\rbrack=121.9$ and $\widehat{\operatorname{MDI}}\lbrack \mathbf{Y}\rbrack=215.5$.
These large values suggest substantial variability in the counts, with the MDI particularly indicating strong overdispersion.

Another key feature of the pollen data set is its high degree of sparsity, with $63.21\%$ of the counts being equal to zero.
To assess whether the data exhibit excess zeros relative to the multinomial sampling distribution, we introduce a multivariate zero-inflation (ZI) index adapted from the binomial ZI index of \citet{Kim2018}.
Specifically, the new index is given by the average of the marginal binomial ZI indices, i.e.,
$\operatorname{ZI}\lbrack \mathbf{Y} \rbrack = d^{-1} \sum_{j=1}^d \operatorname{ZI}_b\lbrack Y_j \rbrack,$
where $\operatorname{ZI}_b\lbrack Y_j \rbrack = p_{0j} - \Pr\lbrack Y_j=0\rbrack$, with $p_{0j}$ denoting the empirical proportion of zeros in the $j$-th margin, and $\Pr\lbrack Y_j=0 \rbrack = n^{-1}\sum_{i=1}^n(1 - \widehat{\pi}_{j})^{N_i}$ representing the expected probability of zero under the marginal binomial distribution. Here, we use the maximum likelihood estimate for the taxa-specific probability, i.e., $\widehat{\pi}_{j}=\sum_{i=1}^n y_{ij} / \sum_{i=1}^n N_i$. Compared with the ratio-based index proposed by \citet{Kim2018} for univariate binomial distribution, our formulation differs in two ways: we accommodate observation-specific totals, $N_i$, and rely on the difference rather than the ratio between the observed and expected zero probabilities.
The latter reduces the effect of large category-specific ratios when averaging across categories to construct the multivariate ZI index for count-compositional data.
For this data set, the empirical value is $\widehat{\operatorname{ZI}}\lbrack \mathbf{Y}\rbrack=0.576$, indicating substantial zero-inflation, with the multinomial distribution markedly underestimating the observed frequency of zeros.

Table \ref{tab:descriptive_stats_pollen} reports taxa-specific descriptive statistics: the abundance (average count), the empirical dispersion index, $\operatorname{DI}\lbrack Y_j\rbrack$, the percentage of zeros, and the empirical zero-inflated binomial index, $\operatorname{ZI}_b\lbrack Y_j\rbrack$.
Notably, all taxa exhibit a high level of overdispersion with the marginal empirical $\operatorname{DI}\lbrack Y_j\rbrack=\operatorname{Var}\lbrack Y_j\rbrack/\mathbb{E}\lbrack Y_j\rbrack$ being greater than one, in agreement with the MCV and MDI indices.
Moreover, the $\operatorname{ZI}_b\lbrack Y_j\rbrack$ values are substantially greater than zero for all taxa, with the lowest value being $0.07$ for the most abundant taxon, \textit{Pinus.D}.

\begin{table}[!ht]
\centering
\caption{Taxa-specific descriptive statistics for the modern pollen-climate data. For each of the $d=28$ taxa in the modern pollen-climate data, the following statistics are given: Abundance, i.e., the average count; $\operatorname{DI}\lbrack Y_j\rbrack$, the empirical dispersion index; \% zeros, the percentage of zero counts; $\operatorname{ZI}_b\lbrack Y_j\rbrack$, the zero-inflated binomial index.}
\label{tab:descriptive_stats_pollen}
\scriptsize
\begin{tabular}{lrrrr}
\hline
Taxa & Abundance & $\operatorname{DI}\lbrack Y_j\rbrack$  & \% zeros & $\operatorname{ZI}_b\lbrack Y_j\rbrack$ \\
  \hline
\textit{Abies} & 10.81 & 122.88 & 66.16 & 0.66 \\
  \textit{Alnus} & 69.92 & 196.69 & 20.51 & 0.21 \\
  \textit{Artemisia} & 38.35 & 309.33 & 37.79 & 0.38 \\
  \textit{Betula} & 140.03 & 193.07 & 25.11 & 0.25 \\
  \textit{Carpinus} & 2.27 & 141.59 & 91.06 & 0.81 \\
  \textit{Castanea} & 1.72 & 309.46 & 92.71 & 0.75 \\
  \textit{Cedrus} & 3.03 & 470.18 & 96.65 & 0.92 \\
  \textit{Chenopodiaceae} & 39.33 & 321.46 & 37.90 & 0.38 \\
  \textit{Corylus} & 7.31 & 125.16 & 78.69 & 0.79 \\
  \textit{Cyperaceae} & 51.24 & 220.80 & 26.16 & 0.26 \\
  \textit{Ephedra} & 1.15 & 161.94 & 94.11 & 0.62 \\
  \textit{Ericales} & 14.90 & 206.28 & 71.82 & 0.72 \\
  \textit{Fagus} & 13.21 & 162.83 & 70.93 & 0.71 \\
  \textit{Gramineae} & 103.08 & 198.62 & 10.27 & 0.10 \\
  \textit{Juniperus} & 17.51 & 258.97 & 59.82 & 0.60 \\
  \textit{Larix} & 1.60 & 144.21 & 93.25 & 0.73 \\
  \textit{Olea} & 10.28 & 194.72 & 84.98 & 0.85 \\
  \textit{Ostrya} & 4.68 & 41.15 & 73.94 & 0.73 \\
  \textit{Phillyrea} & 1.12 & 96.80 & 93.12 & 0.60 \\
  \textit{Picea} & 87.09 & 256.58 & 37.37 & 0.37 \\
  \textit{Pinus.D} & 225.28 & 210.42 & 6.66 & 0.07 \\
  \textit{Pinus.H} & 4.80 & 402.60 & 97.11 & 0.96 \\
  \textit{Pistacia} & 2.02 & 193.75 & 91.41 & 0.78 \\
  \textit{Quercus.D} & 77.80 & 265.59 & 40.04 & 0.40 \\
  \textit{Quercus.E} & 19.77 & 403.79 & 85.71 & 0.86 \\
  \textit{Salix} & 18.10 & 219.01 & 41.47 & 0.41 \\
  \textit{Tilia} & 1.54 & 79.42 & 85.16 & 0.64 \\
  \textit{Ulmus} & 9.52 & 57.46 & 60.20 & 0.60 \\
   \hline
\end{tabular}
\end{table}

\subsection{Comparison of the posterior mean correlations}\label{supp:comparison_correlation}

\begin{figure}[!ht]
    \centering
    \includegraphics[width=1\linewidth]{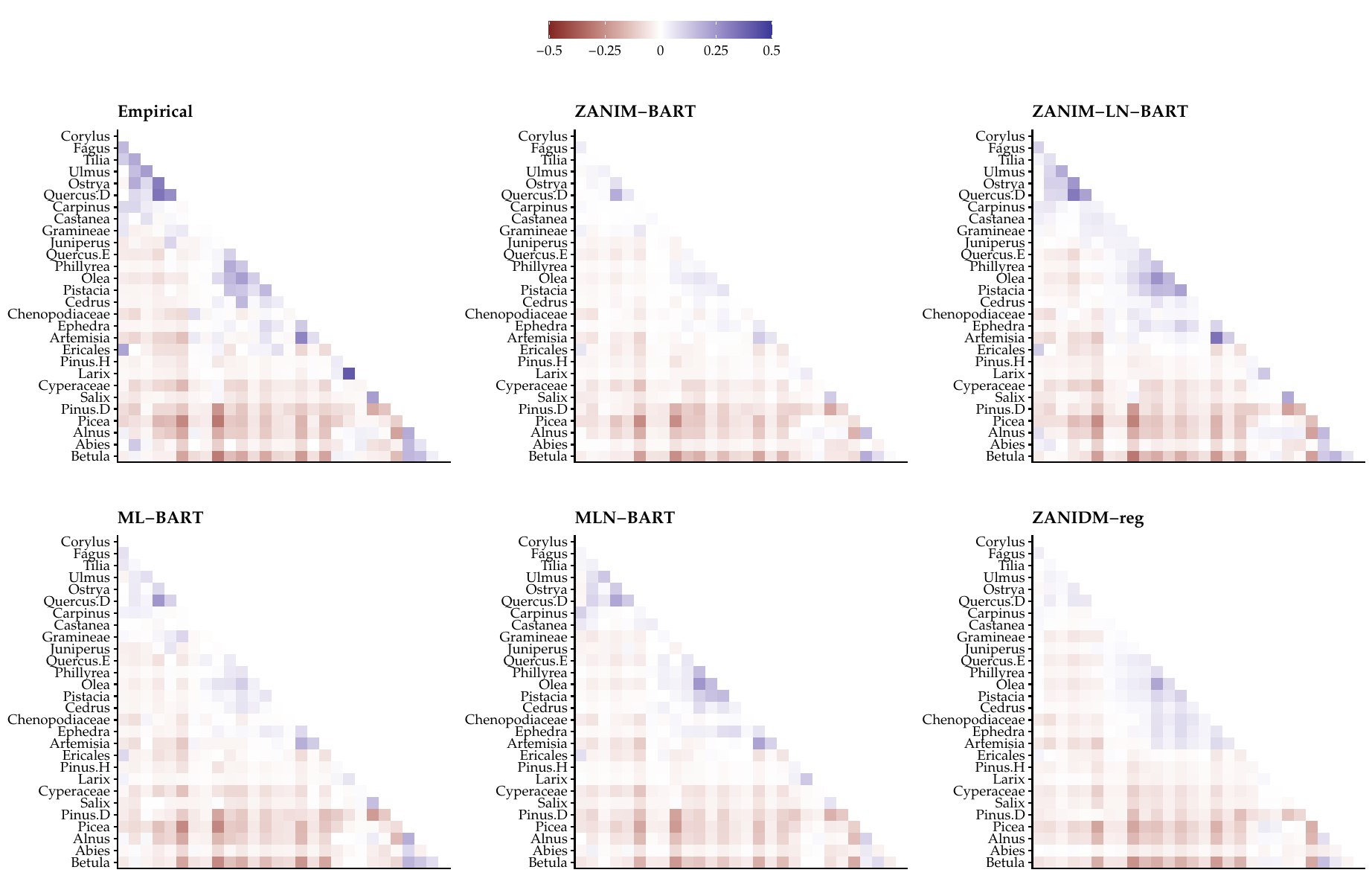}
    \caption{Empirical correlations of the counts (top left), posterior predictive means correlations of of the counts under the ZANIM-BART (top middle), ZANIM-LN-BART (top right) models, ML-BART (bottom left), ML-BART (bottom middle).
    Each row in the heat map corresponds to a taxon in each case.}
    \label{fig:correlations}
\end{figure}

Section \ref{sec:application} performs holdout predictive checks using four statistics that describe the overdispersion, zero-inflation, and compositional structure of the data.
Here, we provide a comparison of models in terms of estimating the correlation matrix of the counts.
\autoref{fig:correlations} shows the empirical correlations of the counts, along with the posterior predictive mean correlations under the ZANIM-BART, ZANIM-LN-BART, ML-BART, and MLN-BART models.
We stress that these correlations are obtained by averaging the sample correlation matrices of each posterior predictive draw of the counts, and do not correspond to the latent $\Sigma_U$ estimated by the models with random effects.
It is notable that the empirical correlation matrix exhibits both negative and positive dependencies.
Although the standard multinomial model can only accommodate negative correlations, all models considered here can also capture positive dependencies.
We can see that the posterior predictive correlation matrix under ZANIM-LN-BART aligns with the empirical one, while the correlations estimated under the other models fail to capture some important dependencies. This is consistent with the better overall performance of ZANIM-LN-BART in describing the pollen data.

To quantify the discrepancy between the empirical correlation and the posterior predictive mean correlations under the models, we use the RV coefficient, which is a measure of similarity between two matrices varying between 0 (strong dissimilarity) and 1 (strong similarity), defined as
\[\operatorname{RV}(\mathbf{A}, \mathbf{B}) = \frac{\operatorname{trace}\left(
\mathbf{A}\mathbf{B}^\top\mathbf{B}\mathbf{A}^\top\right)}{
\sqrt{
\operatorname{trace}\left(\mathbf{A}\mathbf{A}^\top\right)^2
\operatorname{trace}\left(\mathbf{B}\mathbf{B}^\top\right)^2
}}.\]
The results provided in \autoref{tab:metrics_correlation} also include additional metrics to quantify the elementwise deviations of the methods, namely the Frobenius and $\ell_1$ norms of the differences between the estimated correlation matrices and the empirical one.
From \autoref{tab:metrics_correlation}, it is clear that the ZANIM-LN-BART model achieves the highest similarity and lowest discrepancy, indicating the best overall recovery of the empirical correlation structure of the holdout data.

\begin{table}[!ht]
\centering
\caption{Comparison of posterior predictive mean correlation matrices across models using similarity and discrepancy metrics relative to the empirical correlation matrix.
The RV coefficient measures global similarity (values close to one indicate better agreement), while the Frobenius and $\ell_1$ norms quantify elementwise deviations (lower values are better).}
\scriptsize
\label{tab:metrics_correlation}
\begin{tabular}{lrrrr}
  \hline
 Model & RV & Frobenius norm & $\ell_1$-norm \\
  \hline
  ZANIM-BART & 0.971 & 1.401 & 1.459 \\
  ZANIM-LN-BART & \textbf{0.989} & \textbf{0.851} & \textbf{0.723} \\
  ML-BART & 0.980 & 1.162 & 1.124 \\
  MLN-BART & 0.980 & 1.179 & 1.244 \\
  ZANIDM-reg & 0.962 & 1.587 & 1.568 \\
  DM-reg & 0.950 & 1.818 & 1.753 \\
  \hline
\end{tabular}
\end{table}

\subsection{Complementary partial dependence plots}\label{supp:add_pdps}

\begin{figure}[!ht]
    \centering
    \includegraphics[width=1.0\linewidth]{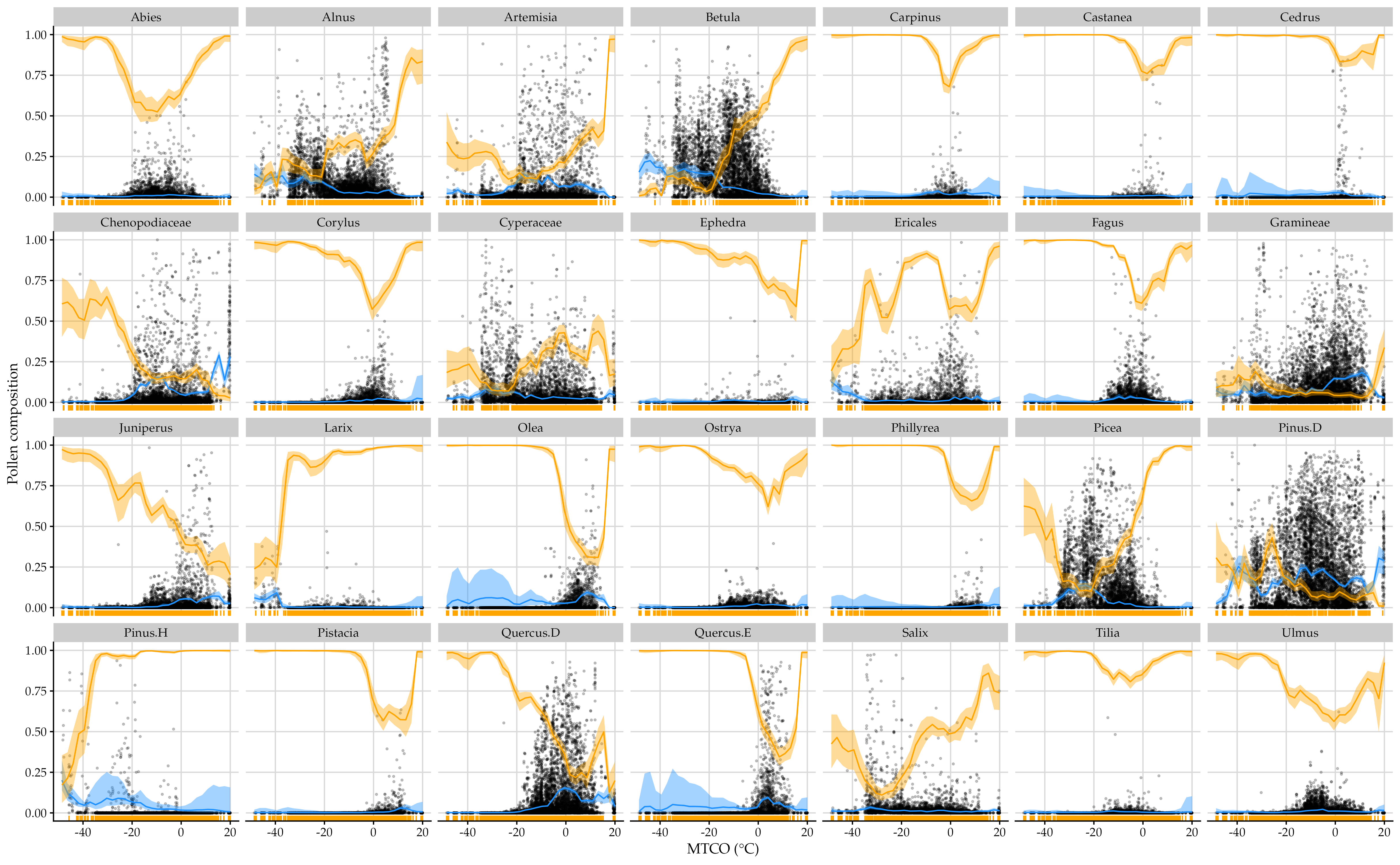}
    \caption{Partial dependence plots, with associated $95\%$ credible interval bands, for all $d=28$ pollen taxa, showing the effects of the MTCO covariate on both the compositional (blue) and structural zero (orange) probabilities under the ZANIM-LN-BART model.
    Black dots represent the observed relative abundances, $y_{ij} / N_i$, and the orange rugs along the $x$-axes represent samples where the observed counts are zero.}
    \label{fig:pdp_effects_mtco}
\end{figure}

\begin{figure}[!ht]
    \centering
    \includegraphics[width=1\linewidth]{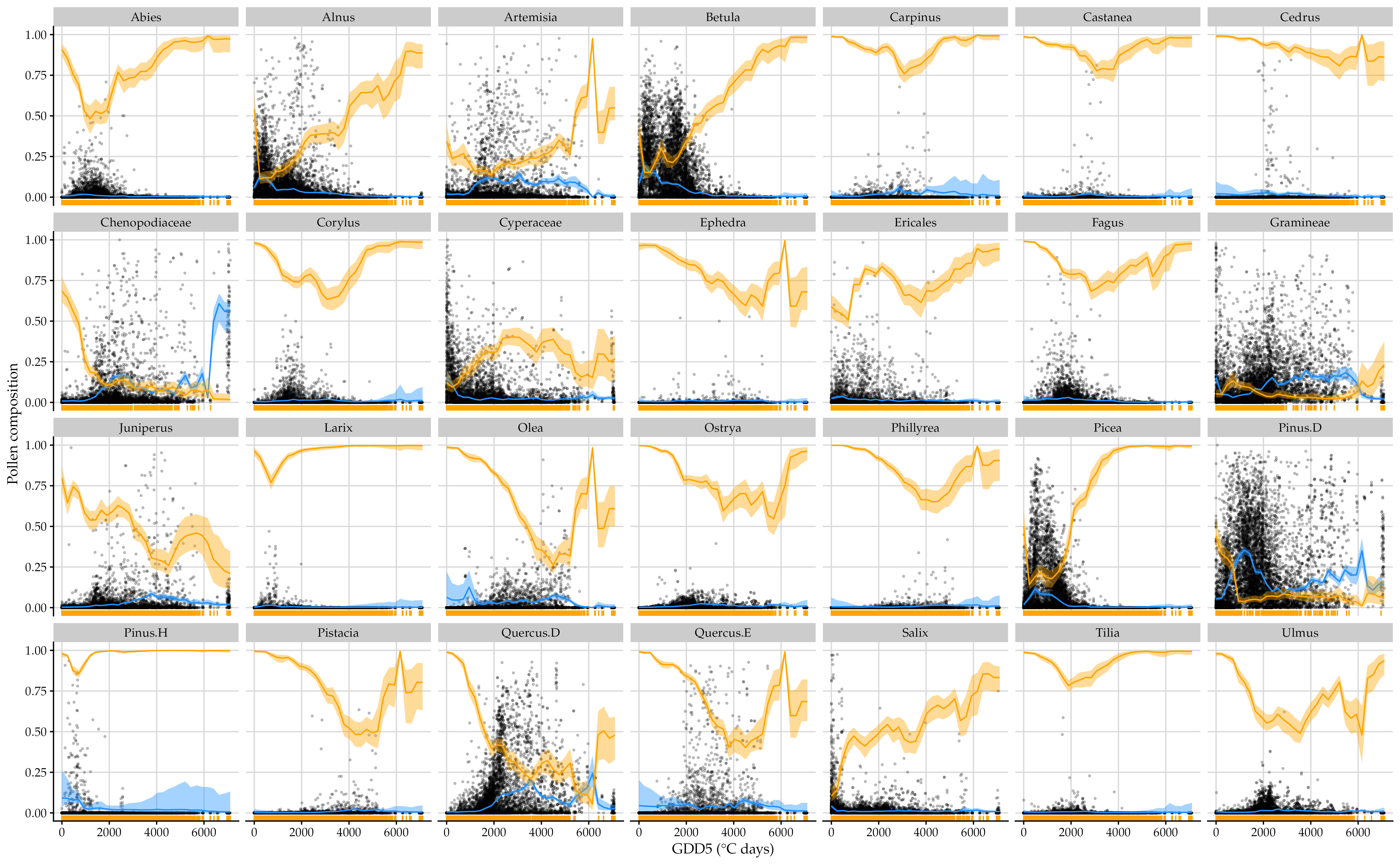}
    \caption{Partial dependence plots, with associated $95\%$ credible interval bands, for all $d=28$ pollen taxa, showing the effects of the GDD5 covariate on both the compositional (blue) and structural zero (orange) probabilities under the ZANIM-LN-BART model.
    Black dots represent the observed relative abundances, $y_{ij} / N_i$, and the orange rugs along the $x$-axes represent samples where the observed counts are zero.}
    \label{fig:pdp_effects_gdd5}
\end{figure}

\begin{figure}[!ht]
    \centering
    \includegraphics[width=1\linewidth]{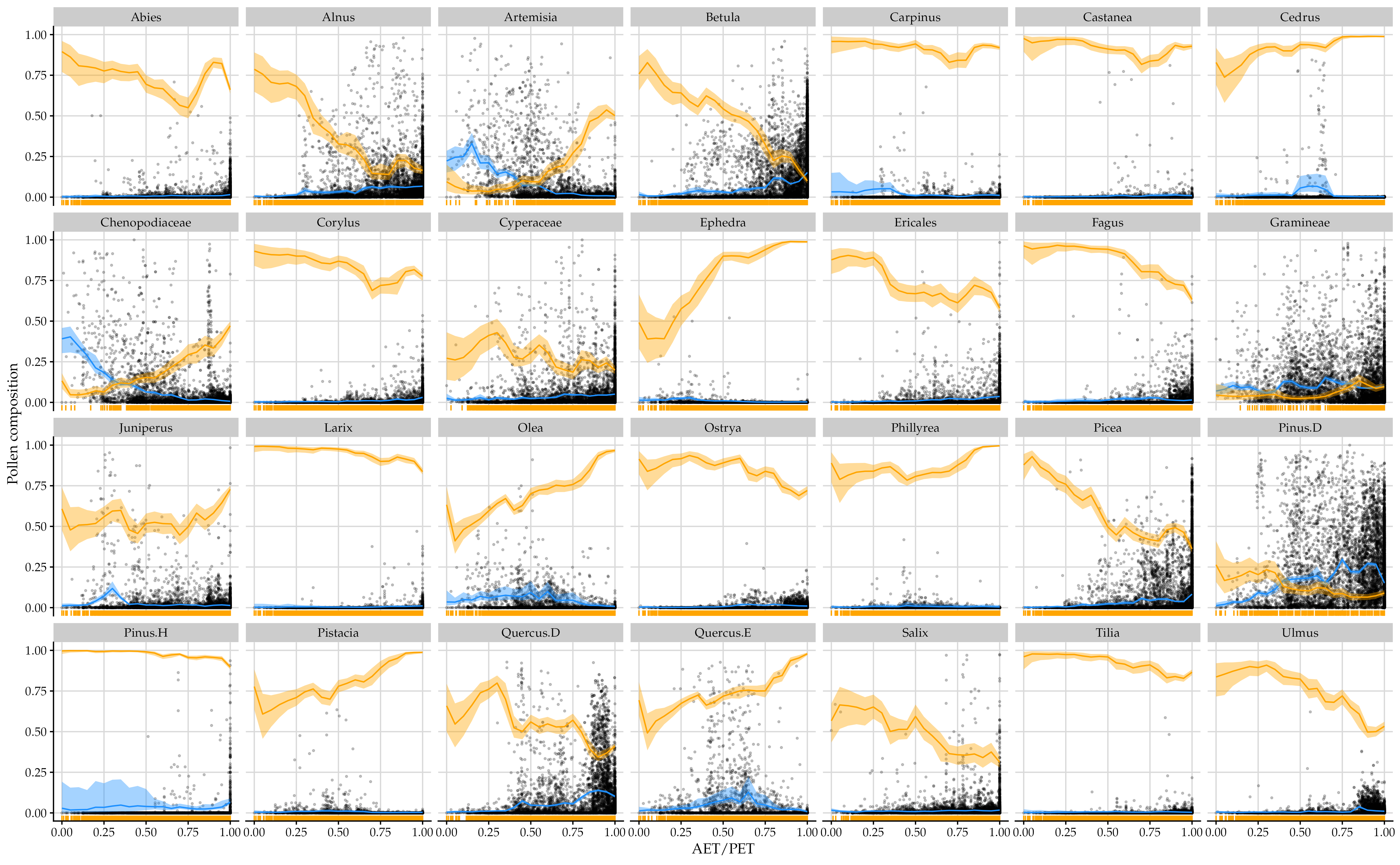}
    \caption{Partial dependence plots, with associated $95\%$ credible interval bands, for all $d=28$ pollen taxa, showing the effects of the AET/PET covariate on both the compositional (blue) and structural zero (orange) probabilities under the ZANIM-LN-BART model.
    Black dots represent the observed relative abundances, $y_{ij} / N_i$, and the orange rugs along the $x$-axes represent samples where the observed counts are zero.}
    \label{fig:pdp_effects_aet.pet}
\end{figure}

\autoref{fig:pdps_picea_pinusd} in the main paper shows partial dependence plots (PDPs) under the ZANIM-LN-BART model for
the \textit{Picea} and \textit{Pinus.D} taxa with the marginal effects of the MTCO, GDD5, and AET/PET covariates.
We now provide complementary PDPs for all $d=28$ taxa in
\autoref{fig:pdp_effects_mtco}, \autoref{fig:pdp_effects_gdd5}, and \autoref{fig:pdp_effects_aet.pet},
with the effects of MTCO, GDD5, and AET/PET, respectively, on the compositional (blue curves) and structural zero (orange curves) probabilities.

 \section{Human gut microbiome data analysis}\label{supp:human_gut_microbiome}

The human gut microbiome data studied by \citet{Wu2011} has been used to motivate the development of several new methods for count-compositional data \citep[see, e.g.,][]{Xia2013,Chen2013,Tang2019,Koslovsky2023,Ascari2025}.
The data consist of faecal samples from $98$ healthy volunteers, along with their demographic data and diet information.
Microbial operational taxonomic units (OTUs) were taxonomically classified up to the genus level and taxa with fewer than two samples were removed, resulting in $80$ genera.
Diet information was collected using two questionnaires: one assessing recent diet (``Recall'') and another assessing habitual long-term diet via a food frequency questionnaire (``FFQ'').
In the supplementary data of \citet{Wu2011}, $154$ variables are reported for the Recall set and $214$ for the FFQ set.
In this analysis, we focus on understanding how long-term diet covariates from the FFQ set are associated with the gut microbiota.
To reduce the number of highly-correlated covariates, we apply a feature screening procedure similar to that of \citet{Chen2013}.
Specifically, for any pair of covariates with absolute correlation greater than $0.9$, we compute the correlations of each covariate with all other covariates and remove the one with the largest mean absolute value.
This procedure reduces the number of FFQ covariates to $p=108$.
We use the taxa count matrix available in the \textsf{R} package \texttt{miLineage} \citep{Zheng2018}, specifically the object \verb|data.real$OTU|, which contains $n=96$ individuals (two individuals have missing data).
We further study the same subset of $d=13$ taxa as in \citet{Ascari2025}.
The resulting data comprises a count matrix $\mathbf{Y}_{96\times 13}$ and a covariate matrix $\mathbf{X}_{96 \times 108}$.

Summary statistics for each taxon are reported in Table \ref{tab:descriptive_stats_microbiome}.
The data are evidently sparse, with $20\%$ of the counts being zeros.
The marginal zero-inflated binomial index of \citet{Kim2018}, ranges from $0.0$ (\textit{Bacteroides}) to $0.62$ (\textit{Prevotella}).
There is also considerable variation in the total counts, $N_i$, which range from $500$ to $13{,}760$, alongside pronounced overdispersion: the empirical marginal dispersion indices vary from $92.78$ for \textit{Odoribacter} to $4789.69$ for  \textit{Prevotella}.
To further characterise variability, overdispersion, and zero-inflation, we compute the multivariate indices discussed in Section \ref{supp:add_palaeoclimate}.
The resulting MCV, MDI, and ZI values of $2374.16$, $1762.82$, and $0.20$, respectively, corroborate the substantial levels of overdispersion and zero-inflation indicated by the marginal indices.

\begin{table}[!ht]
\centering
\scriptsize
\caption{Taxa-specific descriptive statistics for the microbiome data. For each of the $d=13$ taxa in the human gut microbiome data, the following statistics are given: Abundance, i.e., the average count; $\operatorname{DI}\lbrack Y_j\rbrack$, the empirical dispersion index; \% zeros, the percentage of zero counts; $\operatorname{ZI}_b\lbrack Y_j\rbrack$, the zero-inflated binomial index.}
\label{tab:descriptive_stats_microbiome}
\begin{tabular}{lrrrr}
  \hline
Taxa & Abundance & $\operatorname{DI}\lbrack Y_j\rbrack$ & \% zeros & $\operatorname{ZI}_b\lbrack Y_j\rbrack$ \\
  \hline
\textit{Alistipes} & 455.09 & 549.77 & 6.25 & 0.06 \\
  \textit{Bacteroides} & 3707.74 & 1717.73 & 0.00 & 0.00 \\
  \textit{Barnesiella} & 164.34 & 310.81 & 40.62 & 0.41 \\
  \textit{Coprococcus} & 69.33 & 150.69 & 9.38 & 0.09 \\
  \textit{Faecalibacterium} & 337.48 & 367.94 & 4.17 & 0.04 \\
  \textit{Odoribacter} & 83.30 & 92.78 & 21.88 & 0.22 \\
  \textit{Oscillibacter} & 169.62 & 213.92 & 2.08 & 0.02 \\
  \textit{Parabacteroides} & 364.62 & 416.38 & 4.17 & 0.04 \\
  \textit{Parasutterella} & 50.77 & 149.29 & 25.00 & 0.25 \\
  \textit{Phascolarctobacterium} & 63.60 & 152.91 & 46.88 & 0.47 \\
  \textit{Prevotella} & 669.86 & 4789.70 & 62.50 & 0.62 \\
  \textit{Ruminococcus} & 81.11 & 222.00 & 26.04 & 0.26 \\
  \textit{Subdoligranulum} & 141.64 & 504.72 & 11.46 & 0.11 \\
   \hline
\end{tabular}
\end{table}

\subsection{Model comparisons}

Given the large number of covariates in these data, we consider the ZIDM and DM regression models of \citet{Koslovsky2023} and \citet{Wadsworth2017}, respectively.
We note that both the ZIDM and DM regression models have spike-and-slab priors on their regression coefficients in order to perform variable selection.
For the ZIDM model, this prior is assumed for both the count and zero-inflation components of the model.
To fit these models, we use the implementation with default arguments available via the \textsf{R} package \texttt{ZIDM} at \url{https://github.com/mkoslovsky/ZIDM}.
In contrast to the pollen application in Section \ref{sec:application}, where variable selection was not required, we denote these models as ZIDM-bvs and DM-bvs, to emphasise the differences in the implementation and the inclusion of Bayesian variable selection.
Similarly, in order to allow the BART-based models ML-BART, MLN-BART, ZANIM-BART, and ZANIM-LN-BART to perform variable selection for each category-specific set of population-level count and structural zero probabilities, we adopt the sparsity-inducing Dirichlet prior for decision tree splitting rules of \citet{Linero2018}, described previously in Section \ref{sec:priors}.

Default hyperparameters were used for the BART-based models as discussed in Section \ref{sec:priors}.
For all models, the MCMC chains were run with $10{,}000$ iterations and the first $5{,}000$ are discarded as burn-in, resulting in $5{,}000$ valid posterior samples.
Under these settings, the runtime in seconds of the MCMC schemes of the ZANIM-BART, ZANIM-LN-BART, ML-BART, MLN-BART, ZIDM-bvs, and DM-bvs models were $60.57$, $72.84$, $36.67$, $51.57$, $107.23$, and $35.52$ seconds, respectively.

We compare the models by assessing the consistency of their posterior predictive distributions with the observed data.
We employ diagnostic statistics similar to those used in the pollen data analysis (Section \ref{sec:application}), specifically the MDI, entropy, and the multivariate zero-inflation index.
However, we stress that our checks are not conducted on holdout data, as per the pollen data analysis, for the present application.
We omit the proportion of zeros from the figures, as it provides similar information to that given by the zero-inflation index.
This is supported by their nearly identical empirical values for the microbiome data set ($0.20032$ for the proportion of zeros and $0.20029$ for the ZI index), which effectively indicates that none of the observed zeros can be explained solely by a multinomial distribution.

\begin{figure}[!ht]
    \centering
\includegraphics[width=1.0\linewidth]{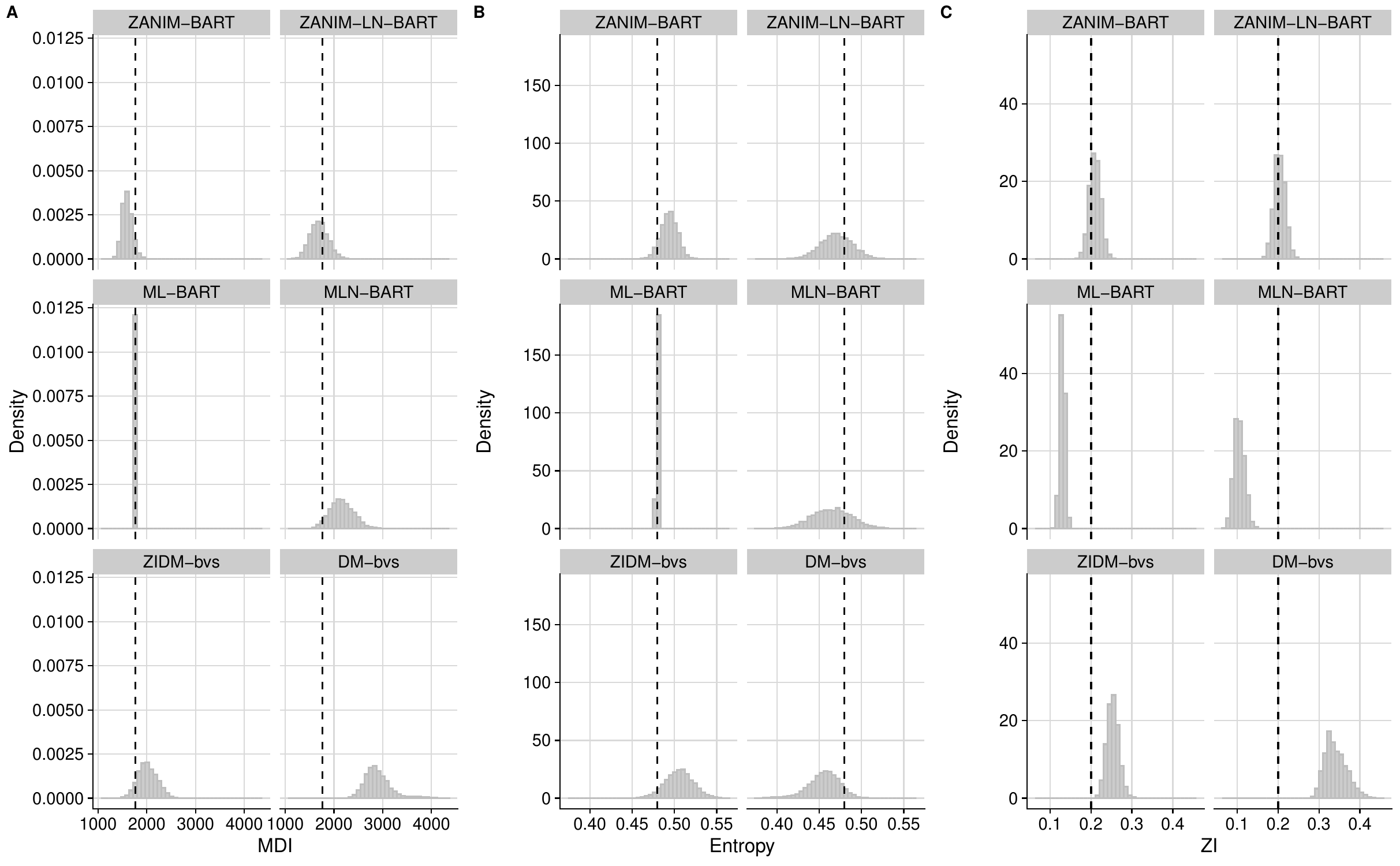}
    \caption{Posterior predictive checks using the MDI (\textbf{A}), entropy (\textbf{B}), and multivariate ZI index (\textbf{C}) from the ZANIM-BART, ZANIM-LN-BART, ML-BART, MLN-BART, ZIDM-bvs, and DM-bvs models. The horizontal dashed lines indicate the empirical values of each statistic.}
    \label{fig:ppcs_microbiome}
\end{figure}

Figure \ref{fig:ppcs_microbiome} gives the posterior predictive distribution of the three diagnostic statistics under the six competing models. Overall, the ZANIM-LN-BART model provides the best balance across all considered diagnostics.
This suggests that the ZANIM-LN-BART model adequately captures the key features of the microbiome data, including the overdispersion, entropy, and zero-inflation.
The ZANIM-BART model remains competitive, but shows slightly worse results relative to the ZANIM-LN-BART model. However, models without explicit structural zero components perform quite poorly in reproducing the observed zero patterns.
In particular, the ML-BART and MLN-BART models markedly underestimate the excess zeros, while the DM-bvs model overestimates them, as indicated by their corresponding distributions of the ZI index.
Similarly, the ZIDM-bvs model, despite including structural zero components, provides a poor characterisation of the zero-inflation present in the data, as seen by its overestimation of the ZI index.
While the ML-BART model produces posterior predictive diagnostics with high probability regions covering the empirical values of the MDI and entropy summaries, it is evidently unable to capture the excess of zeros in the data, as stated above.
In summary, these results underscore the importance of models, such as the proposed ZANIM-LN-BART, which can simultaneously account for structural zeros, overdispersion, and complex dependencies, while flexibly capturing covariate effects without imposing any rigid parametric assumptions.

Given the improvement in fit, it is reasonable to suspect that ZANIM-LN-BART is identifying important dietary covariates and possibly nonlinear effects that the the other models are unable to capture.
We thus first compare the number of dietary covariates identified by the models for each taxa.
To do so, we compute the marginal posterior probability of inclusion (MPPI) of the $d \times p = 1{,}404$ taxa/covariate pairs associated with the population-level count and structural zero probabilities, respectively.
Under the BART-based models, the MPPIs are computed separately for the compositional and structural zero probabilities by averaging the number of times a covariate is used to define a splitting rule over all corresponding trees and iterations.
For the ZIDM-bvs and DM-bvs models, which employ spike-and-slab priors, the MPPI values for each taxa/covariate pair are obtained by averaging the respective inclusion indicators over the MCMC draws.

Subject to the threshold of $\operatorname{MPPI}\geq 0.5$, we have that ML-BART, MLN-BART, ZANIM-BART, and ZANIM-LN-BART identify $300$, $144$, $212$, and $68$ taxa/covariate pairs, respectively, with regard to associations with the compositional probabilities.
Among those, $286$, $115$, $198$, and $46$, respectively, have MPPI values greater than $0.98$, which corresponds to a Bayesian false discovery rate of $0.1$.
In comparison, the ZIDM-bvs and DM-bvs models select $48$ and $86$ covariates, of which just $14$ and $13$, respectively, have MPPI values greater than $0.98$.
Regarding the structural zero probabilities, a similar analysis indicates that ZANIM-BART, ZANIM-LN-BART, and ZIDM-bvs identify $14$, $12$, and $7$ taxa/covariate pairs with $\operatorname{MPPI}\geq 0.5$; among those, $5$ and $9$ for ZANIM-BART and ZANIM-LN-BART have MPPI greater than $0.98$, while none were identified under the ZIDM-bvs model.

As expected, the ML-BART and ZANIM-BART models select substantially more covariates than their counterparts with random effects (MLN-BART and ZANIM-LN-BART).
The inclusion of latent random effects in the latter models allow them to capture unobserved heterogeneity, thereby reducing the need to attribute such variation to observed covariates. This, in turn, reduces the complexity of the corresponding trees.
Furthermore, as discussed in Section \ref{sec:zanim_ln_bart}, the ML-BART model relies solely on the observed covariates by performing inference only on the population-level count probabilities $\theta_{ij}$ via the category-specific regression trees.
Since $y_{ij}/N_i$ is itself a noisy but often accurate proxy for the underlying individual-level probabilities, the ML-BART tends to overfit this quantity because of the BART priors.
Consequently, any unobserved variation not explicitly modelled is then treated as signal rather than noise, which can lead to overfitting and selection of additional covariates with spurious importance.
In light of this, we stress that, unlike the posterior predictive checks performed as part of the pollen data analyses in Section \ref{sec:application}, the posterior predictive checks performed here for the microbiome data are not conducted on holdout data (given that the sample size is merely $n=96$).
These factors help to explain both the large number of taxa-covariate associations identified by ML-BART, and the extremely good performance in the posterior predictive checks, especially for the MDI and entropy diagnostics, driven by overfitting to $y_{ij}/N_i$. In fact, averaged over the posterior draws, ML-BART consistently has more leaves in the trees for the category-specific count components than both ZANIM-LN-BART and MLN-BART, for all taxa, and more leaves than ZANIM-BART for the vast majority of taxa.
This supports the above explanation and further suggests that ML-BART is likely overfitting.
Notably, ZANIM-LN-BART has fewer leaves than ZANIM-BART for all but one taxa, further suggesting that the incorporation of latent random effects mitigates the risk of overfitting.

\subsection{Inferential results}

\begin{figure}[!ht]
    \centering
    \includegraphics[width=1\linewidth]{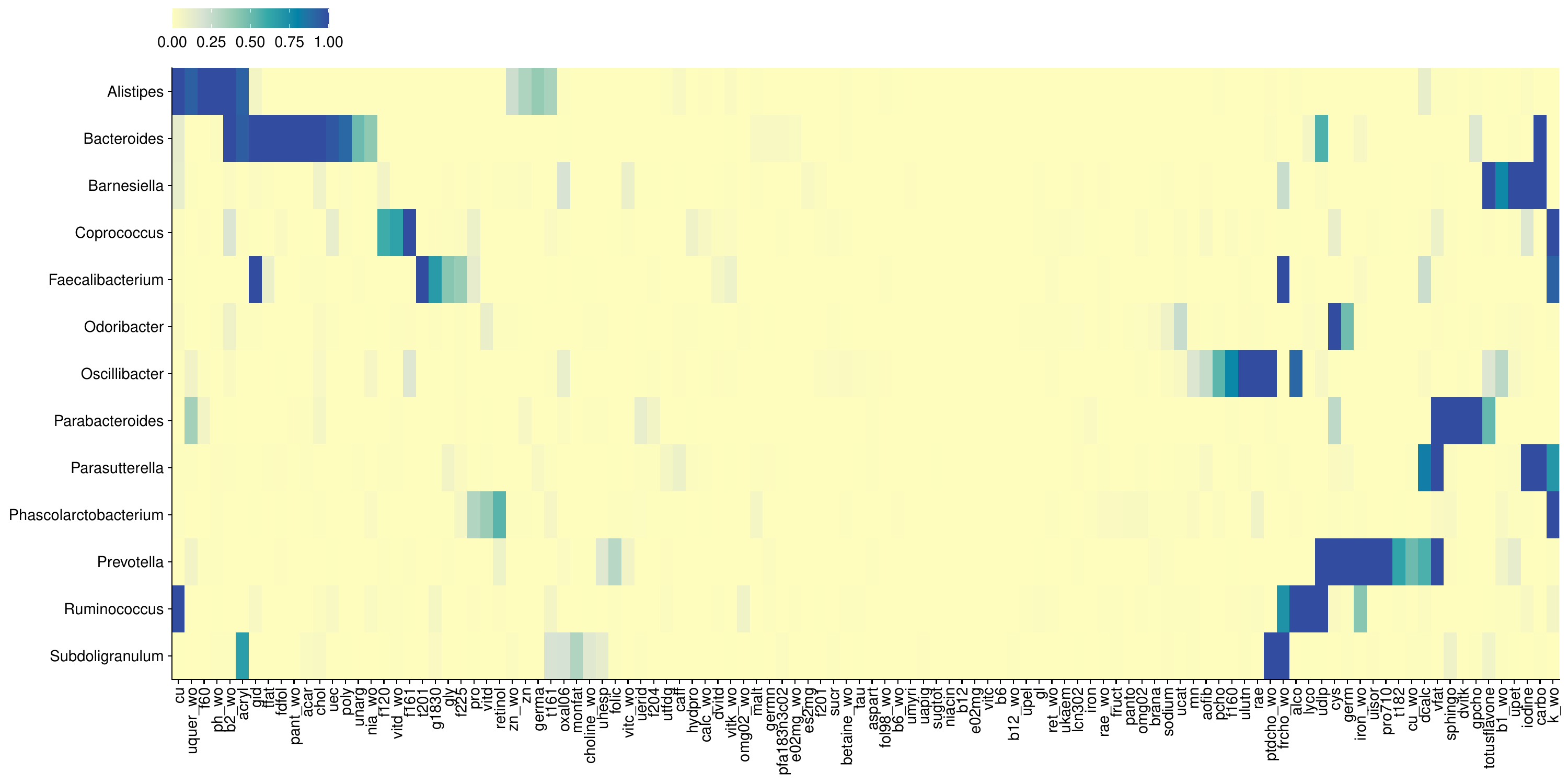}
    \caption{Heat map with the marginal posterior probability of inclusion (MPPI) of the $d \times p = 1{,}404$ taxa/covariate pairs associated with the population-level count probabilities under the ZANIM-LN-BART model.
    The covariates are in the columns, with their abbreviated names, while the rows indicate the taxa.}
    \label{fig:heat_map_mppi_count_zanim_ln}
\end{figure}

\begin{figure}[ht!]
    \centering
    \includegraphics[width=1\linewidth]{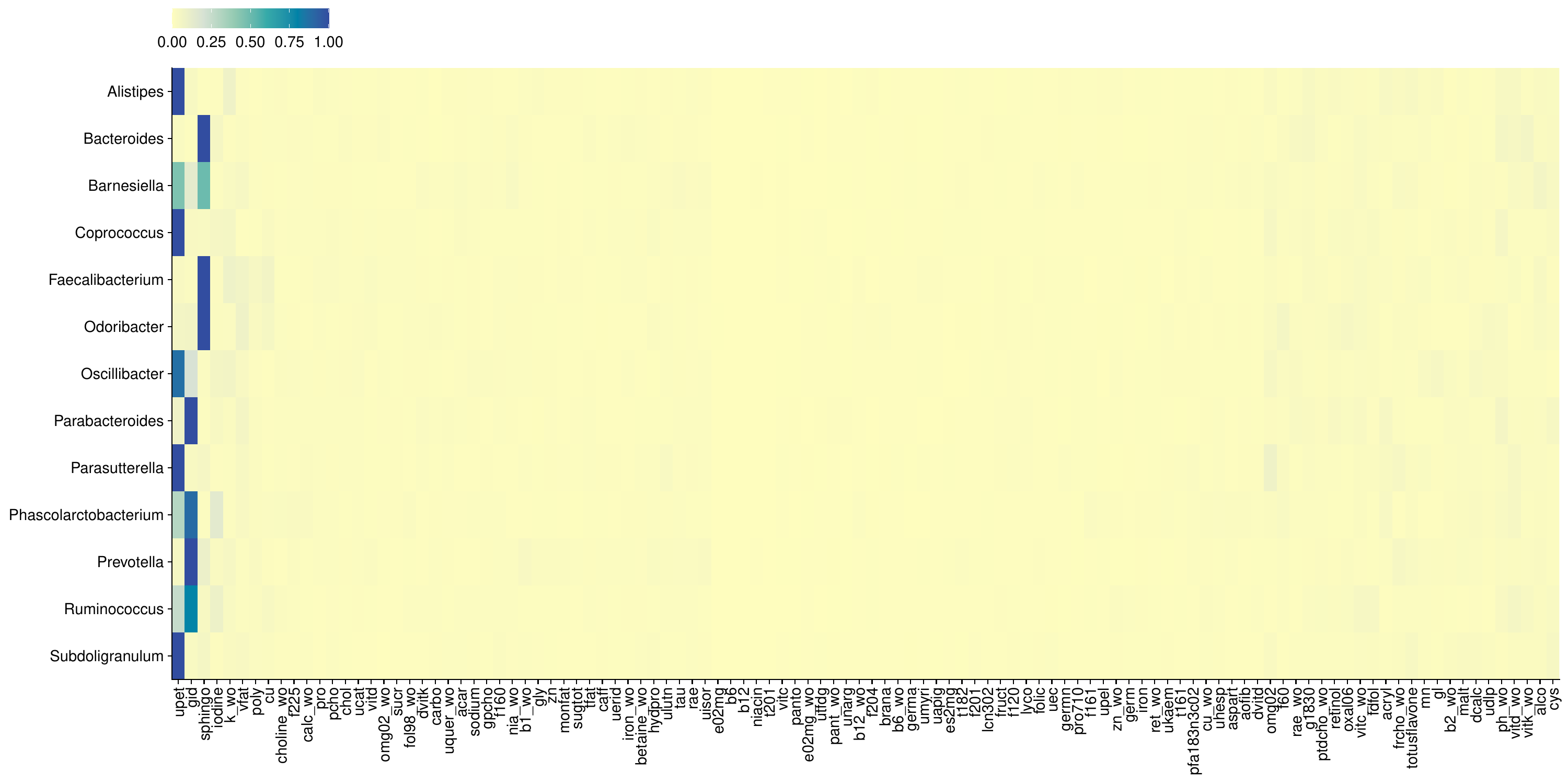}
    \caption{Heat map with the marginal posterior probability of inclusion (MPPI) of the $d \times p = 1{,}404$ taxa/covariate pairs associated with the population-level structural zero probabilities under the ZANIM-LN-BART model.
    The covariates are in the columns, with their abbreviated names, while the rows indicate the taxa.}
    \label{fig:heat_map_mppi_zi_zanim_ln}
\end{figure}

\autoref{fig:heat_map_mppi_count_zanim_ln} and \autoref{fig:heat_map_mppi_zi_zanim_ln} present heat maps with the MPPI values for all taxa/covariate pairs under ZANIM-LN-BART for the population-level count and structural zero components, respectively.
For better visualisation in each case, seriation has been applied to the columns, which represent the covariates, using the \textsf{R} package \texttt{seriation} \citep{Hahsler2008}.
Overall, we note that ZANIM-LN-BART evidently achieves sparsity for both components, identifying considerably fewer important taxa-specific covariate effects than the total number of available covariates.
According to \autoref{fig:heat_map_mppi_count_zanim_ln}, the taxa with large numbers of important covariates associated with the count probabilities were \textit{Bacteroides} and \textit{Prevotella}, with 8 and 7 covariates having MPPI values greater than $0.98$, respectively.
The most frequent covariates were \textit{carbo} (carbohydrates) and \textit{vfat} (vegetable fat): both appear three times each.
The carbohydrates covariate is important for explaining the abundance of \textit{Bacteroides}, \textit{Parasutterella}, and \textit{Barnesiella}, while vegetable fat is associated with \textit{Prevotella}, \textit{Parasutterella}, and \textit{Parabacteroides}.
We further observe more covariate sparsity in the structural zero components given by the MPPI values in \autoref{fig:heat_map_mppi_zi_zanim_ln}, with the majority of the taxa having only one covariate associated with their structural zero probabilities.
It is noteworthy to mention that only three covariates across all the taxa have MPPI greater than 0.98 for any taxa, namely \textit{upet} (Petunidin, anthocyanidin), \textit{sphingo} (Choline, Sphingomyelin), and \textit{gid} (Glycemic Index).
These nutrients are important for four (\textit{Alistipes}, \textit{Coprococcus}, \textit{Parasutterella}, and \textit{Subdoligranulum}), three (\textit{Bacteroides}, \textit{Faecalibacterium}, and \textit{Odoribacter}) and two (\textit{Parabacteroides} and \textit{Prevotella}) taxa, respectively.

We now illustrate a particularly useful feature of our proposed models: their ability to capture nonlinear dietary covariate effects in the microbiome data.
We stress that this is a novel aspect not explored by previous methods which analysed this microbiome data set.
Given that the ZANIM-LN-BART model provides a consistently better fit than the competing models, we consider its posterior distributions of the population-level count and structural probabilities, $\smash{f_j^{(\mathrm{c})}(\mathbf{x}_i)}$ and $\smash{f_j^{(0)}(\mathbf{x}_i)}$, respectively, to obtain estimates of the taxa-specific partial effects of dietary covariates.
Specifically, we employ the posterior projection method of \citet{Woody2021}, which we use to construct summaries of the posterior distribution of $\smash{f_j^{(\mathrm{c})}(\mathbf{x}_i)}$ and $\smash{f_j^{(0)}(\mathbf{x}_i)}$ as follows.
For each posterior sample of $\smash{f_j^{(\mathrm{c})}(\mathbf{x}_i)}$, an optimal summary $\smash{\tilde{f}_j^{(\mathrm{c})}(\mathbf{x}_i)}$ is calculated, considering a given class of summary functions $\mathcal{Q}$, via $\smash{\tilde{f}_j^{(\mathrm{c})}(\mathbf{x}_i)} = \arg \min_{\kappa \in \mathcal{Q}} \lVert f_j^{(\mathrm{c})} - \kappa\rVert_2$.
Here, we specify $\mathcal{Q}$ as the family of generalised additive models.
The specific choice of basis expansion is not central; indeed, \citet{Woody2021} argue that any reasonable basis would be suitable.
Our implementation uses the default settings of the \texttt{gam} function from the \texttt{mgcv} package in \textsf{R} \citep{Wood2017}.
To measure the quality of the summary, \citet{Woody2021} recommended the use of the \textit{summary} $R^2$ defined by
\[\operatorname{summary}\,R^2 = 1 - \frac{\sum_{i=1}^n \left\lbrack f_j^{(\mathrm{c})}(\mathbf{x}_i) - \tilde{f}_j^{(\mathrm{c})}(\mathbf{x}_i) \right\rbrack^2}{\sum_{i=1}^n \left\lbrack f_j^{(\mathrm{c})}(\mathbf{x}_i) - \bar{f}_j^{(\mathrm{c})} \right\rbrack^2},\]
where $\smash{\bar{f}_j^{(\mathrm{c})} = n^{-1} \sum_{i=1}^n f_j^{(\mathrm{c})}(\mathbf{x}_i)}$.
Values of the $\operatorname{summary}\,R^2$ closer to $1$ indicate that the projected posterior distribution is a good approximation of $\smash{f_j^{(\mathrm{c})}(\mathbf{x}_i)}$ at the observed values of the covariates.
A similar procedure is performed for summarising $\smash{f_j^{(0)}(\mathbf{x}_i)}$.

\begin{figure}[!ht]
    \centering
    \includegraphics[width=1.0\linewidth]{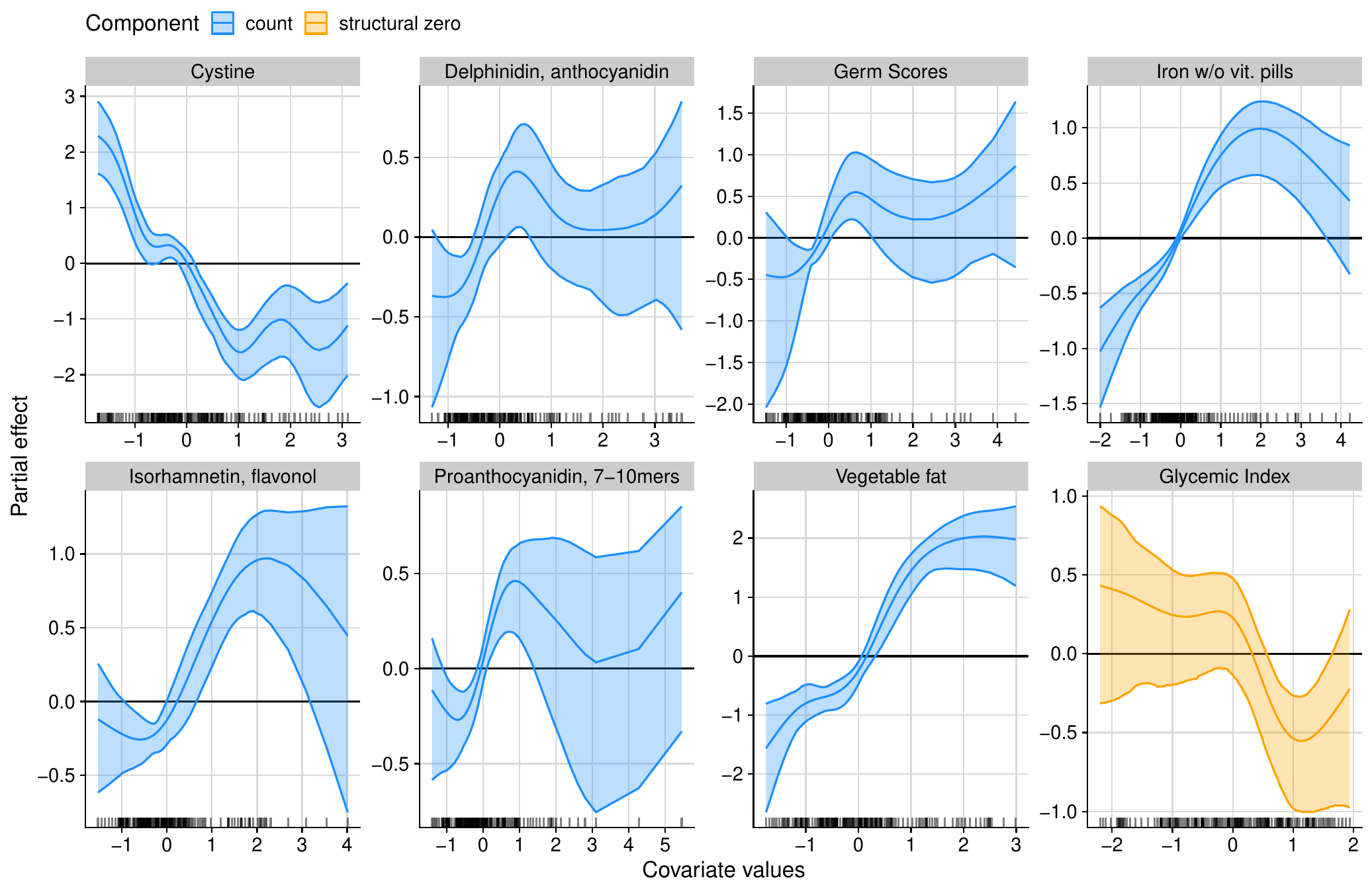}
    \caption{Posterior partial effects summaries for the population-level count (blue) and structural zero (orange) probabilities obtained using the ZANIM-LN-BART model for the \textit{Prevotella} taxon.
    The blue curves show the partial effect of $7$ nutrients with MPPI values greater than 0.98 on the count probabilities (abundance) of \textit{Prevotella}, while the orange curve shows the partial effect of the only one nutrient (Glycemic index) with MPPI value greater than 0.98 on the structural zero probabilities.
    Solid lines denote projected mean partial effects, with the shaded areas representing $95\%$ credible bands.
    The mean of the $\operatorname{summary}\,R^2$ for the count and structural zero projections were $0.95$ and $0.88$, respectively.}
    \label{fig:posterior_projection_prevotella}
\end{figure}

For illustrative purposes, we consider the \textit{Prevotella} taxon, which is the most zero-inflated and overdispersed, with empirical marginal $\operatorname{ZI}_b\lbrack Y_j\rbrack$ and $\operatorname{DI}\lbrack Y_j\rbrack$ indices of $0.62$ and $4789.69$, respectively.
For this taxon, considering MPPI values greater than 0.98, the ZANIM-LN-BART model identifies $7$ covariates associated to the count probabilities, and only the `Glycemic Index' covariate affecting the structural zero probabilities.
\autoref{fig:posterior_projection_prevotella} shows the partial effects obtained from the posterior projection method of \citet{Woody2021} for the seven nutrients associated with the count (blue) component and the sole nutrient associated with the structural zero (orange) component.
We recall that the covariates given by \citet{Wu2011} are standardised with mean zero and variance one, hence the partial effects should be interpreted relative to the average of the nutrients covariates.
We can see that the ZANIM-LN-BART model can effectively capture different nonlinear effects on both components for the \textit{Prevotella} taxon.
For instance, at concentration levels of the `Cystine' covariate below its mean, the partial effect is positive, indicating an increase in the abundance of \textit{Prevotella}.
As the cystine levels increase above its mean, these effects becomes negative in a nonlinear fashion.
In contrast, `Vegetable Fat' shows an opposite pattern with levels below zero, indicating a partial negative effect on the abundance of \textit{Prevotella}, and values above the mean having a positive partial effect on the abundance.
It is noteworthy to see that `Iron without vitamin pills` has a quadratic partial effect on the abundance for this taxon.
In particular, the abundance of \textit{Prevotella} increases with iron intake up to a peak, at values nearly 2 standard deviations above the mean, after which the effect begins to decline (but remain positive).

Finally, the orange curve in the final panel of \autoref{fig:posterior_projection_prevotella} represents the partial effect of the glycemic index on the structural zero probability of \textit{Prevotella}.
This can be interpreted as the probability that \textit{Prevotella} is truly absent rather than just unobserved due to sampling variability.
We can see that the $95\%$ credible intervals are generally wide, reflecting large uncertainty in the estimates of the structural zero partial effect.
Although the $95\%$ credible intervals overlap with zero across most of the range, we can see that there is high probability of \textit{Prevotella} being absent at glycemic index values below zero (i.e., its mean).
This partial effect decreases in a nonlinear fashion with a negative effect for glycemic index values with one standard deviation above the mean.
For completeness, we report convergence assessments for the ZANIM-LN-BART model fit to these data in Section \ref{supp:convergence}.

 \section{Convergence assessments}\label{supp:convergence}

In general, assessing convergence for multivariate models is cumbersome, because of the large number of parameters.
These difficulties are even more pronounced for our novel count-compositional models, as the $\bm{\theta}$ and $\bm{\zeta}$ parameters have category-specific BART priors.
For the standard BART model of \citet{Chipman2010}, convergence assessments  typically rely on monitoring the residual standard deviation parameter.
As there is no such parameter available under the BART-based models proposed and/or evaluated in this paper, we follow the approach discussed by \citet{Sparapani2021} for other BART-based models of monitoring the trace of functions of the parameters.
Specifically, we adapt traditional MCMC diagnostics to our case using the Frobenius norm between the empirical compositions, $y_{ij}/N_i$, and the corresponding model-based predictions given by the posterior draws of the individual-level count-probabilities, $\vartheta^{(t)}_{ij}$, where we recall that the superscript $(t)$ refers to the iteration index.
This quantity is defined by
\begin{equation}\label{eq:frob_vartheta}
\operatorname{FROB}\left(\vartheta^{(t)}\right) = \sqrt{\sum_{i=1}^n\sum_{j=1}^d \left(y_{ij}/N_i - \vartheta^{(t)}_{ij} \right)^2},
\quad t\in\{1,\ldots,R\}.
\end{equation}

Notably, \citet{Koslovsky2023} averaged this quantity as a goodness-of-fit measure, albeit without monitoring its trace as we do.
In addition to the regression-based models, ZANIDM-reg and DM-reg, this quantity is also available and of interest for our proposed ZANIM-BART, ZANIM-LN-BART, and MLN-BART models, thereby facilitating a fair comparison.
An appealing feature of this quantity is that it depends on both population-level parameters $\bm{\theta}$ and $\bm{\zeta}$, for the models which incorporate structural zero components.
It further depends on the latent random effects for the models which have them.
Thus, it encapsulates many aspects of the models in one identifiable quantity.
However, we note that for the ML-BART model, which incorporates neither observation-specific random effects nor structural zero probabilities, $\smash{\theta^{(t)}_{ij}}$ is used in place of $\smash{\vartheta^{(t)}_{ij}}$.
In any case, under the ML-BART model, $\smash{\theta_{ij}}$ tends to estimate the empirical compositions well.

\begin{figure}[!ht]
    \centering
    \includegraphics[width=1\linewidth]{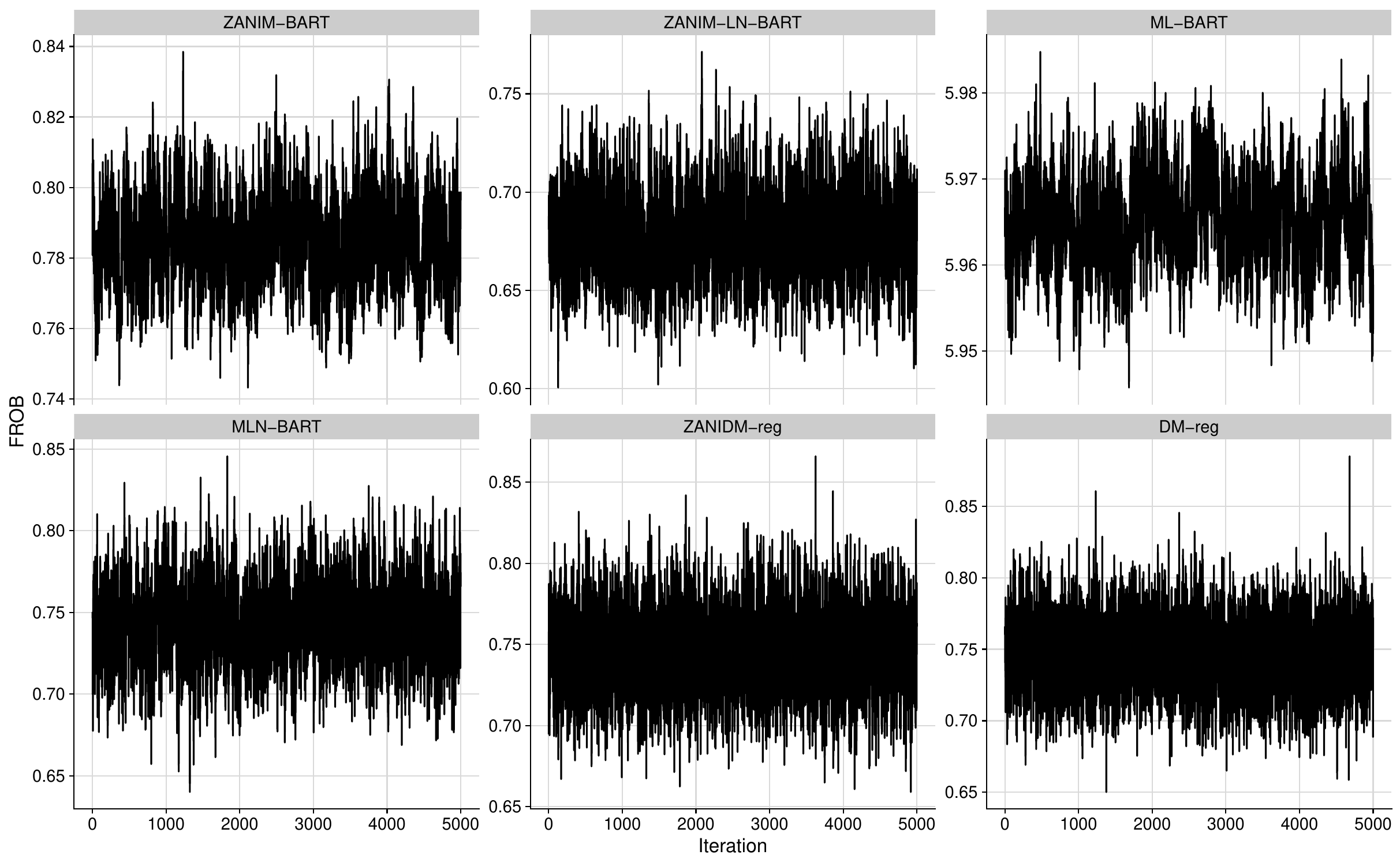}
    \caption{
    Traceplots of the Frobenius norm defined in \eqref{eq:frob_vartheta} for various models under Scenario 1 of the simulation studies, where the counts were generated from the ZANIM DGP.}
    \label{fig:traceplot_frob_scenario_1_dgp_zanim}
\end{figure}

In Section \ref{sec:simulations}, simulation studies were conducted under two scenarios.
In Section \ref{sec:simstudy1} (`Scenario 1'), data were generated with $d=4$, $n=400$, and $p=1$, under different distributional assumptions, with nonlinear functional forms describing the relationships between the covariate and the population-level count and structural zero parameters.
As DGPs, the ZANIM, ZANIDM, ZANIM-LN, and MLN distributions were considered.
As models, the ZANIM-BART, ZANIM-LN-BART, ML-BART, MLN-BART, ZANIDM-reg, DM-reg, and MLN-GP models were considered.
For brevity, Figure \ref{fig:traceplot_frob_scenario_1_dgp_zanim} shows the traceplots of $\operatorname{FROB}\left(\vartheta^{(t)}\right)$ for all models under the ZANIM DGP only.
However, we note that the MLN-GP model is excluded here as its inference scheme is not based on MCMC.
Clearly, satisfactory convergence is achieved for all models.
Similarly adequate convergence behaviour was also observed in traceplots obtained in the cases where the counts were generated using the ZANIDM, ZANIM-LN, and MLN distributions.

\begin{figure}[!hb]
    \centering
    \includegraphics[width=1\linewidth]{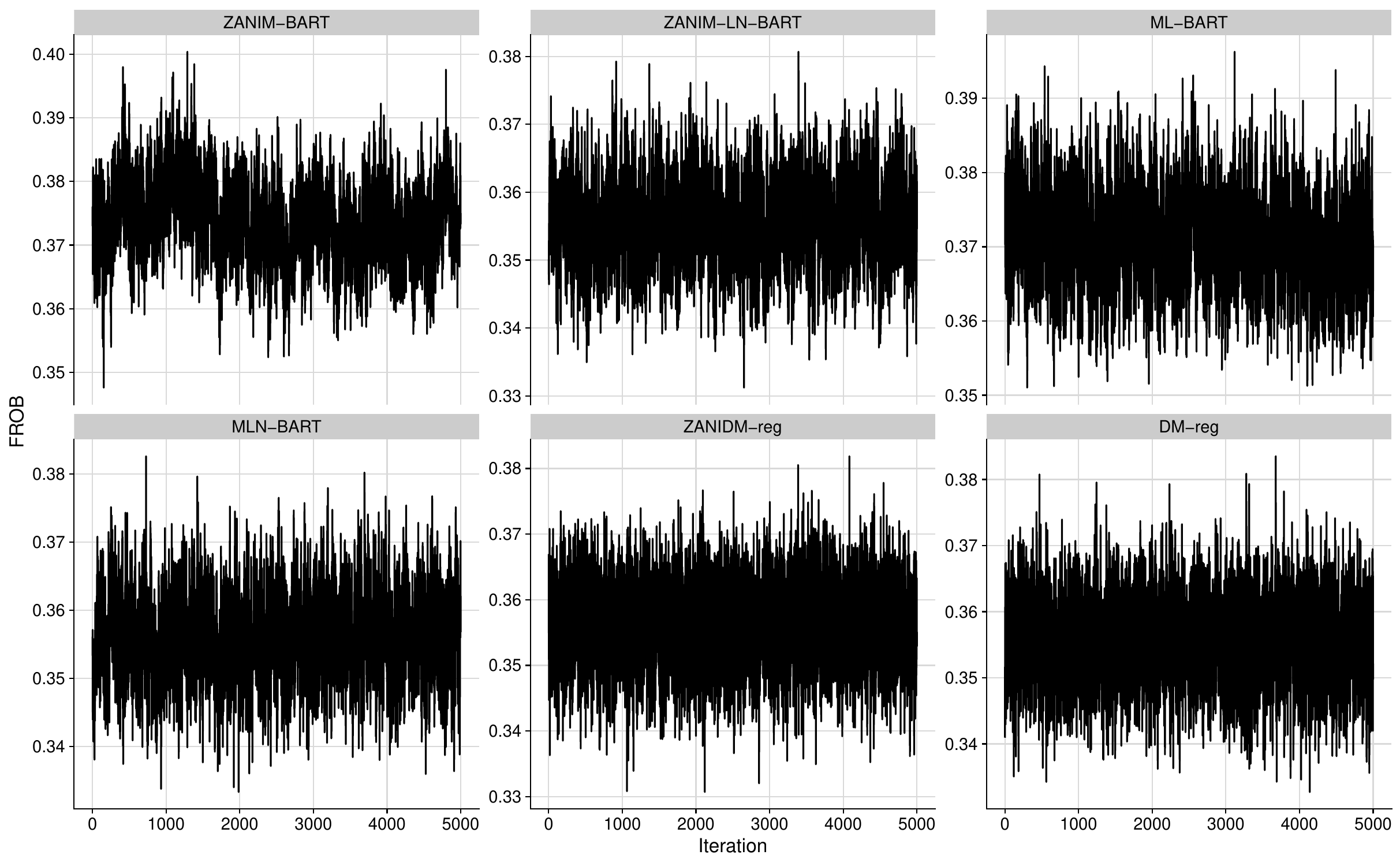}
    \caption{Traceplots of the Frobenius norm defined in \eqref{eq:frob_vartheta} for various models under one replicate of Scenario 2 of the simulation studies, with $n=500$ and $d=20$, where the counts were generated from the ZANIDM DGP.}
    \label{fig:traceplot_frob_scenario_2_dgp_zanidm}
\end{figure}

Subsequently, more challenging simulation experiments were conducted in Section \ref{sec:simstudy2} (`Scenario 2'), with complex functional forms describing the relationships between $p=6$ covariates and the population-level count and structural zero probability parameters.
Furthermore, the number of categories and the sample size were allowed to vary, with $d\in\{20,40\}$ and $n\in\{200,500,1000\}$.
The counts were generated from the ZANIDM and ZANIM-LN distributions, with the results averaged over six replicate data sets per DGP.
Given the large number of fitted models, the number of replicates, the number of $(n, d)$ pairs, and the two DGPs considered, there are many traceplots that could be shown.
For brevity, Figure \ref{fig:traceplot_frob_scenario_2_dgp_zanidm} shows traceplots of $\operatorname{FROB}\left(\vartheta^{(t)}\right)$ for all models fitted to one replicate data set from the $n=500$ and $d=20$ setting under the ZANIDM DGP only.
However, we note that the convergence behaviour of this randomly chosen replicate is representative of the many other traceplots not shown for the other replicates and $(n, d)$ settings.
In each case, adequate convergence is achieved for all models.
The same holds for the results obtained under the ZANIM-LN DGP.

\begin{figure}[!ht]
    \centering
    \includegraphics[width=1\linewidth]{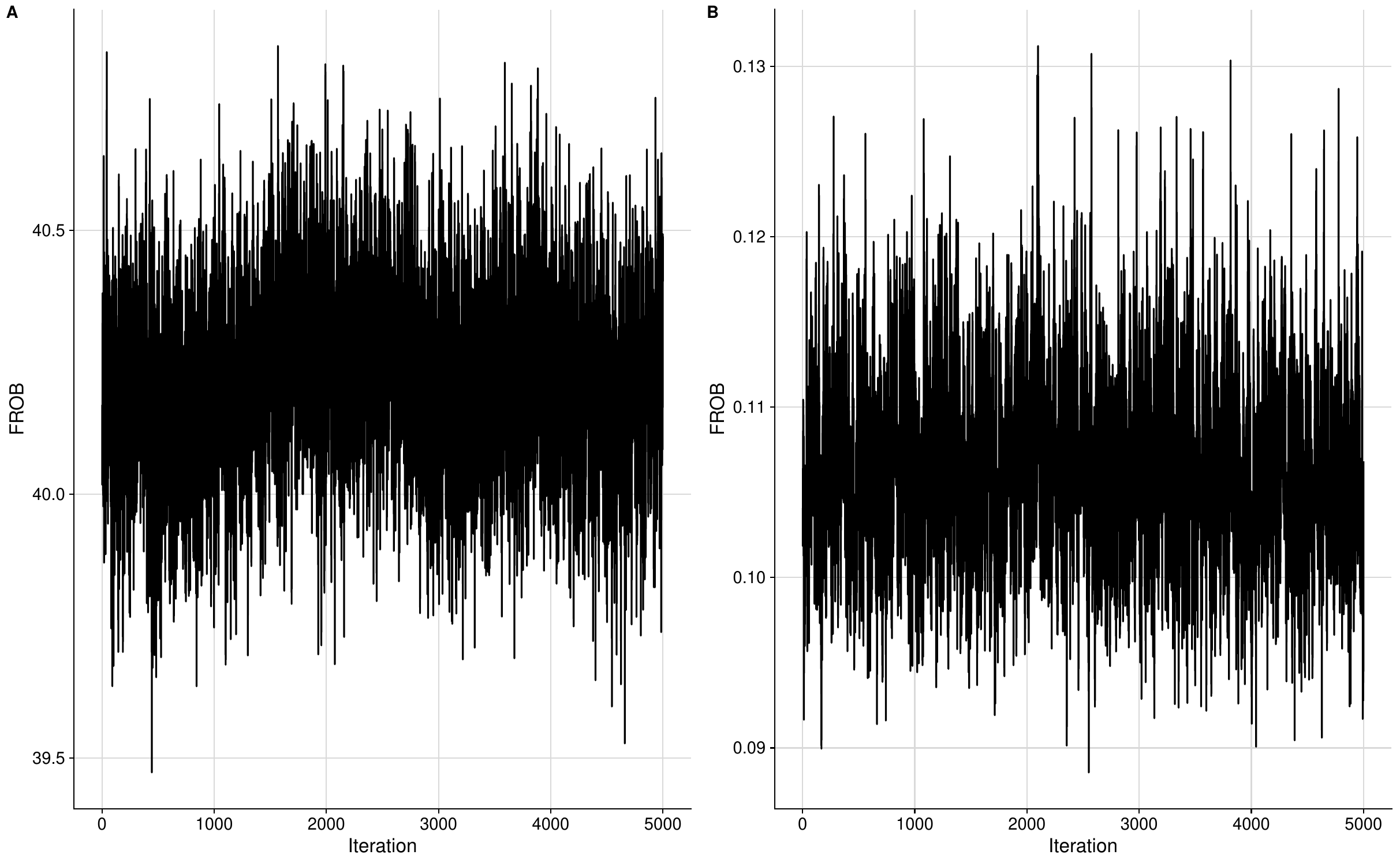}
    \caption{Traceplots of the Frobenius norm defined in \eqref{eq:frob_vartheta} under the ZANIM-LN-BART model for the \textbf{A}: pollen-climate application of Section \ref{sec:application} and \textbf{B}: microbiome application of Section \ref{supp:human_gut_microbiome}.}
    \label{fig:trace_applications}
\end{figure}

Finally, we note that satisfactory convergence is also obtained under both real data applications presented in this paper.
Recall that our proposed ZANIM-LN-BART provided the best fit for both the pollen-climate case study in Section \ref{sec:application} and the additional human gut microbiome analysis in Section \ref{supp:human_gut_microbiome}.
Ultimately, this novel model formed the basis of the inferential results reported in each case. For completeness, Figure \ref{fig:trace_applications} shows traceplots of $\operatorname{FROB}\left(\vartheta^{(t)}\right)$ for the ZANIM-LN-BART model under both applications, although we note that all other considered models again exhibited satisfactory convergence in each case.

\end{document}